\documentclass[12pt,a4]{article}
\usepackage{amsthm,amsmath,latexsym,amssymb,amsfonts,amscd}
\usepackage{graphics,lscape,fancyhdr,array,stmaryrd,euscript}
\usepackage{ifthen,empheq}
\usepackage{sidecap}
\usepackage{ifpdf}
\usepackage{verbatim}
\usepackage{color}
\usepackage{relsize}

\numberwithin{equation}{section}
\usepackage{tikz}

\newcommand{\TikzRect}[2]{\filldraw[color=black,fill=red]  (#1-\R,#2-\R) rectangle (#1+\R,#2+\R);}

\newcommand{\UnfoldedEqFieldMap}{\begin{figure}[h!]
\caption{\it  A picture from \cite{Didenko:2014dwa} showing the position of various fields and their mixing via unfolded equations. The coordinates are the number of undotted/dotted indices carried by a field. The fields connected by links talk to each other via the unfolded equations \eqref{firstordereqs}.}
\label{UnfoldedFigure}
\begin{tikzpicture}[scale=0.8]
\tikzset{%
  >=latex, 
  inner sep=0pt,%
  outer sep=2pt,%
  mark coordinate/.style={inner sep=0pt,outer sep=0pt,minimum size=3pt,
    fill=black,circle}%
}
\def\R{0.15}
\def\mar{0.15}

\draw[->,black,thick] (0,0) -- (0,8) node[left]{dotted};
\draw[->,black,thick] (0,0) -- (8,0) node[right]{undotted} ;

\filldraw[color=black,fill=green]  (9,6) circle (\R);
\node[right] at (9.3,6) {gauge module, one-forms, $\omega$};

\filldraw[color=black,fill=red]  (9-\R,7-\R) rectangle (9+\R,7+\R);
\node[right] at (9.3,7) {Weyl module, zero-forms, $C$};

\filldraw[color=black,fill=green]  (4,0) circle (\R);
\filldraw[color=black,fill=green]  (3,1) circle (\R);
\filldraw[color=black,fill=green]  (2,2) circle (\R);
\filldraw[color=black,fill=green]  (1,3) circle (\R);
\filldraw[color=black,fill=green]  (0,4) circle (\R);
\draw[green,thick] (4,0) -- (0,4);

    \draw[-latex,thick] (3.5,3.5) node[above,scale=1.0]{$\omega^{\alpha(s-1),\dot{\alpha}(s-1)}$}
        to[out=-90,in=45] (2+\mar, 2+\mar);

    \draw[-latex,thick] (.8,.9) node[below,scale=1.0]{$\Fron_{\mm(s)}$}
        to[out=90,in=-135] (2-\mar, 2-\mar);

    \draw[-latex,thick] (4,6) node[right,scale=1.0]{$C^{\dot{\alpha}(2s)}$}
        to[out=180,in=0] (\mar, 5);

    \draw[-latex,thick] (-2,6) node[above,scale=1.0]{$\omega^{\dot{\alpha}(2s-2)}$}
        to[out=-90,in=180] (-\mar, 4);

    \draw[-latex,thick] (6,4) node[above,scale=1.0]{$C^{\alpha(2s)}$}
        to[out=-90,in=90] (5, \mar);

    \draw[-latex,thick] (6,-2) node[right,scale=1.0]{$\omega^{{\alpha}(2s-2)}$}
        to[out=180,in=-90] (4, -\mar);

\TikzRect{0}{5}
\TikzRect{1}{6}
\TikzRect{2}{7}
\TikzRect{3}{8}
\draw[red,thick] (5,0) -- (9,4);

\TikzRect{5}{0}
\TikzRect{6}{1}
\TikzRect{7}{2}
\TikzRect{8}{3}
\draw[red,thick] (0,5) -- (4,9);

    \draw[-latex,thick] (-1.5,-1.5)  to[out=20,in=-90] (4.5, -\R);
    \draw[-latex,thick] (-1.5,-1.5)  to[out=70,in=180] (-\R, 4.5);

\filldraw[color=black,fill=black]  (-\R,4+2*\R) rectangle (\R,5-2*\R);
\filldraw[color=black,fill=black]  (4+2*\R,-\R) rectangle (5-2*\R,\R);

\node[below] at (-1.5,-1.5) {\large $\mathcal{V}(\Omega,\Omega,C)$};

\end{tikzpicture}
\end{figure}}

\newcommand{\putfigureifpdf}[1]{%
\ifpdf%
{#1}%
\else%
\fi%
}%
\usepackage{hyperref}

\ifpdf%
\usepackage[%
	sorting=none,%
	maxnames=99,%
	firstinits=true,%
	doi=false,%
	citestyle=numeric-comp,%
	backend=bibtex%
	]{biblatex}%
\AtEveryBibitem{\clearfield{title}}%
\renewbibmacro{in:}{}%
\else%
\fi%

\newcommand{\pl}{\partial}

\newcommand{\be}{\begin{equation}}
\newcommand{\ee}{\end{equation}}

\newcommand{\mm}{{\ensuremath{\underline{m}}}}
\newcommand{\nn}{{\ensuremath{\underline{n}}}}

\newcommand{\kk}{{\ensuremath{\underline{k}}}}

\newcommand{\aA}{{\ensuremath{\mathcal{A}}}}
\newcommand{\aB}{{\ensuremath{\mathcal{B}}}}

\newcommand{\gad}{{\dot{\alpha}}}
\newcommand{\gbd}{{\dot{\beta}}}
\newcommand{\gdd}{{\dot{\gamma}}}

\newcommand{\gmd}{{\dot{\mu}}}
\newcommand{\gnd}{{\dot{\nu}}}
\newcommand{\pib}{{\bar{\pi}}}
\newcommand{\ga}{\alpha}
\newcommand{\gb}{\beta}
\newcommand{\gc}{\gamma}

\newcommand{\tad}{{\widetilde{\text{ad}}}}
\newcommand{\ad}{{{\text{ad}}}}
\newcommand{\adD}{{\mathsf{D}}}
\newcommand{\tadD}{\widetilde{\mathsf{D}}}

\newcommand{\DO}{{D^{\rm{yz}}}}
\newcommand{\tadDO}{{\widetilde{D}^{\rm{yz}}}}
\newcommand{\homo}[2]{{\Gamma_{#1}\left\langle#2\right\rangle}}

\newcommand{\fud}[2]{{}^{#1}{}_{#2}\,}
\newcommand{\fdu}[2]{{}_{#1}{}^{#2}\,}

\newcommand{\fdudu}[4]{{}_{#1}{}^{#2}{}_{#3}{}^{#4}\,}

\newcommand{\hh}{{\hat{{h}}}}
\newcommand{\bry}{{{\bar{y}}}}
\newcommand{\brz}{{{\bar{z}}}}
\newcommand{\bru}{{{\bar{u}}}}
\newcommand{\brv}{{{\bar{v}}}}

\newcommand{\brxi}{{{\bar{\xi}}}}

\newcommand{\breta}{{{\bar{\eta}}}}
\newcommand{\brzeta}{{{\bar{\zeta}}}}
\newcommand{\bL}{\bar{L}}

\newcommand{\bvarpi}{{{\bar{\varpi}}}}

\newcommand{\klein}{{\varkappa}}
\newcommand{\brklein}{{\bar{\varkappa}}}

\newcommand{\funat}[2]{\left.{#1}\rule{0pt}{12pt}\right|_{#2}}

\newcommand{\besubeqs}{\begin{subequations}}
\newcommand{\esubeqs}{\end{subequations}}

\newcommand{\brJ}{\bar{J}}
\newcommand{\brN}{\bar{N}}

\newcommand{\formJ}{{\boldsymbol{\mathsf{J}}}}
\newcommand{\formL}{{\boldsymbol{\mathsf{L}}}}
\newcommand{\formK}{{\boldsymbol{\mathsf{K}}}}
\newcommand{\formU}{{\boldsymbol{\mathsf{U}}}}

\newcommand{\formP}{{\boldsymbol{\mathsf{P}}}}
\newcommand{\formj}{{\boldsymbol{\mathsf{j}}}}

\newcommand{\formR}{{\boldsymbol{\mathsf{R}}}}
\newcommand{\compR}{{\mathsf{R}}}

\newcommand{\compJ}{{\mathsf{J}}}
\newcommand{\compj}{{\mathsf{j}}}
\newcommand{\compK}{{\mathsf{K}}}

\newcommand{\omegatwo}{\omega^{(2)}}
\newcommand{\Ctwo}{C^{(2)}}
\newcommand{\Frontwo}{\Phi^{(2)}}
\newcommand{\Fron}{{\Phi}}
\newcommand{\Front}{{\phi}}
\newcommand{\xitwo}{\xi^{(2)}}

\newcommand{\Bone}{{B^{(1)}}}
\newcommand{\Aone}{{\aA}}
\newcommand{\Wone}{{W^{(1)}}}
\newcommand{\Btwo}{{B^{(2)}}}

\newcommand{\Wtwo}{{W^{(2)}}}
\newcommand{\Mone}{{M}}
\newcommand{\Mtwo}{{M^{(2)}}}

\newcommand{\Cxi}{{\xi}}
\newcommand{\Ceta}{{\eta}}

\begin{document}
\hfill
\vskip 0.1\textheight
\begin{center}

{\Large\bfseries Higher Spin Interactions in Four Dimensions: \\
\vspace{0.3cm}
Vasiliev vs. Fronsdal} \\

\vskip 0.1\textheight

Nicolas \textsc{Boulanger},${}^1$ Pan \textsc{Kessel},${}^2$ Evgeny \textsc{Skvortsov},${}^{2,3}$ Massimo \textsc{Taronna},${}^{2}$

\vspace{2cm}

{\em ${}^{1}$ Service de M\'ecanique et Gravitation, Universit\'e de Mons --- UMONS\\ 20 Place du Parc, B-7000 Mons, Belgium}\\
\vspace*{5pt}
{\em ${}^{2}$ Albert Einstein Institute, Am M\"{u}hlenberg 1, Golm, D-14476 Germany}\\
\vspace*{5pt}
{\em ${}^{3}$ Lebedev Institute of Physics, Leninsky ave. 53, 119991 Moscow, Russia}

\vskip 0.05\textheight

{\footnotesize \texttt{nicolasboul@gmail.com; pan.kessel, evgeny.skvortsov, massimo.taronna@aei.mpg.de}}

\vskip 0.05\textheight

	{\bf Abstract }

\end{center}
\begin{quotation}
\noindent
We consider four-dimensional Higher-Spin Theory at the first nontrivial order corresponding to the cubic action. All Higher-Spin interaction vertices are explicitly obtained from Vasiliev's equations. In particular, we obtain
the vertices that are not determined solely by the Higher-Spin algebra structure constants. The dictionary between the Fronsdal fields and Higher-Spin connections is found  and the corrections to the Fronsdal equations are derived. These corrections turn out to involve derivatives of arbitrary order. We observe that the vertices not determined by the Higher-Spin algebra produce naked infinities, when decomposed into the minimal derivative vertices and improvements. Therefore, standard methods can only be used to check a rather limited number of correlation functions within the HS AdS/CFT duality. A possible resolution of the puzzle is discussed.
\end{quotation}

\newpage

\tableofcontents

\section{Introduction}

Our study is motivated by the Higher-Spin AdS/CFT duality \cite{Sezgin:2002rt, Klebanov:2002ja} that identifies the $4d$ Higher-Spin (HS) theory \cite{Vasiliev:1990en, Vasiliev:1999ba} as an AdS/CFT dual of both free and critical vector models, depending on the boundary conditions imposed on the scalar field of the HS multiplet.

Remarkable tests of the duality from the bulk side were performed in \cite{Giombi:2009wh, Giombi:2010vg, Giombi:2012he,Giombi:2013fka, Tseytlin:2013jya, Giombi:2014yra, Giombi:2014iua,Beccaria:2014xda,Beccaria:2014zma,Beccaria:2015vaa}. However, all these highly nontrivial tests are based either on the particle content of the HS algebra multiplet \cite{Giombi:2013fka, Tseytlin:2013jya, Giombi:2014yra, Giombi:2014iua,Beccaria:2014xda,Beccaria:2014zma,Beccaria:2015vaa} or on the structure constants of the HS algebra \cite{Giombi:2009wh, Giombi:2010vg,Giombi:2012he} and therefore do not rely on the existence of a fully non-linear theory. As we explain in Section \ref{sec:GeneralStructure}, the HS algebra determines only a part of the interaction vertices, but there are also other vertices that contribute already to the three-point functions. These are the vertices that have led to formally divergent results in \cite{Giombi:2009wh, Giombi:2010vg}. Therefore, it is important to study the bulk physics that extends beyond pure symmetry arguments and to resolve these puzzles as well.

As a first step it is necessary to work out the Higher-Spin theory to the first nontrivial order that corresponds to a cubic action or equations of motion with quadratic corrections due to interactions around an $AdS$ background. In the present paper, for the description of interacting HS fields we consider the Vasiliev equations \cite{Vasiliev:1990en, Vasiliev:1999ba} that provide a set of first-order differential equations on a space extended by certain additional, non-commutative, variables. By solving the differential equations with respect to these additional variables one can derive the equations of motion for HS fields including gravity, propagating on an AdS background. This is what we do in the present paper, see \cite{Sezgin:2002ru,Kristiansson:2003xx,Sezgin:2003pt,Giombi:2009wh, Giombi:2010vg,Giombi:2012he} for related results.\footnote{In \cite{Kristiansson:2003xx} the authors attempted to derive the stress-energy tensor built out of the scalar field that sources Einstein equations at the second order in perturbations around $AdS_4$. We find that this expression is incorrect. We are grateful to Per Sundell for early discussions about this issue. }

The equations for HS fields that come out of Vasiliev Theory are naturally expressed in the form of unfolded equations \cite{Vasiliev:1988xc, Vasiliev:1988sa} in terms of two master fields --- a gauge connection $\omega$ and a (twisted)-adjoint matter field $C$ of the HS algebra. These two fields encode a real scalar field together with infinite tower of Fronsdal fields \cite{Fronsdal:1978rb} $\Fron_{\mm_1...\mm_s}$ that are gauge fields with transformations given by
\begin{align}
\delta \Fron_{\mm_1...\mm_s}&=\nabla_{\mm_1} \xi_{\mm_2...\mm_s}+\text{permutations}\,,
\end{align}
as well as the on-shell nontrivial derivatives thereof of arbitrary order. They are used to construct interaction vertices, which we extract in this work. Furthermore, the unfolded equations contain the equations for $\Fron_{\mm_1...\mm_s}$ and the differential consequences thereof that the derivatives of $\Fron_{\mm_1...\mm_s}$  have to obey. Therefore, in terms of $\omega$ and $C$ the equations involve an infinite number of derivatives by construction. In order to know if Vasiliev's equations describe a local field theory, one needs to work out the equations for the Fronsdal fields $\Fron_{\mm_1...\mm_s}$. To this effect one has to solve for all the auxiliary components in $\omega$ and $C$, projecting finally onto the Fronsdal equations. We will elaborate on this step in significant detail. Among other things, we will observe that one should not identify linearized HS Weyl tensors with $C$ beyond the free level.

Having obtained the equations of motion for the physical fields  $\Fron_{\mm(s)}\equiv\Fron_{\mm_1...\mm_s}$, one observes that they are explicitly pseudo-local for fixed spins, i.e. they display infinite series of terms with unbounded number of derivatives, compensated by appropriate negative powers of the cosmological constant $\Lambda$:
\begin{align}
\square \Fron_{\mm(s)}+...&=\sum_{k,l} a_{k,l}\Lambda^{-l}\nabla_{\mm(s-k)}\nabla_{\nn(l)} \Fron\, \nabla_{\mm(k)}\nabla^{\nn(l)} \Fron+...\,.
\end{align}
Above we sketched the contribution of the scalar field to the right-hand side of the spin-$s$ equations.

At least at the level of the cubic action it is known that given three spins there is a finite number of cubic vertices with the number of derivatives bounded by the sum of the spins \cite{Metsaev:2005ar}. All vertices with more than the necessary number of derivatives can be reduced to the {\it standard} ones removing higher-derivative improvements with a field-redefinition.

Therefore, in order to bring the equations into a standard form, i.e. with the lowest possible number of derivatives, certain field redefinitions are needed. Such redefinitions are necessary for those vertices that are not determined directly by the HS algebra. Another interpretation is that any bulk coupling can be decomposed into its standard part and improvements and we can determine the coefficient of the standard part. One of the most important conclusions of the present paper is that in doing so one accumulates an infinite prefactor in front of the standard vertices. This possibility was mentioned in \cite{Boulanger:2008tg} and further discussed in \cite{Bekaert:2010hw}. Also, this result is in accordance with the observations in \cite{Giombi:2009wh} and with the infinity appearing in \cite{Giombi:2010vg}. As such our result is first of all about the theory in the bulk. Let us stress however that the appearance of this infinity prevents one from applying standard AdS/CFT methods since it is the coefficient of the bulk vertex that diverges rather than the value of the integral evaluated on boundary-to-bulk propagators. In particular, only a small subset of the correlation functions that are entirely fixed by the HS algebra can be rigorously derived, which was done in \cite{Giombi:2009wh,Giombi:2012he}.  In the Discussion Section we comment on the origin of these infinities and on possible resolutions.
\vspace{0.2cm}
\noindent Summarizing, our results are the following:
\begin{itemize}
\item The unfolded equations for higher-spin fields are obtained from the Vasiliev equations at the second order in weak field expansion;

\item The dictionary between the unfolded equations and Fronsdal's formulation is worked out. A subtlety is found for the HS Weyl tensors, which results in additional interaction vertices on top of those considered in \cite{Giombi:2009wh}. The HS stress-tensors are found to be pseudo-local, i.e. containing an unbounded number of derivatives;

\item Having obtained the second-order (in weak field expansion) corrections to the free Fronsdal equations, we perform the required redefinitions in order to reduce the pseudo-local expressions to local ones. As it turns out, the degree of pseudo-locality of the equations leads to divergent coefficients for the vertices.
\end{itemize}

The paper is self-contained but we do not provide a detailed review of Vasiliev's HS theories. For a comprehensive review of the $4d$ Vasiliev theory and the unfolded approach we refer to \cite{Vasiliev:1999ba, Didenko:2014dwa}.

The outline of the paper is as follows. In Section \ref{sec:GeneralStructure} we briefly review two approaches to the HS problem: Fronsdal's equations with a still unknown nonlinear completion, and unfolded equations provided by Vasiliev Theory. In Section \ref{sec:SecondOrder} we present the unfolded equations for HS fields at the second order in perturbation theory and discuss briefly the properties that can be seen already there. In Section \ref{sec:Dictionary} we relate our results on the unfolded equations to the Fronsdal equations. In particular we derive the second-order corrections to the latter. Section \ref{sec:Resummation} addresses the problem of locality of the Fronsdal equations as obtained from Vasiliev's equations. Perturbation theory of Vasiliev's equations is summarized in Section \ref{sec:SolvingVasiliev}. The discussion of the results and the conclusions are in Section \ref{sec:Conclusions}. More technical aspects of our calculation can be found in the appendices together with a summary of our notation.

\section{General Structure of Higher-Spin Theories}\label{sec:GeneralStructure}

In this section we briefly review the Fronsdal approach \cite{Fronsdal:1978rb}, the unfolded approach \cite{Vasiliev:1988xc, Vasiliev:1988sa} and specialize the latter to the HS theory of interest, which is the $4d$ bosonic HS theory \cite{Vasiliev:1992av}. The bosonic theory is the simplest one and there are also super-symmetric extensions, see \cite{Sezgin:2012ag} for a review. Then, we discuss how the free HS fields can be described in the unfolded approach and end by sketching the general structure of interaction vertices that appear at second order in perturbation theory.

The theory that we are looking for should provide a nonlinear completion for free HS fields. These can be described by the Fronsdal fields\footnote{The summary of our notation can be found in Appendix \ref{app:notation}.} $\Fron_{\mm_1...\mm_s}\equiv\Fron_{\mm(s)}$, whose free propagation in $AdS_d$ with cosmological constant $\Lambda$ is governed by linear equations:
\begin{align}\label{FronsdalEq}
F_{\mm(s)}&=\Box \Fron_{\mm(s)}-\nabla_{\mm}\nabla^\nn\Fron_{\nn\mm(s-1)}+\frac12\nabla_{\mm}\nabla_{\mm}\Fron^{\nn}{}_{\nn\mm(s-2)}-m_s^2\Fron_{\mm(s)}+2\Lambda g_{\mm\mm}\Fron\fdu{\mm(s-2)\nn}{\nn}=0\,,
\end{align}
with $m_s^2=-\Lambda((d+s-3)(s-2)-s)$. The Fronsdal equations enjoy the gauge symmetry
\begin{align}
\delta\Fron_{\mm(s)}&=\nabla_\mm\epsilon_{\mm(s-1)}\,.\label{freesymm}
\end{align}
The Fronsdal field and the gauge parameter obey certain algebraic constraints:
\begin{align}
\Fron\fud{\nn\kk}{\nn\kk\mm(s-4)}\equiv0\,,&& \epsilon\fud{\nn}{\nn\mm(s-3)}\equiv0\,.
\end{align}
The gauge symmetry \eqref{freesymm} should be a free theory limit of some non-abelian HS symmetry, likewise \eqref{freesymm} is the free limit of diffeomorphisms and Yang-Mills transformations for $s=2$ and $s=1$, respectively.
Interactions are known to require an infinite multiplet of fields of all integer spins \cite{Vasiliev:1999ba}, $s=0,1,2,3,4,...$.\footnote{There is also a truncation to even spins, which still keeps the multiplet infinite.}

\paragraph{Unfolded equations.} The only form in which examples of interacting HS theories are known at present is the unfolded formulation \cite{Vasiliev:1988xc, Vasiliev:1988sa}, that perfectly captures the peculiarity of HS theories to have the largest symmetry possible. Unfolded equations are of the form
\begin{align}
&dW^\aA=F^{\aA}(W)\,, && F^{\aA}(W)=\sum_k F^\aA_{\aB_1...\aB_k}W^{\aB_1}\wedge ...\wedge W^{\aB_k}\label{Unfolded}\,,
\end{align}
where $d=dx^\mu \frac{\pl}{\pl x^\mu}$ is the exterior derivative, $\wedge$ is the exterior product and the fields $W^\aA(x)$ are a set of differential forms of various degrees, labeled by $\aA$. The Taylor coefficients $F^\aA_{\aB_1...\aB_k}$ are the structure constants of the unfolded system and are space-time independent. Any set of PDE's can be brought into this form possibly by introducing auxiliary fields. Let us furthermore note that unfolded equations are automatically diffeomorphism invariant as they are expressed in terms of differential forms.  The system is required to obey the Frobenius integrability condition, i.e. be formally consistent with $dd\equiv0$, which results in algebraic constraints that are quadratic in the structure constants $F^\aA_{\aB_1...\aB_k}$:
\begin{align}\label{GrandJacobi}
0\equiv ddW^\aA&=dF^{\aA}(W)=dW^\aB \wedge\frac{\overrightarrow{\pl} F^\aA(W)}{\pl W^\aB} && \Longrightarrow&& F^\aB\wedge\frac{\overrightarrow{\pl} F^\aA(W)}{\pl W^\aB}\equiv0\,.
\end{align}
As a consequence of integrability, the unfolded equations possess gauge symmetries
\begin{align}
\delta W^\aA=d\xi^\aA+\xi^\aB\frac{\overrightarrow{\pl} F^\aA(W)}{\pl W^\aB}\label{unfldgaugesymmetry}\,,
\end{align}
with gauge parameters $\xi^\aA$ taking their values in the same space as $W^\aA$ but
having a form degree of one unit less than that of $W^\aA$. Zero-forms do not have their own gauge parameters and in their gauge variation the term of the form $d\xi^\aA$ is absent. A simple example of unfolded equations is given by flatness condition $dA=\frac12[A,A]$ for a connection $A$ of some Lie algebra. Then, the integrability constraint is the Jacobi identity.

\paragraph{HS algebra.} For HS theories involving only totally-symmetric Fronsdal fields,  $\Fron_{\mm(s)}$ in the metric-like formulation, the set of $W^\aA$ consists of one-forms $\omega$ and zero-forms $C$. The index $\aA$ runs over an infinite-dimensional set and will be omitted hereafter. Both $\omega$ and $C$ take values in the HS algebra that is the global symmetry algebra of a free conformal boson on the boundary of $AdS$. In the case of four-dimensional $AdS$ space, the HS algebra \cite{Vasiliev:1986qx} is simply the Weyl algebra with two pairs of canonical variables or the algebra of  operators acting on the phase-space of the two-dimensional harmonic oscillator. This simplification is achieved due to the isomorphism $so(3,2)\simeq sp(4,\mathbb{R})$.

To realize the HS algebra one takes a quartet $\hat{Y}^A$, $A=1,...,4$ of operators obeying the canonical commutation relations
\begin{align}
[\hat{Y}^A,\hat{Y}^B]&=2iC^{AB}\,,
\end{align}
with $C^{AB}$ being the $sp(4)$ invariant tensor (charge conjugation matrix). Then, the bilinears deliver an oscillator realization of $sp(4)$:
\begin{align}
T^{AB}&=-\frac{i}4\{\hat{Y}^A,\hat{Y}^B\}\,, &&[T^{AB},T^{CD}]=T^{AD}C^{BC}+\text{3 more}\,.
\end{align}
The relevant HS algebra is defined as the algebra of all (even) functions $f(\hat{Y})$ in $\hat{Y}^A$. It is an associative algebra and the product is conveniently realized by the Moyal star-product:
\begin{align}\label{expform}
(f\star g)(Y)=\exp i\big({\overleftarrow{\pl}_A C^{AB} \overrightarrow{\pl}_B}\big)\,,
\end{align}
where we passed from the operators $\hat{Y}^A$ to their symbols $Y^A$, which are ordinary commuting variables to be multiplied by the Moyal product. In most cases the integral representation of the star-product is more useful\footnote{Symplectic indices are raised and lowered as $Y^A=C^{AB}Y_B$, $Y_B=Y^AC_{AB}$. }
\begin{equation}\label{Ystarproduct}
(f\star g)(Y)=\int dU dV f(Y+U) g(Y+V) e^{iU_A V^A}\,.
\end{equation}
The exponential formula \eqref{expform} can be derived by integrating \eqref{Ystarproduct} by parts and dropping the boundary terms.

Therefore, the HS theory operates in terms of fields $\omega=\omega_\mm(Y|x)\, dx^\mm$ and $C=C(Y|x)$ that are even in $Y^A$. We need to provide the structure constants $F^\aA_{\aB_1...\aB_k}$ of \eqref{Unfolded} in order to build the equations of motion.

\paragraph{Unfolded equations of HS theories.} The unfolded equations take the following form \cite{Vasiliev:1988sa}:
\besubeqs
\label{xpaceseq}
\begin{align}
&d\omega=F^{\omega}(\omega,C)\label{xpaceseqXA}\,,\\
&dC=F^C(\omega, C)\,,\label{xpaceseqXB}
\end{align}
\esubeqs
where the power of $\omega$ is fixed by the form degree and the structure functions $F^{\omega,C}$ admit an expansion in powers of the zero-forms $C$
\besubeqs
\label{Cexpansion}
\begin{align}
&F^{\omega}(\omega,C)=\mathcal{V}(\omega,\omega)+\mathcal{V}(\omega,\omega,C)+
\mathcal{V}(\omega,\omega,C,C)+...\,,\\
&F^C(\omega, C)=\mathcal{V}(\omega,C)+\mathcal{V}(\omega,C,C)+\mathcal{V}(\omega,C,C,C)+...\,.
\end{align}
\esubeqs
The first vertices provide the initial data for the deformation problem and are fixed by the HS algebra
\begin{align}\label{trivialcocycles}
\mathcal{V}(\omega,\omega)&=\omega\star\omega\,,
&\mathcal{V}(\omega, C)&=\omega\star C-C\star \pi(\omega)\,,
\end{align}
where $\star$ denotes the (associative) product in the relevant HS algebra and $\pi$ is an automorphism of the HS algebra that is induced by the reflection of $AdS$ translation generators $P_a\rightarrow -P_a$.

\paragraph{HS algebra vs. Dynamics.} The problem of classifying vertices is a cohomological problem, i.e. consistent deformations modulo trivial ones induced by field-redefinitions. For example, if we drop the $\pi$-map, the vertex
\begin{align}
\mathcal{V}(\omega,\omega,C)&=\omega\star \omega\star C
\end{align}
is consistent, but it is induced by a redefinition $\omega\rightarrow\omega+\omega\star C$. All consistent vertices expressed as products in the HS algebra are induced by such redefinitions except for \eqref{trivialcocycles}. The question of whether star-product polynomials of fields correspond to admissible redefinitions is subtle since, for instance, $C\star C$ is pseudo-local and therefore contains an infinite number of derivatives. As we will show, the interactions that are very close to the HS algebra product $C\star C$ do not admit an interpretation in terms of local field theory vertices in AdS, see \cite{ST} for a more detailed discussion. In any case, nontrivial higher vertices cannot be written as pure star-products of the fields.

Given some algebra the vertices \eqref{trivialcocycles} exist and are consistent on their own, but the existence of higher vertices is not guaranteed.\footnote{Many different HS algebras and truncations thereof were found in \cite{Vasiliev:1986qx}. However, it turned out that only few of those admit higher vertices \eqref{Cexpansion} and the rest are obstructed. For example, while one can try to define some HS algebra in flat space, its deformations are obstructed, which follows from the fact that the limit $\Lambda\rightarrow0$ is singular in Vasiliev's equations, but is well-defined at the level of HS algebra. The reason is that while the $AdS$ algebra can be contracted to the Poincar\'{e} algebra by sending $\Lambda\rightarrow0$, the corresponding unfolded equations may not admit such a contraction. Moreover, under certain assumptions one can prove that the HS algebra is essentially unique \cite{Boulanger:2013zza}.} Therefore, tests of the AdS/CFT duality have to involve \eqref{Cexpansion} beyond \eqref{trivialcocycles}, although for some AdS/CFT computations it can be sufficient to consider the spectrum of the free theory or to use the structure constants of the HS algebra, for instance, via \eqref{trivialcocycles}. The computations based on \eqref{trivialcocycles} are highly nontrivial \cite{Giombi:2009wh,Giombi:2012he} and are by no means guaranteed to give the correlators compatible with AdS/CFT. At present, all computations beyond \eqref{trivialcocycles} have faced the difficulty of getting naked infinities, \cite{Giombi:2009wh, Giombi:2010vg}. To be precise, it is $\mathcal{V}(\omega,C,C)$ that was found to be problematic in \cite{Giombi:2009wh, Giombi:2010vg}, but as we will show the same is true for all the vertices that are at least bilinears in $C$. We will analyze this issue in Section \ref{sec:Resummation}.

\paragraph{AdS as an HS background.} The unfolded system \eqref{xpaceseq} has a natural family of vacua given by a flat connection $\Omega$ of the HS algebra,
\begin{align}
&d\Omega=\Omega\star\Omega\,,\label{adsbckgrnd}
\end{align}
provided that $C=0$, i.e. all higher vertices vanish and the equation for $C$ trivializes. Any such vacuum has the full HS algebra as a global symmetry, while vacua with non-vanishing $C$ would break the HS symmetry. The simplest vacuum with $C=0$ is $AdS_4$ space:
\begin{align}\label{Omegaconn}
\Omega=\frac12 \varpi^{\ga\ga}L_{\ga\ga}+ h^{\ga\gad} P_{\ga\gad}+\frac12 \varpi^{\gad\gad}\bar{L}_{\gad\gad}\,,
\end{align}
where we split the $sp(4)$ generators $T^{AB}$ into Lorentz generators $L_{\ga\ga}$ and $\bar{L}_{\gad\gad}$ of $sl(2,\mathbb{C})_{\mathbb{R}}$ and translation generators $P_{\ga\gad}$\,:
\begin{align*}
L_{\ga\ga}&=T_{\ga\ga}=-\frac{i}4\{y_\ga,y_\ga\}\,, &&  P_{\ga\gad}= T_{\ga\gad}=-\frac{i}4\{y_\ga,\bry_\gad\} \,, &&\bar{L}_{\gad\gad}=T_{\gad\gad}=-\frac{i}4\{\bry_\gad,\bry_\gad\}\,,
\end{align*}
with $Y_A=(y_\ga,\bry_\gad)$. The (anti)self-dual components of the $4d$ spin-connection are $\varpi^{\ga\ga}=\varpi^{\ga\ga}_\mm\, dx^\mm$ and $\varpi^{\gad\gad}=\varpi^{\gad\gad}_\mm\, dx^\mm$ while $h^{\ga\gad}=h^{\ga\gad}_\mm\, dx^\mm$ is the invertible vierbein of $AdS_4$. This is the background around which we expand \eqref{xpaceseq}.

\paragraph{First order.} The flat connection \eqref{Omegaconn} provides an exact solution $\omega=\Omega$, $C=0$ of the unfolded system \eqref{xpaceseq}, which describes $AdS_4$. Let us consider linearized fluctuations around it  \cite{Vasiliev:1988xc}. To this effect, we expand $\omega\rightarrow \Omega+ \omega$ and pick the part that is linear in $\omega$ and $C$. One finds:
\besubeqs\label{firstordereqs}
\begin{align}
&d\omega=\{\Omega,\omega\}_\star+\mathcal{V}(\Omega,\Omega,C)\label{xpaceseqEA}\,,\\
&dC=\Omega\star C-C\star \pi(\Omega)\label{xpaceseqEB}\,.
\end{align}
\esubeqs
Hereafter we reserve the symbols $\omega$ and $C$ for the first order fluctuations. The appearance of $\mathcal{V}(\Omega,\Omega,C)$ corresponds to a deviation from a flat connection. The components of the curvature that are allowed not to be vanishing are parameterized by $\mathcal{V}(\Omega,\Omega,C)$. Despite the slightly involved structure due to the presence of several ``vertices" the equations are linear and describe free fields propagating in anti-de Sitter space.

The rough structure of \eqref{firstordereqs} is that to each gauge field there corresponds a jet, i.e. the space of all its derivatives. There is an infinite-dimensional subspace of those derivatives that are gauge invariant. In this subspace there is a subspace that is set to zero by the equations of motion. The complement, i.e. the space of all gauge invariant derivatives that are left free after imposing the equations of motion, is parameterized by $C$. It starts with the HS Weyl tensor $C^{a(s),b(s)}$ that is the order-$s$ curl of the Fronsdal field:\footnote{While the Fronsdal field $\Fron_{\mm(s)}$ is naturally a world-tensor, $\omega$ and $C$, when expanded in $Y^A$, correspond to tensors in the tangent space. The transfer between world and tangent tensors is performed by the vierbein $h^a_\mm$, which is $h^{\ga\gad}_\mm$ in the $sl(2,\mathbb{C})$ base, and its inverse. The formulas here-below are sketchy and are meant to indicate the order of derivatives rather than precise expressions. }
\begin{align}
\begin{aligned}
C^{\ga(2s)}&\\
C^{\gad(2s)}
\end{aligned}
&: && C^{a(s),b(s)}\sim[\mbox{anti-symmetrized in $b$ and $a$}]\,\,\nabla^{b_1}...\nabla^{b_s}\Fron^{a_1...a_s} -\text{traces}\,,\label{LinWeylTensors}
\end{align}
where we used that in the spinorial language of $4d$ the HS Weyl tensors decompose into the (anti)self-dual components $C_{\ga(2s)}$ and $C_{\gad(2s)}$ that can be found in the expansion of  $C(y,\bry=0)$ and $C(y=0,\bry)$, respectively. Other components of $C$ contain on-shell nontrivial derivatives of the Weyl tensors:
\begin{align}
\begin{aligned}
C^{\ga(2s+k),\gad(k)}&\\
C^{\ga(k),\gad(2s+k)}
\end{aligned}&: && [\mbox{symmetrized in all $a$}]\,\, \nabla^{a_{s+1}}...\nabla^{a_{s+k}} C^{a_1...a_s,b_1...b_s}-\text{traces}\,, &&k=0,...,\infty\,.
\end{align}
The $C$ field also contains a scalar field, $\Fron_0=C(y=0,\bry=0)$, together with all of its on-shell nontrivial derivatives $C^{\ga(k),\gad(k)}\sim \nabla...\nabla \Fron_0$.

The gauge connection $\omega$ contains the Fronsdal field $\Fron_{\mm(s)}$ that is a totally-symmetric component of the HS vielbein and, in the $4d$ spinorial language, is identified with the component $\omega(y,\bry)$ that has equal number of dotted and undotted indices
\begin{align}\label{FronVielLinear}
\Fron_{\mm(s)}&=\omega^{\ga(s-1),\gad(s-1)}_\mm\, h_{\mm|\ga\gad} ...h_{\mm|\ga\gad}\,.
\end{align}
The rest of $\omega(y,\bry)$ contains some gauge-noninvariant derivatives of the Fronsdal field up to order-$(s-1)$:
\begin{align}
\begin{aligned}
\omega^{\ga(s-1-k),\gad(s-1+k)}&\\
\omega^{\ga(s-1+k),\gad(s-1-k)}
\end{aligned}&: && [\mbox{anti-symmetrized in $b$ and $a$}]\,\,\nabla^{b_1}...\nabla^{b_k}\Fron^{a_1...a_s}\,, &&k=0,...,s-1\,.
\end{align}
This identification works at the free field level only, while there are nonlinear corrections at higher orders. The links between different fields are illustrated by fig. \ref{UnfoldedFigure}. \putfigureifpdf{\UnfoldedEqFieldMap}

The two-form $\mathcal{V}(\Omega,\Omega,C)$ relates the order-$s$ curl of the Fronsdal field to the first Weyl tensor. If it had not been for $\mathcal{V}(\Omega,\Omega,C)$, \eqref{xpaceseqEA} would have been a linearization of the flatness condition for $\Omega+\omega$, which locally has pure gauge solutions only. Another nontrivial fact is that there is Fronsdal's equation hidden inside \eqref{xpaceseqEA} as we will review in detail in Section \ref{subsec:FronsdalCurrents}.

It is useful to rewrite the free equations and gauge transformations as:
\begin{align}
&\adD \omega=\mathcal{V}(\Omega,\Omega,C)\,,\label{xpaceseqQBA}&&\delta \omega=\adD\xi\,,  &
&\tadD C=0\,, && \delta C=0\,,
\end{align}
where we have used the background covariant derivatives $\adD$ and $\tadD$. Note that $C$ is gauge invariant to the lowest order in accordance with  \eqref{unfldgaugesymmetry}. Using the explicit form of the HS algebra product one finds:
\begin{align}
\adD\bullet&=d\bullet -\Omega\star \bullet\pm\bullet\star \Omega=\nabla-h^{\ga\gad}(y_\ga\pl_\gad+\bry_\gad\pl_\ga)\,,\label{AdjointDer}\\
\tadD\bullet&=d\bullet -\Omega\star \bullet\pm\bullet\star \pi(\Omega)=\nabla+ih^{\ga\gad}(y_\ga\bry_\gad-\pl_\ga\pl_\gad)\,,\label{TwistedDer}\\
\nabla&=d-\varpi^{\ga\ga}y_\ga \pl_\ga-\varpi^{\gad\gad}\bry_\gad \pl_\gad\,,\label{LorentzDer}
\end{align}
where $\pm$ accounts for a graded commutator and $\nabla$ is the Lorentz-covariant derivative around $AdS_4$. The vertex that glues $C$ to the $\omega$-equations reads:
\begin{align}\label{OMSTcocycle}
\mathcal{V}(\Omega,\Omega,C)&=-\frac12 H^{\ga\ga}\pl_\ga\pl_\ga C(y,\bry=0)-
\frac12 H^{\gad\gad}\pl_\gad\pl_\gad C(y=0,\bry)\,,
\end{align}
where $H^{\ga\ga}=h\fud{\ga}{\gnd}\wedge h^{\ga\gnd}$, and analogously for $H^{\gad\gad}$. Here we choose a normalization that results from the Vasiliev equations of Section \ref{sec:SolvingVasiliev}. Setting $y=0$ or $\bry=0$ is a way to project onto the HS Weyl tensors. We also note that only the background vierbein $h^{\ga\gad}$ part of $\Omega$ contributes to $\mathcal{V}(\Omega,\Omega,C)$, as it is manifestly Lorentz-covariant.

\paragraph{Second order.}
In the present paper we are interested in the structure of interactions up to the second order, where we expect to find:
\besubeqs\label{xpaceseqBQQ}
\begin{align}
&\adD \omegatwo-\mathcal{V}(\Omega,\Omega,\Ctwo)=\omega\star\omega+
\mathcal{V}(\Omega,\omega,C)+
\mathcal{V}(\Omega,\Omega,C,C)\label{xpaceseqBA}\,,\\
&\tadD \Ctwo=\omega\star C-C\star \pi(\omega)+\mathcal{V}(\Omega,C,C)\label{xpaceseqBB}\,.
\end{align}
\esubeqs
We singled out the operators $\adD$ and $\tadD$ that appear already at the first order and govern the free propagation around $AdS_4$. The equations of motion for the second order fields $\omegatwo$ and $\Ctwo$ are sourced by some ``currents" that are bilinear in the first order fields. If we drop these sources we get free equations \eqref{firstordereqs} for $\omegatwo$ and $\Ctwo$. As was already mentioned, there are two vertices that are expressed solely in terms of the HS algebra structure constants. For example, correlation functions of \cite{Giombi:2009wh, Giombi:2012he} were extracted from $\omega\star C-C\star \pi(\omega)$, which does not require any knowledge of the nonlinearities of the Vasiliev equations. Here, in particular, we intend to determine the full structure at the second order, i.e.    $\mathcal{V}(\Omega,\omega,C)$, $\mathcal{V}(\Omega,\Omega,C,C)$ and $\mathcal{V}(\Omega,C,C)$.

It is also useful to work out the gauge transformations \eqref{unfldgaugesymmetry} to the second order
\besubeqs\label{xpaceseqBQG}
\begin{align}
\delta \omegatwo&=\adD\xitwo -[\omega,\xi]_\star+\xi\frac{\overrightarrow{\pl}}{\pl\omega}\mathcal{V}(\Omega,\omega,C)\label{xpaceseqBAG}\,,\\
\delta \Ctwo&=\xi\star C-C\star \pi(\xi)\label{xpaceseqBBG}\,,
\end{align}
\esubeqs
which includes the deformation $\mathcal{V}(\Omega,\omega,C)$ that goes beyond the HS algebra.

Vasiliev's equations allow us to extract the unfolded equations for the HS theory around $AdS_4$. These equations
are over-determined and over-complete. First of all, the set of variables is over-complete since the only dynamical variables are known to be the Fronsdal fields $\Fron_{\mm(s)}$ that occupy certain components of $\omega$ for $s>0$ and $\Fron_0=C(y=0,\bry=0)$ for $s=0$. The rest of $\omega$ and $C$ are either pure gauge or are expressed as derivatives of the Fronsdal fields by virtue of \eqref{firstordereqs}.

Secondly, the unfolded equations are also over-complete because the only dynamical equations can be chosen to be the Fronsdal equations with sources that are schematically of the form:
\begin{align}\label{onlyequations}
\square \Frontwo_{s}+...=g \sum_{s_1,s_2} f_{s,s_1,s_2}\nabla...\nabla\Fron_{s_1}\nabla...\nabla\Fron_{s_2}\,,
\end{align}
where $\Fron_{s}$ is an index-free abbreviation for $\Fron_{\mm(s)}$. We also introduced by hand $g$ as the unique coupling constant of the HS theory while $f_{s,s_1,s_2}$ are certain numbers (we set $\Lambda=1$ here) in front of all possible bilinears in the Fronsdal fields that are fixed by the theory up to redefinitions. The rest of the equations are differential consequences of \eqref{onlyequations}, i.e. the equations for the derivatives of $\Frontwo_s$.

Equations \eqref{onlyequations} are the same one would find in the metric-like formalism from a hypothetical Lagrangian of the form
\begin{align}
S&=\frac12\sum_s  \int \big[(\nabla\Fron_s)^2+\ldots\big] +\frac{g}{3}\sum_{s,s_1,s_2} b_{s,s_1,s_2} \int (\nabla...\nabla\Fron_{s_1}\nabla...\nabla\Fron_{s_2}\Fron_{s}) +O(g^2)\,,
\end{align}
which we assume to exist for HS theories. The quadratic piece is given by a sum of Fronsdal Lagrangians over all spins. The cubic piece contains a unique coupling constant $g$, while the coefficients in front of all possible cubic couplings are fixed to be certain numbers $b_{s,s_2,s_3}$, which are related to $f_{s,s_1,s_2}$, assuming that \eqref{onlyequations} are integrable.

The same Lagrangian can be rewritten in the frame-like formalism \cite{Vasiliev:1980as} where the variables are HS connection $\omega$, HS Weyl tensors and descendants thereof packed into $C$. This formulation is closer to the unfolded equations since it makes use of the same set of variables. The action, which is partly known, reads
\begin{align}\label{fraction}
S&= \sum_s a_s\int R_s\wedge R_s +g \sum_{s_1,s_2,s_3}\int\big(  d_{s_1,s_2,s_3}C_{s_1}C_{s_2}C_{s_3}+ d'_{s_1,s_2,s_3} C_{s_1}C_{s_2}\omega_{s_3}\big)+ O(g^2)\,,
\\ R&=\adD\omega -g\, \omega\star\wedge\omega\,.\notag
\end{align}
The first $RR$-term is of Yang-Mills type as it makes use of the usual Yang-Mills field-strength $R$ that is expanded over the AdS vacuum. It contains the kinetic term for the HS fields, which is equivalent to Fronsdal Lagrangian upon using the frame-like vs. metric-like dictionary, as well as some of the cubic vertices, which were originally constructed by Fradkin and Vasiliev in \cite{Fradkin:1986qy}. The coefficients $a_s$ are fixed by the Yang-Mills type of gauge symmetry $\delta \omega=D\xi-g[\omega,\xi]_\star$ and are related to the trace on the HS algebra. As opposed to the previous vertices, the $C^3$-vertices are abelian and the coefficients $d_{s_1,s_2,s_3}$ cannot be determined by the first nontrivial deformation $\delta_1 \omega=-g[\omega,\xi]_\star$ of the gauge transformations. One has to proceed to quartic vertices or use the unfolded equations of motion consistent to all orders. Therefore, in order to extract the other coefficients, $d_{s_1,s_2,s_3}$, one can compare the equations coming from \eqref{fraction} with the unfolded ones. Another way of fixing $d_{s_1,s_2,s_3}$ is to compare with the three-point functions in the free boson theory. On the other hand, the vertices $d'_{s_1,s_2,s_3}$ correspond to current interactions, i.e. to a gauge field contracted with a conserved tensor current, and can be fixed by the HS algebra, see \cite{Kessel:2015kna} for the $3d$ example.

The r.h.s of \eqref{xpaceseqBA} can be thought of as HS stress-tensors. They consist of two parts: gauge-invariant and gauge-noninvariant. In the following we will associate $\mathcal{V}(\Omega,\Omega,C,C)$ with the gauge-invariant HS stress-tensors. Indeed, it is bilinear in $C$ that is gauge-invariant to the lowest order. While $\mathcal{V}(\Omega,\omega,C)$ is fixed by the HS symmetry at the given order, $\mathcal{V}(\Omega,\Omega,C,C)$ and $\mathcal{V}(\Omega,C,C)$ are not. This is analogous to the situation with the cubic action \eqref{fraction}, where the abelian part cannot be fixed without having any information about the quartic action.
Vasiliev's equations, which can in principle give the unfolded equations to all orders, yield specific expressions for $\mathcal{V}(\Omega,\Omega,C,C)$ and $\mathcal{V}(\Omega,C,C)$, which we will analyze in significant detail.

\section{Higher-Spin Theory at the Second Order}\label{sec:SecondOrder}
In this section we discuss the explicit results obtained by solving Vasiliev's equations to the second order. Let us display once again the general structure one gets to the second order:
\besubeqs\label{xpaceseqCQ}
\begin{align}
\adD \omegatwo-\mathcal{V}(\Omega,\Omega,\Ctwo)&=\omega\star\omega+
\mathcal{V}(\Omega,\omega,C)+
\mathcal{V}(\Omega,\Omega,C,C)\label{xpaceseqCA}\,,\\
\tadD \Ctwo&=\omega\star C-C\star \pi(\omega)+\mathcal{V}(\Omega,C,C)\,,\label{xpaceseqCB}
\end{align}
\esubeqs
where we isolated on the left-hand side the structures that would govern the free propagation of $\omegatwo$ if there were no sources on the right-hand side. We refer to the source on the right-hand-side of the equations as {\it backreaction}.

We find it convenient to work with Fourier transforms of $\omega$ and $C$:
\begin{equation}
C(Y|x)=C(y,\bar y|x)=\int d^2\xi\, d^2\bar\xi\, e^{iy^\ga\xi_\ga+i\bar y^\gad\bar\xi_\gad}\,C(\xi,\bar\xi|x)\,,
\end{equation}
and analogously for $\omega(Y|x)$. In the following, the $x$-dependence is left implicit. The vertices that are determined by the HS algebra are trivial to evaluate even without knowing Vasiliev's equations\footnote{While $\xi,\brxi$ and $\eta,\breta$ denote the holomorphic and anti-holomorphic parts of the spinors, the full $sp(4)$ spinors $\xi^A,\eta^A$ are also abbreviated by $\xi$ and $\eta$. }
\besubeqs
\begin{align}
\omega\star\omega&= \int  d^4\xi \, d^4\eta\, e^{i((y-\eta)(y+\xi)+(\bry-\breta)(\bry+\brxi))}\omega(\Cxi|x)\,\omega(\Ceta|x)\,,\\
\omega\star C-C\star \pi(\omega)&= \int  d^4\xi \, d^4\eta\, \notag \big(e^{i((y-\eta)(y+\xi)+(\bry-\breta)(\bry+\brxi))} \omega(\Cxi|x) C(\Ceta|x)\\&\qquad\qquad\quad-e^{i((y+\eta)(y+\xi)+(\bry-\breta)(\bry+\brxi))}C(\Cxi|x) \omega(\Ceta|x)\big)\,.
\end{align}
\esubeqs
There is one more term that is easy to evaluate, since it comes from the free equations
\begin{align}\label{omstfourier}
\begin{aligned}
\mathcal{V}(\Omega,\Omega,\Ctwo)&= \int  d^2\xi \, [H^{\gad\gad} T_{\gad\gad}(Y,\brxi)+H^{\ga\ga} T_{\ga\ga}(Y,\xi)]\,\Ctwo(\xi|x)\,, \\
T_{\gad\gad}(Y,\brxi)&=\frac12 \brxi_\gad\brxi_\gad e^{i(\bry\brxi+\theta)}\,,
\qquad\qquad T_{\ga\ga}(Y,\xi)=\frac12 \xi_\ga\xi_\ga e^{i(y\xi-\theta)}\,.
\end{aligned}
\end{align}
There is also a free parameter, $\theta$, which is inherited from Vasiliev's equations. The theory at $\theta=0$ and $\Delta=1$ boundary conditions should be dual to the free boson, \cite{Sezgin:2003pt, Klebanov:2002ja}.
The nontrivial part of the HS interactions resides in the rest of the vertices. The formulas for the vertices may look cumbersome at first glance, but they are relatively simple taking into account that they encode interactions of an infinite multiplet of fields. With details left to Section \ref{sec:SolvingVasiliev}, the Vasiliev equations yield the following expressions for the vertices:
\besubeqs
\begin{align}
\mathcal{V}(\Omega,C,C)&= \int  d^2\xi \, d^2\eta\,  \formK(Y,\xi,\eta)\,C(\xi|x) C(\eta|x)\,,\\
\mathcal{V}(\Omega,\omega,C)&= \int  d^2\xi \, d^2\eta\,  \left(\formL(Y,\xi,\eta)\,\omega(\xi|x) C(\eta|x)+\bar{\formL}(Y,\xi,\eta)\,C(\xi|x) \omega(\eta|x)\right)\,,\\
\mathcal{V}(\Omega,\Omega,C,C)&= \int  d^2\xi \, d^2\eta\, \formJ(Y,\xi,\eta)\,C(\xi|x) C(\eta|x)\,,
\end{align}
\esubeqs
where all the nontrivial information resides in the kernels $\formK$, $\formL$ and $\formJ$,
\besubeqs\label{SimplifiedVertices}
\begin{align}
&\formK(Y,\xi,\eta)=\int_0^1 dt\,\Big(h^{\ga\gad}\left[(\bry_\gad t-(1-t)\brxi_\gad)\eta_\ga R_2-(\bry_\gad t+(1-t)\breta_\gad)\xi_\ga S_1\right]+h.c.\Big)\,,\\
&\formL(Y,\xi,\eta)=\int_0^1 dt\,\Big(h^{\ga\gad}R_1\xi_\ga(\breta_\gad+t\brxi_\gad)+h.c.\Big)\,,\\
&\bar{\formL}(Y,\xi,\eta)=\int_0^1 dt\,\Big(h^{\ga\gad}R_2\eta_\ga(\brxi_\gad+t\breta_\gad)+h.c.\Big)\,,\\
&\begin{aligned}
&\formJ(Y,\xi,\eta)=\int_0^1 dt\, \int_0^1 dq\, \Big(H^{\ga\ga}(y+\xi)_\ga(y+\eta)_\ga Q_1\Big(iq^2t^2+(\brxi\breta)\frac{qt(1-qt)}2\Big)\\
&\qquad\qquad-\frac{i}2 H^{\gad\gad} \brxi_\gad\breta_\gad Q_1+\frac{i}2(1-t) H^{\gad\gad} \brxi_\gad\breta_\gad P_1+\frac{i}2 H^{\gad\gad}\pl_\gad\pl_\gad K_0+h.c.\Big)\,,
\end{aligned}\label{fullJbackreaction}
\end{align}
\esubeqs
while the phases $R_1,R_2,S_1,Q_1,P_1$ and $K_0$ are:
\besubeqs
\begin{align}
R_1&=\exp{i\left( (y(1-t)-t\eta)\xi+(\bry-\breta)(\bry+\brxi)+\theta\right)}\,,\\
R_2&=\exp{i\left( (y(1-t)-t\xi)\eta+(\bry-\breta)(\bry+\brxi)+\theta\right)}\,,\\
S_1&=\exp{i\left( (y(1-t)+t\eta)\xi+(\bry-\breta)(\bry+\brxi)+\theta\right)}\,,\\
Q_1&=\exp{i\left((qt(y+\eta)(y+\xi)+(\bry-\breta)(\bry+\brxi)+2\theta\right)}\,,\\
P_1&=\exp{i\left((t(y+\eta)(y+\xi)+(\bry-\breta)(\bry+\brxi)+2\theta\right)}\,,\\
K_0&=\exp{i\left(t\eta\xi+(\bry-\breta)(\bry+\brxi)+2\theta\right)}\,.
\end{align}
\esubeqs
The meaning of the $h.c.$-operation is to exchange barred and unbarred variables as well as to flip the sign of $\theta$ but not to conjugate any complex numbers. The unfolded equations above can be directly checked to obey the Frobenius integrability constraint \eqref{GrandJacobi}, see Appendix \ref{app:Consistency} for the details.

\subsection{Comments on the Results}\label{subsec:CommentsResults} Despite the technical nature of these results one can still draw some comments at the level of the unfolded equations and then translate all the results in the language of Fronsdal fields, see eq. \eqref{onlyequations}, which will allow us to quantify the locality properties of the unfolded equations written here-above.

In checking the consistency of the unfolded equations one observes how various vertices relate to each other. In particular, one can see that $\formJ$ without the $K_0$-term, which we will denote as $\formJ^{s.t.}$, is $\adD$-conserved and is independent of the other vertices:
\begin{align}\label{HSstBackreaction}
&\begin{aligned}
\formJ^{s.t.}=\int_0^1 dt\, \int_0^1 dq\,& \Big[H^{\ga\ga}(y+\xi)_\ga(y+\eta)_\ga Q_1\Big(iq^2t^2+(\brxi\breta)\frac{qt(1-qt)}2\Big)+\\
&\qquad\qquad-\frac{i}2 H^{\gad\gad} \brxi_\gad\breta_\gad \Big(Q_1-(1-t) P_1\Big)+h.c.\Big]\,,\\
\adD\formJ^{s.t.}&\approx0\,,
\end{aligned}
\end{align}
where $\approx$ means that the conservation holds on-shell --- $\tadD C=0$. Therefore, it is tempting to refer to $\formJ^{s.t.}$, which is gauge-invariant by itself, as stress-tensors. In particular, the contribution of such stress-tensors to the correlation functions can be separated from the rest. Moreover, the scalar field appears only in $\formJ^{s.t.}$ since $\omega$ does not contain the $s=0$ component of the HS multiplet. Therefore, by \eqref{xpaceseqCA} the $s-0-0$ correlation function is accounted for solely by $\formJ^{s.t.}$.

It is tempting to associate $\Ctwo$ with the HS Weyl tensors\footnote{One can try to generalize the definition of HS Weyl tensors to the interaction level, but unlike in gravity for which there exists a nonlinear analog of the usual Weyl tensor, there is no simple way to do so. For example, any definition will be affected by field-redefinitions, which are naturally absent in the gravity case, in the sense that there is no uniquely defined metric tensor in Vasiliev Theory. Therefore, we prefer to talk about {\it linearized HS Weyl tensors} that are given by $s$ curls of the Fronsdal field and loosely refer to $\Ctwo$ as HS Weyl tensors sometimes.} for the Fronsdal fields $\Frontwo$ that sit inside $\omegatwo$ as was the case at the free level, but there is a subtlety. The $K_0$-term contributes to exactly the same component of the equations as the so-called Weyl tensors $\Ctwo$ do. Moreover, other terms on the right-hand-side of \eqref{HSstBackreaction} also have a non-vanishing projection onto the same component, building up a nonlinear deformation of $\Ctwo$. Therefore, one has to redefine $\Ctwo$ so as to eliminate all other contributions to the linearized Weyl tensors. Such a redefinition, when substituted into the equation for $\Ctwo$, will produce additional vertices in \eqref{xpaceseqCB}. This observation may be important for the method of extracting correlators from the boundary behavior of $\Ctwo$ developed in \cite{Giombi:2009wh} where the free level relation between $\Frontwo$ and $\Ctwo$ was implied. We will give more details on the redefinition in Section \ref{subsec:WeylTensors} devoted to HS Weyl tensors.

Technically, the expressions for the vertices in the $4d$ theory are much simpler than those in the three-dimensional case considered in \cite{Kessel:2015kna}. The reason is two-fold. Firstly, the $3d$ theory contains additional fields that have to be truncated away order by order in perturbations. Such redefinitions yield additional vertices. Secondly, the $4d$ theory does not produce prefactors quartic in $Y,\xi,\eta$ at the second-order.

As is also evident from Vasiliev Equations and our explicit results, the interaction vertices are of similar form as the pure HS algebra product, $C\star \pi(C)$. Technically, in order to get the interaction vertices we need to take the star-product of two first-order fields, add $t,q$ integrals and act with simple differential operators that are contracted with $h^{\ga\gad}$. Therefore, the vertices are seemingly non-local because $C$ encodes the derivatives of $\Fron_s$ of unbounded order and $C\star \pi(C)$ contains a sum over all derivatives for each spin. Note that $\mathcal{V}(\Omega,\omega,C)$, $\omega\star\omega$ and $\omega\star C-C\star \pi(\omega)$ are local because $\omega$ contains a finite number of derivatives for a given spin.

This non-locality can be fake because the unfolded equations contain not only the equations for $\Frontwo_s$, but
also for $\nabla...\nabla\Frontwo_s$ and hence they have to be non-local in the naive sense of having an unbounded number of derivatives. In order to quantify the degree of non-locality one has to project onto the equation of motion for $\Frontwo_s$, \eqref{onlyequations}, and see if it is local or not when the spins $s,s_1,s_2$ of the three fields are fixed. This step requires working out a dictionary between unfolded equations and Fronsdal fields, which we discuss in the next section.

\section{Unfolded vs. Fronsdal Dictionary}\label{sec:Dictionary}
The HS theory is known within the unfolded approach and therefore in terms of $\omega$ and $C$. The only nontrivial content of these equations are the corrections to the Fronsdal equations \eqref{onlyequations}, while the rest are pure gauge or auxiliary fields and equations for the derivatives thereof. While passing from the HS symmetry friendly variables $\omega$ and $C$ to the Fronsdal fields $\Fron_s$ makes the underlying symmetry not manifest, it can still be useful to work out the dictionary and see how the equations look like in terms of the Fronsdal fields. We would like to determine \eqref{onlyequations}. We refer to the source on the right-hand side of \eqref{onlyequations} as {\it Fronsdal currents}. Among other things this would allow one to apply standard AdS/CFT methods. However, we will show this to be impossible due to a high degree of non-locality of the unfolded HS equations, which is one of the conclusions of our study.

\subsection{Fronsdal Currents} \label{subsec:FronsdalCurrents}
Unfolded equations provide a first-order form of HS equations. The relation between $\omega$ and $\Fron_{\mm(s)}$ for $s>2$ is similar to that of the vielbein $e^a_\mm$ and spin-connection $\omega^{a,b}_\mm$ variables to the metric $\Fron_{\mm\nn}$. It is well-known that in the example of gravity the Einstein equations are encoded in the torsion constraint and the Riemann two-form:
\begin{align}
de^a-\omega\fud{a,}{b} \wedge e^b&=T^a\,, && R^{a,b}=d\omega^{a,b}-\omega\fud{a,}{c}\wedge \omega^{c,b}=J^{a,b}\,,
\end{align}
where we allow for nonvanishing torsion that can be generated by the matter fields or, as in the HS case,  is induced by all the fields from the HS multiplet. Firstly, one solves for the spin-connection $\omega=\omega(e,\pl e,T)$ in terms of the vielbein, its first derivative and the torsion, $T$, if any. Secondly, the solution for $\omega$ is plugged into the Riemann two-form to recover the Einstein equations with the source containing contributions from $J$ and $T$. In the spinorial language of $4d$ the vierbein is $e^{\ga\gad}=e^a \sigma_a^{\ga\gad}$ and the Riemann two-form $R^{a,b}$ splits into two (anti)-selfdual components $R^{\ga\ga}$ and $R^{\gad\gad}$.

The part of the unfolded equations of Section \ref{sec:SecondOrder} that contains the Fronsdal equations with nonlinear corrections \eqref{onlyequations} is:
\begin{align}\label{omegaeqQ}
(\nabla+Q)\omegatwo&=\formJ\,, && (\nabla+Q)\formJ=0\,,
\end{align}
where it is useful to split $\adD$ into the Lorentz-covariant derivative $\nabla$, \eqref{LorentzDer}, and operators $Q_\pm$ that act algebraically on $\omegatwo$
\begin{align}
Q&=y^\ga h\fdu{\ga}{\gad} \pl_\gad +\bry^\gad h\fud{\ga}{\gad}\pl_\ga\equiv Q_++Q_-\,.
\end{align}
Note that we can drop the contribution of the HS Weyl tensors $\Ctwo$ as it has no effect on the Fronsdal equations.

As in SUGRA there is a non-zero torsion generated by all the fields from the HS multiplet. Torsion and the right-hand side of the HS analog of the equation for the Riemann two-form come as different components of the backreaction $\formJ$.  The torsion constraint is the $N=\brN$ component, where $N$ and $\brN$ are the number operators for $y$ and $\bry$, respectively, e.g. $N=y^\nu \pl_\nu$. The Riemann two-form splits into (anti)-selfdual parts corresponding to the $N=\brN\pm 2$ components, c.f. fig. \ref{UnfoldedFigure}. Both selfdual and anti-selfdual components contain the same information and we will consider only the self-dual one, $N=\brN- 2$. Therefore, $\formJ$ can be taken to be the $N=\brN$ and $N=\brN\pm 2$ components of the full backreaction of Section \ref{sec:SecondOrder}. If we are interested in the correction to the Fronsdal equations that is due to the stress-tensors built out of the scalar field, then only $\formJ^{s.t.}$, \eqref{HSstBackreaction}, needs to be taken into account as the scalar entirely sits into $C$ and has no components in $\omega$.

In this section we will sometimes avoid using $\omegatwo$ and replace it with $\omega$ for the sake of conciseness and when no confusion can arise. We also use $\omega_{\pm k}$ as a short-hand notation for the components $\omega^{\ga(s-1\pm k),\gad(s-1\mp k)}$ of $\omega$, so that $k=0$ corresponds to the vierbein, $e=\omega_0$. The fields $\omega_{\pm1}$ are the (anti)-self dual components of the HS spin-connection.

The Fronsdal field $\Frontwo_{\mm(s)}$ consists of two traceless tensors of ranks $s$ and $s-2$ and hence in the spinorial language is represented by some $\Front_{\ga(s),\gad(s)}$ and $\Front'_{\ga(s-2),\gad(s-2)}$, which, in the spirit of the approach we follow, can be collected for all spins into
\begin{align}\label{spinorialFrfield}
\Front&=\sum_s \frac{1}{s!s!}\Front_{\ga(s),\gad(s)}y^{\ga(s)}\bry^{\gad(s)}\,, &
\Front'&=\sum_s \frac{1}{(s-2)!(s-2)!}\Front'_{\ga(s-2),\gad(s-2)}y^{\ga(s-2)}\bry^{\gad(s-2)}\,.
\end{align}
An analogous decomposition holds true for the Fronsdal tensors $F_{\mm(s)}$, \eqref{FronsdalEq},
\begin{align}\label{spinorialFroperator}
f&=\sum_s \frac{1}{s!s!}f_{\ga(s),\gad(s)}y^{\ga(s)}\bry^{\gad(s)}\,, &
f'&=\sum_s \frac{1}{(s-2)!(s-2)!}f'_{\ga(s-2),\gad(s-2)}y^{\ga(s-2)}\bry^{\gad(s-2)}\,.
\end{align}

In what follows we will implicitly use the results of the $\sigma_-$-analysis that sorts out pure gauge and auxiliary components of $\omega$, see \cite{Shaynkman:2000ts,Bekaert:2005vh}. First of all, it is easy to see that the Fronsdal field is embedded into the HS vielbeins $e=\omega_0$ as
\begin{equation}
e=h^{\ga\gad}\pl_\ga\pl_\gad \Front+y^\ga\bry^\gad h_{\ga\gad}\Front'+Q_-(\bullet)+Q_+(\bullet)\,,
\end{equation}
where the last two terms are pure gauge by virtue of the linearized gauge symmetry $\delta \omegatwo=(\nabla+Q)\xitwo$ and we set them to zero by choosing the appropriate $\xitwo$. The equations for $e=\omega_0$ and $\omega_{+1}$ can be easily obtained from \eqref{omegaeqQ} and read
\begin{subequations}
\begin{align}
R_0=\nabla e+Q_+ \omega_{-1} +Q_- \omega_{+1}&=\formJ_0\,,\label{realTorsion}\\
R_{+1}=\nabla \omega_{+1}+Q_+ e +Q_- \omega_{+2}&=\formJ_{+1}\,,
\end{align}
\end{subequations}
where on the right-hand side one has to pick up the components $\formJ_0$ and $\formJ_{+1}$ of $\formJ$ with $N=\brN$ and $N=\brN+2$, respectively. There is a similar expression for $R_{-1}$, but it contains exactly the same information as $R_{+1}$.

In order to recover the Fronsdal equations one has to solve for $\omega_{\pm1}$ in terms of $\formJ_0-\nabla e$, then plug the solution in $R_{+1}$ and project onto the right component (for example, for the spin-two case in addition to the Einstein equations the two-form $R_{\pm1}$ contains also a Weyl tensor, which we do not need to recover \eqref{onlyequations}). By a simple index counting argument one can see how the Fronsdal tensor is embedded into $R_{+1}$:
\begin{align}\label{FronsdalEmbedding}
R_{+1}\Big|_F&:= H^{\gad\gad}\pl_\gad\pl_\gad f +y^\ga y^\ga H_{\ga\ga} f'\,,
\end{align}
where $|_F$ tells that we ignore all components except for those indicated explicitly on the right-hand side. As a result, one finds the Fronsdal operator for $\Frontwo_s$ with a source $\formj$, which we call the Fronsdal current, expressed in terms of $\formJ$ as:
\begin{align}\label{currentdef}
\Big[-\nabla (Q_-)^{-1}\nabla e +Q_+ e\Big]_{F}&=\formj\,,  && \formj=\big[(I-\nabla (Q_-)^{-1})\formJ\big]\Big|_{+1,F}\,.
\end{align}
Note that the combination $\nabla (Q_-)^{-1}\nabla e|_F$ makes the two-derivative terms in the Fronsdal operator while $Q_+ e$ contributes to the mass-like term.\footnote{See Appendix \ref{app:FronsdalOperator} for the derivation of the Fronsdal operator, which allows us to fix the normalization between the spinorial and vectorial bases.} Also, the term $Q_-\omega_{+2}$ drops out because of the $|_F$ projection. The Fronsdal current $\formj$ has the same decomposition as $F_{\mm(s)}$ and $\Fron_{\mm(s)}$, c.f. \eqref{spinorialFrfield}. Eventually we will recover its traceless component $j$ and its trace $j'$.

The first step is to invert $Q$. It is useful to expand $\omega$ into irreducible components. As any one-form it can be decomposed as
\begin{align}\label{omegadecomp}
\omega(y,\bry)&=h^{\ga\ga}\pl_\ga\pl_\gad \omega^{\pl\pl} +Q_+ \omega^{Q_+}+ Q_- \omega^{Q_-}+ y^\ga \bry^\gad h_{\ga\gad}\omega^{y\bry}\,,
\end{align}
where we wrote all possible terms that can make a one-form. Applying $Q_\pm$ to it we get
\begin{align}
\begin{aligned}
Q_+\omega&=\frac{N}{2}H^{\gad\gad}\partial_\gad\partial_\gad\omega^{\pl\pl}
+\frac12 \left[-N H^{\gad\gad}\bry_\gad\partial_\gad+(\bar{N}+2)H^{\ga\ga}y_\ga\partial_\ga\right]\textbf{}\omega^{Q_-}
-\frac{\bar{N}+2}{2}H^{\ga\ga}y_\ga y_\ga\,\omega^{y\bry}\,,\\
Q_-\omega&=\frac{\bar{N}}{2}H^{\ga\ga}\partial_\ga\partial_\ga\omega^{\pl\pl}
+\frac12 \left[-\bar{N}H^{\ga\ga}y_\ga\partial_\ga+(N+2)H^{\gad\gad}\bry_\gad\partial_\gad\right]\omega^{Q_+}
-\frac{N+2}{2}H^{\gad\gad}\bry_\gad\bry_\gad\,\omega^{y\bry}\,,
\end{aligned}
\end{align}
where $(Q_{\pm})^2=0$ was used as well as the identities for the vierbein one-form $h^{\ga\gad}$ collected in Appendix \ref{app:notation}. This expansion should be matched with the analogous expansion of $\formJ$ as a two-form
\begin{align}\label{Jdecomp}
\begin{aligned}
\formJ&=H^{\ga\ga}\pl_\ga\pl_\ga J^{\pl\pl}+H^{\ga\ga}y_\ga\pl_\ga J^{y\pl}+ H^{\ga\ga}y_\ga y_\ga J^{yy}\\&\qquad\qquad
+H^{\gad\gad}\pl_\gad\pl_\gad \bar{J}^{\pl\pl}+H^{\gad\gad}\bry_\gad\pl_\gad \bar{J}^{y\pl}+ H^{\gad\gad}\bry_\gad \bry_\gad \bar{J}^{yy}\,.
\end{aligned}
\end{align}
Therefore, solving the torsion constraint corresponds to inverting some number operators:
\begin{align}\label{omegacomponents}
\omega_{+1}^{\pl\pl}&=\frac2{\bar{N}}J^{\pl\pl}\,, &
\omega_{+1}^{Q_+}&=\frac{1}{N+\bar{N}+2}(NJ^{y\pl} +(\bar{N}+2)\bar{J}^{y\pl})\,, &
\omega_{+1}^{y\bry}&=\frac{-2}{(N+2)}\bar J^{yy}\,.
\end{align}
It is important that the number operators never degenerate except for the case of spin-one where there is no torsion constraint to be solved. Next, we need to apply $\nabla$, which we cannot fully do without knowing the structure of $\formJ$,
so we keep $\nabla=h^{\ga\gad}\nabla_{\ga\gad}$ as it is for a moment. Lastly, we need to determine $\formj$ from  $\formJ$ as in \eqref{currentdef} and project onto the components indicated on the right-hand side of \eqref{FronsdalEmbedding}, which can be done with the help of the following identity
\begin{equation}
f_{\ga}(y)\equiv\pl_\ga[N^{-1}y^\gb f_\gb]+y_\ga[(N+2)^{-1}\pl_\gb f^\gb]\,,\label{niceidentity}
\end{equation}
that allows to split any $f_\ga$ into a gradient and $y$-parallel components.  Moreover, given some $H^{\ga\ga} g_{\ga\ga}(y)$ one can apply \eqref{niceidentity} twice to get
\begin{equation}\label{niceidentityT}
\begin{aligned}
H^{\ga\ga}g_{\ga\ga}(y)&=H^{\ga\ga}
\Big(\pl_\ga\pl_\ga\frac{1}{N(N-1)}y^\gb y^\gb g_{\gb\gb}+y_\ga\pl_\ga \frac{2}{N(N+2)}y^\gb \pl_\gc g\fdu{\gb}{\gc}\\&\qquad\qquad\qquad\qquad+y_\ga y_\ga\frac{1}{(N+2)(N+3)}\pl_\gb\pl_\gb g^{\gb\gb}\Big)\,.
\end{aligned}
\end{equation}
Finally, combining \eqref{currentdef}, \eqref{omegacomponents} and \eqref{niceidentity} we get the expressions for the components $j$ and $j'$ of the Fronsdal current $\formj=H^{\gad\gad}\pl_\gad\pl_\gad \compj+H^{\ga\ga}y_\ga y_\ga \compj'$ in terms of the unfolded backreaction $\formJ$:
\besubeqs\label{FronsdalCurrentsDef}
\begin{align}
\compj&=-\frac{1}{\brN(\brN-1)}\pl_\nu \nabla^{\nu\gnd}\bry_\gnd J^{\pl\pl}-\frac1{2\brN(\brN+N)}y_\nu \bry_\gnd \nabla^{\nu\gnd}J^\diamond+\brJ^{\pl\pl}\,,
\\
\compj'&=-\frac{1}{(N+2)(N+3)}\pl_\nu \nabla^{\nu\gnd} \bry_\gnd \brJ^{yy}+\frac{1}{2(N+2)(\brN+N+4)} \pl_\nu\pl_\gnd \nabla^{\nu\gnd}J^\diamond+J^{yy}\,,
\end{align}
\esubeqs
where $J^\diamond:=N J^{y\pl}+(\brN+2)\bar{J}^{y\pl}$. As expected, all the six components \eqref{Jdecomp} of $\formJ$ contribute to the Fronsdal current. These expressions, when evaluated on the backreaction $\formJ$ obtained from Vasiliev Equations, solve the problem of reconstructing corrections to the Fronsdal equations at the second order, i.e. we determine \eqref{onlyequations}. The explicit formulas are a bit involved and we put them in Appendix \ref{app:FronsdalCurrents}. In the sections below we discuss examples and general properties.

It is also useful to see how the conservation identity looks in terms of $\formj$. It has to be the consequence of the Bianchi identity that the Fronsdal tensor \eqref{FronsdalEq} obeys:
\begin{align}\label{FronsdalBianchi}
B_{\mm(s-1)}:=\nabla^\nn F_{\nn\mm(s-1)}-\frac12 \nabla_\mm F\fdu{\mm(s-2)\nn}{\nn}\equiv0\,,
\end{align}
where we used the symmetrization convention with a minimal number of terms necessary to make an expression symmetric. Introducing a spinorial representation of $B_{\mm(s-1)}$
\begin{align}\label{BianchiVectorial}
B&=\sum_s \frac{1}{(s-1)!(s-1)!}B_{\ga(s-1),\gad(s-1)}y^{\ga(s-1)}\bry^{\gad(s-1)}\,,
\end{align}
we see by the index counting argument that the $y_\ga\hat{h}^{\ga\gad}\pl_\gad B$ component\footnote{$\hat{h}^{\ga\gad}=h\fud{\ga}{\gbd}\wedge H^{\gbd\gad}$, see Appendix \ref{app:notation} for more detail.} of $\nabla R_{+1}$, which we denote as $\nabla R_{+1}|_{B}$, plays the role of the Bianchi identity for the Fronsdal tensor that is embedded into $R_{+1}$. Therefore, the conservation identity for $\formj$ is
\begin{align}
\nabla \formj\Big|_{B}&=0\,,
\end{align}
that holds provided $\adD \formJ=0$, which in turn is true on the free mass-shell where $\tadD C=0$. It can be rewritten in terms of the trace, $\compj'$, and the trace-free, $\compj$, components as
\begin{align}\label{BianchiSpinorial}
\brN \pl_\nu\pl_\gnd \nabla^{\nu\gnd}\compj+(N+2) y^\nu \bry^\gnd\nabla_{\nu\gnd}\compj'=0\,.
\end{align}

\subsection{Canonical Currents}\label{subsec:Canonical}
It turns out that the Fronsdal currents can be split into many independent components, some of which can be dropped, simplifying the expressions a lot. Any conserved current consists of three parts: the trivial part called improvement, i.e. the part that is conserved without using the equations of motion, the on-shell part that is proportional to the free equations of motion and the nontrivial part that makes an on-shell conserved current. What we want to do is to decompose the stress-tensors $\formJ^{s.t.}$ \eqref{HSstBackreaction} into nontrivial currents and improvements.

The minimalistic stress-tensors, which we call {\it primary}, contain the minimal number of derivatives and give rise to a subset of the standard vertices that are known as current interactions. In particular, the rank-$s$ primary stress-tensor built out of the scalar field contains $s$-derivatives:
\begin{align}
j^{can}_{\mm(s)}&= \Fron \overleftrightarrow{\nabla}_\mm...\overleftrightarrow{\nabla}_\mm\Fron+O(\Lambda)\,.
\end{align}
It is also possible to built a rank-$(2s+k)$ conserved tensor that is bilinear in the spin-$s$ Weyl tensors and has $k$-derivatives. For example, there is the Bel-Robinson tensor $C_{\ga(4)}C_{\gad(4)}$, which is bilinear in the usual Weyl tensors. There also exist super-currents that are built from fields of different spins. There is a simple generating function that encodes conserved currents with the minimal number of derivatives \cite{Gelfond:2006be}\footnote{In \cite{Gelfond:2006be} the generating function is written as $C(y,\bry|x)C(iy,i\bry|x)$, but for our purpose it is more convenient to replace $C(iy,i\bry|x)$ with $C(-y,\bry|x)$ that behaves in  exactly the same way since the wave operator $\tadD$ is bilinear in spinors. Such form corresponds to $C\pi(C)$.}
\begin{align}\label{canprimgenfunc}
\begin{aligned}
\compj^{\text{can}}= C(y,\bry|x)C(-y,\bry|x)&=\sum_{m,n,p,q} \frac{(-)^{p}}{n!m!p!q!}\,C_{\ga(n),\gad(m)}C_{\ga(p),\gad(q)}\, y^{\ga(n+p)} \bry^{\gad(m+q)}\\
&=\sum_{l,k} \frac{1}{l!k!} \,\compj^{\text{can}}_{\ga(l),\gad(k)}\,y^{\ga(l)} \bry^{\gad(k)}\,.
\end{aligned}
\end{align}
There is also an associated two-form that can be used as a source for $R_{\pm1}$ in the unfolded approach:
\begin{align}
\formj^{can}&= (H^{\ga\ga}\pl_\ga\pl_\ga +H^{\gad\gad}\pl_\gad\pl_\gad)C(y,\bry|x)C(-y,\bry|x)\,.
\end{align}
Conserved tensors $\compj^{\text{can}}$ are in fact conformal primaries of $so(4,2)$, which explains our terminology. The conformal symmetry is however broken in the $4d$ HS theory. One can extend the list of conserved tensors by:
\begin{align}\label{cangenfuncdes}
\left(\frac{\pl}{\pl u^\ga}\frac{\pl}{\pl v_{\ga}}\frac{\pl}{\pl\bar{u}^\gad}\frac{\pl}{\pl{\bar{v}}_{\gad}}\right)^l C(u,\bru|x)C(-v,\brv|x)\Big|_{U=V=Y}\,.
\end{align}
In Minkowski space this corresponds to the fact that on top of the conserved tensor $\Fron \overleftrightarrow{\pl}_a...\overleftrightarrow{\pl}_a\Fron $ there is a family of conserved
tensors $\pl_{c(l)}\Fron \overleftrightarrow{\pl}_a...\overleftrightarrow{\pl}_a\pl^{c(l)}\Fron $ that have $2l$ derivatives contracted. We will call such currents, i.e. those with $l>0$, {\it successors}. The term {\it improvement} is reserved for pure improvements, i.e. currents that are conserved without using the equations of motion. It is important to stress that successors do have an overlap with primaries, i.e. any successor is a sum of a primary and an improvement. Therefore, we combine primaries and successors into {\it canonical} currents, each of those has a nonvanishing projection onto a primary and therefore can be used to construct nontrivial interactions. Primary currents directly lead to standard Lagrangian vertices.

The Fronsdal currents that we derived from Vasiliev's equations contain a sum over such families with certain coefficients. In Fourier space the generating function for the primaries and successors thereof \eqref{cangenfuncdes} is very close to the $Q_1$ function of Section \ref{sec:SecondOrder}
\be\label{cangenfunc}
Q^{can}=\exp{i\left(\beta y(\xi-\eta)+\gamma \bry(\breta+\brxi)+\alpha_1 \xi\eta+\alpha_2 \brxi\breta\right)}\, C(\xi|x)C(\eta|x)\,, \ee
where $\alpha_1, \alpha_2$ count the number of contracted derivatives and $\beta, \gamma$ count the spin of the current. The presence of $\xi\eta$ and $\brxi\breta$ in $Q_1$ and hence in the Fronsdal currents indicates that the currents are pseudo-local.

Also, there is a part of the Fronsdal currents that contains only pure improvements. While improvements can contribute to exchange diagrams, they can be projected out for the purpose of computing the simplest Witten diagram \scalebox{1.5}{$\Yup$} that makes a three-point function. Such a decomposition simplifies a lot the expressions: canonical currents, i.e. primaries and successors, are traceless and all the traces can be attributed to pure improvements. In order to project onto the canonical sector we note that the form of the generating function \eqref{cangenfunc} implies that all the exponents that contribute to $\formJ^{s.t.}$, i.e. $Q_1$ and $P_1$, \eqref{fullJbackreaction}, are already canonical. Therefore, we need to extract $\xi-\eta$ and $\brxi+\breta$ from the prefactors in \eqref{HSstBackreaction}, which is done by\footnote{In \cite{Gelfond:2010pm} the tensor product $C\otimes C$ was studied by decomposing it into irreducible components. Those correspond to primaries together with successors thereof. Here, we rewrite everything in terms of the Fourier components that are useful for the computations in the perturbation theory of Vasiliev's equations, extending the analysis of \cite{Kessel:2015kna}.}
\begin{align*}
(y+\eta)_\ga (y+\xi)_\ga Q_1= -\frac14 (\xi-\eta)_\ga(\xi-\eta)_\ga Q_1+\text{improvement}= +\frac14 \pl_\ga\pl_\ga Q_1+\text{improvement}\,,\\
\breta_\gad\brxi_\gad Q_1= +\frac14 (\brxi+\breta)_\gad(\brxi+\breta)_\gad Q_1+\text{improvement}= -\frac14 \pl_\gad\pl_\gad Q_1+\text{improvement}\,,
\end{align*}
where, for example, in the second formula we decomposed $\brxi_\gad \breta_\gad$ as $\frac14 (\brxi+\breta)_\gad(\brxi+\breta)_\gad-\frac14 (\brxi-\breta)_\gad(\brxi-\breta)_\gad$ and the second term does not
contribute to canonical currents since it is not purely holomorphic in $(\brxi+\breta)$ and $(\xi-\eta)$, see \cite{Kessel:2015kna}. Finally, the canonical part $\formJ^{can.}$ of the stress-tensors $\formJ^{s.t.}$ is
\begin{align}\label{HSstCanonical}\begin{aligned}
\formJ^{can.}=\int_0^1 dt\, \int_0^1 dq\, \Big[&\frac1{4} H^{\ga\ga}\pl_\ga\pl_\ga Q_1\Big(i+(\brxi\breta)\frac{(1-qt)}{2qt}\Big)\\&\qquad+\frac{i}8 H^{\gad\gad} \pl_\gad\pl_\gad \Big(Q_1-(1-t) P_1\Big)+h.c.\Big]\,.\end{aligned}
\end{align}

\subsection{Corrections to Fronsdal Equations: Examples}\label{subsec:SpinTwo} In this section we give an explicit form for the Fronsdal current built out of the scalar field $\Fron$.\footnote{This expression was claimed to have been found in \cite{Kristiansson:2003xx}. However, it is easy to see  without comparing with the coefficients we found that their result is inconsistent. Indeed, the asymptotic fall-off has to be faster than $1/(l!l!)$ because of the homotopy integrals. See \cite{Kessel:2015kna} for the $3d$ examples along the same lines.} Considering the spin-two example, the generic source reads:
\begingroup\allowdisplaybreaks\begin{align}\notag
\square \Fron_{\mm\mm}+...&= 2\cos 2\theta \sum_l \left(\tilde{a}_{l,1}\nabla_{\mm\nn(l)} \Fron \nabla\fdu{\mm}{\nn(l)}\Fron+2\tilde{a}_{l,0}\nabla_{\mm\mm\nn(l)} \Fron \nabla^{\nn(l)}\Fron+\tilde{c}_{l,0} g_{\mm\mm}\nabla_{\nn(l)} \Fron \nabla^{\nn(l)}\Fron\right)\,,
\end{align}
where we introduced symmetrized and trace-projected derivatives that naturally appear upon transferring components of $C$ to the language of world tensors:
\begin{align}
\nabla_{\mm(s-k)\nn(l)} \Fron \nabla\fdu{\mm(k)}{\nn(l)}\Fron=h_{\mm}^{\ga\gad}...h_{\mm}^{\ga\gad} C_{\ga(s-k)\nu(l),\gad(s-k)\gnd(l)}C\fdudu{\ga(k)}{\nu(l)}{,\gad(k)}{\gnd(l)}\,.
\end{align}
The coefficients we find take the form:\footnote{There is one step that we do not want to bother the reader with. The simplest normalization of the Fronsdal field \eqref{spinorialFrfield} and operator \eqref{spinorialFroperator} in the spinorial language leads to certain spin-dependent rescalings that need to be done in order to map $\compj$ and $\compj'$ into the trace and trace-free components of the Fronsdal equation in the vectorial language. While we give the coefficients $a,c$ in the spinorial form, $\tilde{a},\tilde{c}$ denote the rescaled coefficients. The dictionary can be found in Appendix \ref{app:FronsdalOperator}.}
\begin{align}
a_{l,0}&=\frac{1}{l!l!}\left(-\frac{3}{(2+l)^2}+\frac{7}{2 (2+l)}-\frac{4}{3+l}+\frac{1}{2 (4+l)}\right)\,, \qquad a_{l,0}l!l!\xrightarrow{l\rightarrow\infty} -\frac{2}{l^3}\,,\\
a_{l,1}&=\frac{1}{l!l!}\left(\frac{1}{2 (2+l)^2}-\frac{1}{4 (2+l)}+\frac{1}{4 (4+l)}\right)\,,
\qquad \qquad \qquad a_{l,1}l!l!\xrightarrow{l\rightarrow\infty} \frac{1}{l^3}\,,\\
c_{l,0}&=\frac{1}{l!l!}\left(\frac{1}{12 (1+l)^2}-\frac{3}{8 (1+l)}+\frac{1}{2+l}-\frac{1}{8 (3+l)}\right)\,,
\qquad c_{l,0}l!l!\xrightarrow{l\rightarrow\infty} \frac1{2l}\,,
\end{align}\endgroup
The canonical projection of the Fronsdal current above, i.e. \eqref{HSstCanonical}, gives:
\begin{align}
a_{l,k}&=\frac{4 (-)^k}{(l+2)^2 (l+3) (k!)^2 (l!)^2 (2-k)!^2}\,,\qquad k=0,1\,, & c_{l,0}&=0\,.
\end{align}
It makes no difficulty to extract the contributions of fields with other spins out of the generating functions we derived, see Appendix \ref{app:Coefficients}, but this simple example suffices to make important statements about the locality properties of the theory as a whole. Moreover, we do not expect explicit coefficients to be more efficient to work with than  generating functions that allow us to deal with all the fields at once. As for the spin-$s$ current built of the two scalar fields, we find:
\begin{align}
\square \Fron_{\mm(s)}+...&= 2\cos 2\theta \sum_{l,k} \big(\tilde{a}_{l,k}\,\nabla_{\mm(s-k)\nn(l)} \Fron \nabla\fdu{\mm(k)}{\nn(l)}\Fron+\tilde{c}_{l,k}\, g_{\mm\mm}\nabla_{\mm(s-2-k)\nn(l)} \Fron \nabla^{\mm(k)\nn(l)}\Fron\big)\,,\\
a_{l,k}&=\frac{(-)^k s!s!}{l!l!k!k!(s-k)!(s-k)!}\frac{ \left(2 k^2-2 k s+s^2\right)}{2 s l^3}+O\Big(\frac{1}{l^4l!l!}\Big)\,, \\
c_{l,k}&=\frac{(-)^k (s-2)!(s-2)!}{l!l!k!k!(s-k-2)!(s-k-2)!}\frac{(2 s+3)}{l (s+2) (s+3)}+O\Big(\frac{1}{l^2l!l!}\Big)\,.
\end{align}

The coefficients for the canonical part of the spin-$s$ Fronsdal current, as extracted from $\formJ^{can.}$, are much simpler:
\begin{align}
\square \Fron_{\mm(s)}+...&= 2\cos 2\theta \sum_{l,k} \tilde{a}_{l,k}\nabla_{\mm(s-k)\nn(l)} \Fron \nabla\fdu{\mm(k)}{\nn(l)}\Fron\,,\\
a_{l,k}&=\frac{(-)^k s!s!}{l!l!k!k!(s-k)!(s-k)!}\frac{s (2 l (s-1)+s (2 s-1))}{8 (s-1) (l+s)^2 (l+s+1)^2}
\end{align}
They have the same $(l!l!l^3)^{-1}$ decay at $l\rightarrow\infty$ as the full currents. The canonical currents are traceless and the dependence on $k$ is fully covered by factorials, which is the signature of the canonical currents.

We would like to point out few features of the Fronsdal currents as displayed above.

(i) The full Fronsdal currents are not traceless, which can be cured by dropping improvements and taking the canonical projection. Note that in $4d$ the equations for HS Weyl tensors are conformal, but the Fronsdal equations are not, see \cite{Bracken:1982ny,Siegel:1988gd} and more recently \cite{Barnich:2015tma}. The conformal symmetry of the HS theory appears to be broken by the gauge-noninvariant part of the stress-tensors, i.e. $\omega\star\omega+\mathcal{V}(\Omega,\omega,C)$, which involves $\omega$;

(ii) The currents vanish for $\theta=\pi/4$. Also, we note that the stress-tensor changes its sign at $\theta=\pi/4$. In particular, as it was already noticed in \cite{Sezgin:2003pt}, the signs of the stress-tensor are the opposite for type A, $\theta=0$, and type B, $\theta=\pi/2$, models that are dual to the free boson and free fermion, respectively \cite{Sezgin:2002rt, Klebanov:2002ja,Sezgin:2003pt}. Such $\theta$-dependence looks puzzling in view of the slightly-broken HS symmetry \cite{Maldacena:2012sf};

(iii) Obviously, the Fronsdal currents are pseudo-local expressions, while we know from the Lagrangian approach that the only nontrivial coupling $0-0-s$ of two scalars with a HS field has $s$-derivatives. A chain of redefinitions is needed to remove the improvements reducing everything to the standard form.

Let us now comment on the pseudo-locality of the Fronsdal currents found from Vasiliev Theory. As it was rigourously proven in $3d$ \cite{Prokushkin:1999xq}, the primary $s$-derivative stress-tensor built out of the scalar field is formally exact, i.e. can be removed by a field-redefinition of the spin-$s$ Fronsdal field. Such redefinition, however, involves a pseudo-local series and has to be forbidden on physical grounds.

A more powerful result of \cite{Kessel:2015kna} shows that in $3d$ any conserved two-form $\formJ$ that is bilinear in $C$, which one can put on the right-hand side of $\adD \omegatwo=\formJ+...$, is exact in the class of pseudo-local expressions:
\begin{align}
\formJ&=\formJ(h,h,C,C): && \adD \formJ=0 && \longrightarrow && \formJ=\adD\formU\,, && \formU=\formU(h,C,C)\,.
\end{align}
This result is the AdS counterpart of \cite{Barnich:1993vg} obtained in flat space. In the flat space limit indeed the above improvements reconstruct $\square^{-1}$ singularities of the type considered in \cite{Taronna:2011kt}. The same results are expected to be true in any dimension. In particular, $\formJ^{s.t.}$ and its canonical projection $\formJ^{can}$ have to be exact forms and it should be possible to redefine away $\formj^{\text{can}}$ by a pseudo-local field redefinition. Therefore a better understanding of allowed field redefinitions in the context of HS theories is of paramount importance, see e.g. \cite{Vasiliev:2015mka, ST}.

Finally, let us note that in principle there can be a subtlety in how to relate the HS connection $\omega$ to the Fronsdal fields. For example, one can take the vielbein component $\omega_0$ of $\omega$ and consider trace-invariants $\tilde{\Fron}_s=tr(\omega_0\star...\star \omega_0)$ as a possible different definition of the Fronsdal tensor that is still compatible with \eqref{FronVielLinear} upon linearization. Beyond the free theory level one can also consider sums of appropriate powers of such invariants with yet unknown coefficients, see e.g. \cite{Sezgin:2011hq, Campoleoni:2011hg}. On top of this one can involve the zero-form $C$ into such expressions. This just results into certain field-redefinitions for $\omegatwo$ which would be also pseudo-local when $C$ is included. In this paper, we have made the simplest choice as to identify $\Frontwo$ with the same component of $\omegatwo$ as at the free level, i.e. \eqref{FronVielLinear}.

\subsection{Weyl Tensors}\label{subsec:WeylTensors}
The equations for the HS Weyl tensors are also of interest for the following reasons. First of all, the equations for the $s=0$ and $s=1$ fields of the HS multiplet reside in $\tadD \Ctwo=...$ equation, \eqref{xpaceseqCB}. Secondly, the equations for the Weyl tensors are much simpler. Thirdly, the equations for the Weyl tensors bear the same amount of gauge-invariant information and can be used for the AdS/CFT computation as in \cite{Giombi:2009wh}, where the boundary behavior of the Fronsdal fields was extracted from that of the Weyl tensors. We warn the reader again that $\Ctwo$ does not correspond to the linearized HS Weyl tensors and it should be clarified whether this affects the $AdS/CFT$ computations. At the end of the section we discuss the corrections for the linearized Weyl tensors.

The reason why equations for HS Weyl tensors are expected to be simpler can be seen by applying $s$ curls to the Fronsdal equation, \eqref{FronsdalEq}, with some current $F_{\mm(s)}= j_{\mm(s)}$
\begin{align}
\nabla_{\mm(s)}F_{\nn(s)}|_W&= \nabla_{\mm(s)} j_{\nn(s)}|_W\,,
\end{align}
where the projection onto the Weyl tensor, i.e. a traceless tensor with the symmetry of $(s,s)$-Young diagram is denoted by $|_W$. We see that all terms in $F_{\mm(s)}$, \eqref{FronsdalEq}, except for the Klein-Gordon part are projected to zero:
\begin{align}
(\square -m^2_W)\nabla_{\mm(s)}\Fron_{\nn(s)}|_W&=  \nabla_{\mm(s)} j_{\nn(s)}|_W\,,
\end{align}
where $m^2_W$ is the mass of the Weyl tensor, see e.g. \cite{Iazeolla:2008ix}. At the same time, if the current is built out of the scalar field only, it is easy to see that only one term can survive after the projection:
\begin{align}
\nabla_{\mm(s)} j_{\nn(s)}|_W=\nabla_{\mm(s)} \big(\Fron \overleftrightarrow{\nabla}_\nn...\overleftrightarrow{\nabla}_\nn\Fron+O(\Lambda)\big)|_W=
\nabla_{\mm(s)}\Fron \nabla_{\nn(s)}\Fron|_W\,.
\end{align}
In free theory the Weyl tensors obey the Klein-Gordon equation with a specific mass-like term. For example, the anti-holomorphic Weyl tensors obey
\begin{align}
(\square -2(2+\bar{N}))C(0,\bry|x)&=0\,,
\end{align}
which can be derived by computing $g^{\mm\nn}\nabla_\mm \nabla_\nn C$ where $\nabla_\mm$ needs to be replaced with its on-shell value in accordance with $\tadD C=0$, \eqref{TwistedDer}.

At the second order one finds a source on the right-hand side, which we gave in Section \ref{sec:SecondOrder},
\begin{align}\label{DCtwoP}
\tadD \Ctwo(y,\bry|x)&=\formP(y,\bry|x)\,, && \formP=\omega\star C-C\star \pi(\omega)+\mathcal{V}(\Omega,C,C)\,.
\end{align}
The procedure now is very similar to solving for torsion, but it is much simpler since $\formP$ is a one-form $\formP=h_{\gc\gdd}P^{\gc\gdd}$. After using \eqref{niceidentity} to solve for  $h^{\ga\gad}\pl_\ga\pl_\gad C$ in terms of $\formP-\nabla C$ from \eqref{DCtwoP}, plugging the solution back and projecting onto the Klein-Gordon equation we get
\begin{align}
(\square -2(2+\brN))\Ctwo(0,\bry)&= \frac{2}{\brN+2} (\pl_\ga\nabla^{\ga\gad}\pl_\gad) (y_\gc \bry_\gdd P^{\gc\gdd} )\Big|_{y=0}+ \frac{2i(\brN+1)}{\brN+2}(\pl_\gc\pl_\gdd P^{\gc\gdd})\Big|_{y=0}\label{WeylTorsion}\,,
\end{align}
which computes the source for the HS Weyl tensors, including the scalar and the spin-one fields, given any backreaction $\formP$.

Let us consider the $\formK=h_{\gc\gdd}\compK^{\gc\gdd}=\mathcal{V}(\Omega,C,C)$ part of the backreaction. For example, this is the only part that contains the contribution of the scalar to the spin-$s$ Weyl tensor equations, i.e. contributes to $\langle j_0j_0j_s\rangle$ correlator. First of all, $ \pl_\gc\pl_\gdd K^{\gc\gdd}=0$. Next, the first term of \eqref{WeylTorsion} can be straightforwardly computed and simplified a bit by integrating it by parts:
\begin{align}
(\square -2(2+\bar{N}))\Ctwo(0,\bry)&= \int F(\bry,\xi,\eta)e^{i\theta+i\brxi\breta} C(\xi)C(\eta)\,,\label{WeylTensorEq}
\end{align}
where the interaction kernel is
\begin{align*}
F(\bry,\xi,\eta)&=\int_0^1 dt\, \Big[2(\bry(\brxi+\breta)) e^{ i[t\eta\xi+\bry(\breta+\brxi)]}
+2(\brxi\breta) e^{ i[\bry(\breta+\brxi)]}-4t(\brxi\breta)e^{ i[t\bry(\breta+\brxi)]}\Big]\,.
\end{align*}
Let us note that there is no usual $h.c.$ at the end of the last formula.

There are few remarks one can make by looking at \eqref{WeylTensorEq}. The source vanishes if we set $\bry=0$, i.e.
there are seemingly no corrections to the free propagation of the scalar field coming from $\formK$. In particular, it implies the absence of the $\Fron^3$ coupling \cite{Sezgin:2003pt}. Therefore, correlator $\langle j_{s_1} j_{s_2} j_0\rangle$ is accounted for by the HS algebra structure constants via $\omega\star C-C\star \pi(\omega)$ as observed in \cite{Giombi:2009wh}.

Also the vanishing of the pseudo-local part in the source for the scalar field can be confronted with the currents in Section \ref{subsec:FronsdalCurrents}, which shows that Vasiliev's equations in this gauge and field frame are not Lagrangian. Indeed the source to the scalar and to the spin-$s$ Fronsdal field come from the same $0-0-s$ coupling and hence should be related to each other, while we observe that one is local and the other is pseudo-local.

\paragraph{Linearized Weyl tensors.} Let us give more details on the fact that $\Ctwo$ in \eqref{xpaceseqCQ} are not the linearized HS Weyl tensors for $\Frontwo$, i.e. they are not related via simple $s$-curls like in \eqref{LinWeylTensors}. Let us first rewrite \eqref{xpaceseqCQ} by collecting all the vertices that are bilinear in the first order fields into a two-form $\formR$ for $\omegatwo$ and into a one-form $\formP$ for $\Ctwo$:
\begin{align}\label{xpaceseqCQA}
\adD \omegatwo&=\formR+\mathcal{V}(\Omega,\Omega,\Ctwo)\,, &\tadD \Ctwo&=\formP\,,
\end{align}
where we remind that $\adD=\nabla+Q_{-}+Q_{+}$ and:
\begin{align}\label{omstcoc}
\mathcal{V}(\Omega,\Omega,\Ctwo)=-\frac12 H^{\gad\gad}\pl^y_\gad\pl^y_\gad \Ctwo(0,\bry)e^{i\theta}-\frac12 H^{\ga\ga}\pl^y_\ga\pl^y_\ga \Ctwo(y,0)e^{-i\theta}\,.
\end{align}
It is not difficult to see that all the terms in the decomposition of $\formR$ according to
\begin{align}\label{Rdecomp}
\begin{aligned}
\formR&=H^{\ga\ga}\pl_\ga\pl_\ga \compR^{\pl\pl}+H^{\ga\ga}y_\ga\pl_\ga \compR^{y\pl}+ H^{\ga\ga}y_\ga y_\ga \compR^{yy}\\&\qquad\qquad
+H^{\gad\gad}\pl_\gad\pl_\gad \bar{\compR}^{\pl\pl}+H^{\gad\gad}\bry_\gad\pl_\gad \bar{\compR}^{y\pl}+ H^{\gad\gad}\bry_\gad \bry_\gad \bar{\compR}^{yy}\,.
\end{aligned}
\end{align}
are nonzero (we have already used this type of decomposition in Section \ref{subsec:FronsdalCurrents} to derive the Fronsdal currents from $\formJ^{s.t.}$). In particular, $\compR^{\pl\pl}$ at $\bry=0$ contributes to exactly the same component as the second term of \eqref{omstcoc} and analogously for $\bar{\compR}^{\pl\pl}$ at $y=0$, but there are more contributions.

If we are, for example, interested in the holomorphic Weyl tensors $\Ctwo(y,0)$ we can extend the same procedure as was used in Section \ref{subsec:FronsdalCurrents} to solve the torsion constraint to higher levels, i.e. to $\omegatwo_{+k}$ with $k\geq1$. Indeed, the general formula for $Q_{-}^{-1}$ can be applied as:
\begin{align}
\omegatwo_k=(Q_-)^{-1}(\formR|_{k-1}-\nabla \omegatwo_{k-1}-Q_+\omegatwo_{k-2})\,.
\end{align}
For the purpose of extracting the relation between the Fronsdal fields and $\Ctwo$ we can drop the $Q_+$-term. Iterating the above formula we get
\begin{align}\label{WeylProj}
\Big[\mathcal{G}(\nabla\omegatwo_{0})&=\mathcal{V}(\Omega,\Omega,\Ctwo)+\mathcal{G}\,\formR\Big]\Big|^{\pl\pl}_{\bry=0}\,,
\end{align}
where we define the resolvent $\mathcal{G}^{-1}=(I+\nabla Q_-^{-1})$. Also, taking the $\bullet^{\pl\pl}$-component of the above expression and setting $\bry=0$ projects onto the $\Ctwo(y,0)$, c.f. \eqref{omstcoc}. Indeed,
\begin{align}
\Big[\mathcal{V}(\Omega,\Omega,\Ctwo)\Big]\Big|^{\pl\pl}_{\bry=0}=-\frac12\Ctwo(y,0)e^{-i\theta}\,.
\end{align}
On inspecting the action of $Q_{-}$ and $\nabla$ on various components of \eqref{Rdecomp}-like decomposition, we see that it is sufficient to work out the action of $(-\nabla Q_-^{-1})$ on the $\bullet^{\pl\pl}$-components as it is only these components that eventually contribute to $\Ctwo(y,0)$:
\begin{align}
(-\nabla Q_-^{-1})H^{\ga\ga}\pl_\ga\pl_\ga f^{\pl\pl}=H^{\ga\ga}\pl_\ga\pl_\ga\left( -\frac{1}{N(\brN+1)}(y\nabla \bar{\pl})f^{\pl\pl}\right)\,,
\end{align}
where $(y\nabla \bar{\pl})=y^\nu\nabla_{\nu\gnd}\pl^\gnd$. Therefore, we can define the $\bullet^{\pl\pl}$-projection $\mathcal{G}^{\pl\pl}$ of $\mathcal{G}$:
\begin{align}
\mathcal{G}^{\pl\pl}=\left(1+\frac{1}{N(\brN+1)}(y\nabla \bar{\pl})\right)^{-1}\,.
\end{align}
Also we need $(\nabla\omegatwo_{0})^{\pl\pl}=(2N)^{-1} (y\nabla \bar{\pl})\Front$. Finally, we get for \eqref{WeylProj}
\begin{align}
\mathcal{G}^{\pl\pl}\frac{1}{2N}(y\nabla \bar{\pl})\Front\Big|_{\bry=0}=-\frac12\Ctwo(y,0)e^{-i\theta}+\compR_{eff}\Big|_{\bry=0}
\,, && \compR_{eff}=\mathcal{G}^{\pl\pl}\,\compR^{\pl\pl}\,,
\end{align}
where $\compR_{eff}$ accounts for the difference between $\Ctwo$ and linearized HS Weyl tensors. If we drop $\compR_{eff}$ and project onto the spin-$s$ component then the free level relation is
\begin{align}
\frac{(-)^{s-1}}{4(2s-1)!}(y\nabla \bar{\pl})^s\phi_s=-\frac12\Ctwo_s(y,0)e^{-i\theta}\,,
\end{align}
which, up to some factor, identifies $C_{\ga(2s)}$ as the trace-free part of the order-$s$ curl of the traceless component $\phi$ of the Fronsdal field. In order to make the same rule work at the second order we have to shift $\Ctwo$ by $\compR_{eff}$:
\begin{align}
\Ctwo\longrightarrow \Ctwo+2 e^{i\theta} \compR^{\pl\pl}_{eff}\big|_{\bry=0}+2 e^{-i\theta} \bar{\compR}_{eff}^{\pl\pl}\big|_{y=0}=\tilde{C}^{(2)}\,,
\end{align}
where we also added the anti-holomorphic component $\bar{\compR}_{eff}^{\pl\pl}$, which is obtained analogously. Such a redefinition produces an additional contribution to $\formP$:
\begin{align}
\formP\longrightarrow \formP-2\tadD\left(e^{i\theta} \compR_{eff}^{\pl\pl}\big|_{\bry=0}+ e^{-i\theta} \bar{\compR}_{eff}^{\pl\pl}\big|_{y=0}\right)\,.\label{Pcorrections}
\end{align}
Notice that such redefinition cannot affect the scalar field of $\Ctwo$. Now, $\formP$ will contain new vertices both of type $\mathcal{V}(\Omega,C,C)$ and $\mathcal{V}(\omega,C)$ starting from $s=1$.

Lastly, let us stress here that the above contribution to the HS Weyl tensors must sum up, by consistency of the unfolded equation (see Appendix \ref{app:Consistency}), in such a way that the equation for $\Ctwo$ becomes just a consequence of the equation for the Fronsdal fields. In particular, the analysis of the $\Ctwo$-equation cannot alter our conclusions drawn directly from the Fronsdal equations. The only cases in which $\Ctwo$ contains nontrivial information on the dynamics are $s=0,1$. For example, we found that the Fronsdal current built out of the scalar field vanishes at $\theta=\pi/4$, which seems to be in conflict with \eqref{WeylTensorEq}. This discrepancy has to be cured by new terms coming from \eqref{Pcorrections}.

\subsection{Purely Star-Product Vertices}\label{subsec:SimplestExample} In this section we consider the simplest
example of the backreaction $\formJ$ that can be put on the right-hand side of $\adD \omegatwo=\formJ$. As before we may ignore $\mathcal{V}(\Omega,\Omega,\Ctwo)$ since it does not contribute to the Fronsdal current. If the right-hand side, $\formJ$, was a zero-form, then the simplest option would be to take $C\star \pi(C)$, which transforms in the adjoint of the HS algebra and hence obeys $\adD (C\star \pi(C))=0$. A small modification is needed to lift $C\star \pi(C)$ to a two-form:
\begin{align}
\adD \omegatwo&=\formJ^{smp}\,, &&
\begin{aligned} \formJ^{smp}&=(H^{\ga\ga}\pl_\ga\pl_\ga+H^{\gad\gad}\pl_\gad\pl_\gad) Q\,,\\ Q&=\exp i((y+\eta)(y+\xi)+(\bry-\breta)(\bry+\brxi))\,,
\end{aligned}
\end{align}
where $Q$ is the Fourier transform of $C\star \pi(C)$, i.e. $Q=e^{iY\xi}\star \pi(e^{iY\eta})$, and we omit the Fourier space integral along with $C(\xi)C(\eta)$. Such a source is very close to the one that comes from Vasiliev's equations, but it is much simpler. Moreover, it can be thought of as the Lorentz-covariant analog of the purely star-product vertex $\mathcal{V}(\omega,\omega,C,C)=\omega\star\omega\star C\star \pi(C)$ that is known to be consistent. Like any $\adD$-conserved two-form that is bilinear in $C$, $\formJ^{smp}$ is exact
\begin{align}
\formJ^{smp}=-\adD h^{\ga\gad}\pl_\ga \pl_\gad Q\,.
\end{align}
The general formula \eqref{FronsdalCurrentsDef} for the Fronsdal current gives
\begin{align}
\formj^{smp}&=H^{\gad\gad}\pl_\gad\pl_\gad \compj^{smp}\,, &&\compj^{smp}=\frac{-1}{(\brN-1)\brN} [-2i+(y(\xi-\eta))](\bry(\brxi+\breta))Q+Q\,,
\end{align}
where the actual Fronsdal current is the $N=\brN$ component of $\compj^{smp}$. Note that the current is traceless and hence canonical. The current is $\nabla$-conserved, i.e. its $AdS_4$-covariant divergence vanishes, which follows from
\begin{align}
\pl^\nu\pl^\gnd \nabla_{\nu\gnd}\compj^{smp}=0\,.
\end{align}
The coefficients can be easily extracted from the generating function of $\compj^{smp}$ and read simply:
\begin{align}\label{starproductcurrents}
\begin{aligned}
\square \Fron_{\mm(s)}+...&= \sum_{l,k} \tilde{a}_{l,k}\nabla_{\mm(s-k)\nn(l)} \Fron \nabla\fdu{\mm(k)}{\nn(l)}\Fron\,,\\
a_{l,k}&=\frac{(-)^k s!s!}{l!l!k!k!(s-k)!(s-k)}\frac{(1+2s)}{s-1}\,.
\end{aligned}
\end{align}
This example shows a slightly worse decay in the $l\rightarrow\infty$ limit than the Fronsdal currents obtained from Vasiliev Theory.

\section{Resummation of Fronsdal Currents and Locality}\label{sec:Resummation}
In this section we extract the primary canonical current component of the HS equations as derived from Vasiliev's theory. We have already observed that the HS symmetry, which is embedded into the unfolded equations, favors pseudo-local expressions arranged as primaries, which are canonical currents with the least number of derivatives, \eqref{canprimgenfunc}, with a tower of successors, \eqref{cangenfuncdes}. Considering cubic and higher order ansatz for an action, or second and higher order terms for field equations, one finds an infinite number of possible structures even if the spins of the three, four etc. fields are fixed. In flat space many of these structures, which correspond to successors, do not overlap with canonical currents and contribute only to improvements. On the contrary, in $AdS$ space, they contain a nonvanishing projection onto the primary canonical part. Whenever a pseudo-local tail, called Born-Infeld tail in \cite{Boulanger:2008tg}, can be resumed into a finite multiple of a primary, it is well-defined, which is not the case for the unfolded equations discussed in Section \ref{sec:SecondOrder} as we will explain now.

The simplest example is given by a $\Fron^3$-vertex, which can also be represented for any $l>0$ as
\begin{align}
 a_l\int \Fron \nabla_{\mm(l)}\Fron\nabla^{\mm(l)} \Fron \sim a_l C_l\int \Fron^3\,. \label{oversimpex}
\end{align}
The reason is that a pair of contracted derivatives can be eaten by $\square$ and $\square \Fron= m^2_\Fron \Fron$ will produce the same $\Fron^3$ vertex but with some overall coefficient $C_l$ in front of it. The same result is obtained by performing a field-redefinition directly in the equations of motion.\footnote{Similar techniques to study current interactions have been developed in \cite{Gelfond:2010pm} in terms of Howe dual algebras.} Therefore, given a possibly infinite number of terms \eqref{oversimpex} one can reduce them to the single $\Fron^3$ vertex with $a=\sum_l a_l C_l$ in front of it and the question is whether $a$ is finite or not.

While above we just gave the idea of the resummation let us now explain what one does in practice. One takes the most general redefinition $\omegatwo\rightarrow \omegatwo+\Delta \omegatwo$ such that $\Delta \omegatwo$ is bilinear in $C$ and it yields only canonical structures. It is easy to see that the most general such $\Delta \omegatwo$ is
\begin{align}
\Delta \omegatwo = h^{\ga\gad}\pl_\ga\pl_\gad Q^{can}\,,
\end{align}
where $Q^{can}$ is the generating function of canonical structures \eqref{cangenfunc}. After such a redefinition
the second-order equations read:
\begin{align}
\adD \omegatwo=\formJ^{can.}- \adD \Delta \omegatwo\,,
\end{align}
where $\formJ^{can.}$ is the canonical projection \eqref{HSstCanonical} of the full backreaction. Note that we can ignore the contribution of the Weyl tensors $\Ctwo$ as it does not affect the Fronsdal equations. Also we can drop any improvement pieces that are not relevant at this order. Moreover, we can omit $\omega\star \omega$ and $\mathcal{V}(\Omega,\omega,C)$ since they are explicitly local and cannot lead to any problems upon resummation, although they are not written in the standard form.

Next one computes $\adD \Delta \omegatwo$ and then the corresponding Fronsdal current $\formj(\adD \Delta \omegatwo)$ according to \eqref{FronsdalCurrentsDef} with the aim of identifying all improvement contributions to $\formJ^{can.}$. Let us denote the successors of the primary canonical currents that have $l$ pairs of indices contracted as $\square^l$-terms. The $\square^0$-terms are the primaries. In the Fourier space the $\square^l$ terms \eqref{cangenfuncdes} read: \begin{align}
j(y(\xi-\eta),\bry(\brxi+\breta))[(\xi\eta)(\brxi\breta)]^l\sim \square^l\,.
\end{align}
For any $\square^l$-term with coefficient $1$ in front of it in a given Fronsdal current it is possible to find a redefinition $\Delta \omegatwo$ such that (i) it contains only the terms $\square^0,...,\square^{l-1}$; (ii) it cancels the given $\square^l$-term; (iii) the result of such redefinition has only certain $\square^0$-term left  with a nontrivial coefficient $C_l$ in front of it:
\begin{align}
\square^l-\formj(\adD \Delta \omegatwo)=C_l\square^0\,.
\end{align}
It is a property of the recurrent system that relates $\Delta\omegatwo$ to $\formj(\adD \Delta \omegatwo)$ that such redefinition exists and the coefficient $C_l$ is a well-defined number that we find in Appendix \ref{app:Resummation} using more refined generating function techniques. Knowing $C_l$ allows one to project all $\square^l$-terms one by one, reducing the pseudo-local tail to its primary canonical part, which has the least number of derivatives, plus an improvement with no overlap on the latter.

There are at least three interpretations of the ideas described here-above. The first one is that one can perform redefinitions in the equations as to reduce every successor to $C_l$ times a primary. It is equivalent to integrating by parts in the action, as in \eqref{oversimpex}. Another interpretation is that we simply project every successor onto its primary component, i.e.
\be\langle \square^l\text{-successor}|\text{primary}\rangle=C_l\,,\ee
i.e. we go to the orthogonal base of structures. The latter interpretation is more general since it does not require an action --- we extract the improvements from a given $\adD$-closed structure and read off the coefficient in front of the only nontrivial part given by a primary canonical current. Therefore, our result does not rely on any subtle choice of decomposition, etc. In this regard it is worth stressing that there exists a basis of currents such that (i) each base vector is a local and conserved expression; (ii) it diagonalizes the action of $D$. In particular, every pseudo-local current can be decomposed into local independently conserved pieces and there exist no truly pseudo-local conserved expressions, which allows us to deal with them term by term. An example of a non-orthogonal base is given by primaries and successors.

Applying this algorithm to the spin-two current, the details being in Appendix \ref{app:Resummation}, the result of resummation is the primary two-derivative current:
\begin{equation}\label{spintwoinfinity}
\compj_{s=2}=-\frac{i}{12} \cos(2 \theta)\left(\sum_{l=1}^\infty l\right)\frac{(\bry(\brxi+\breta))^2 (y(\xi-\eta))^2}{4}+\text{improvements}\,,
\end{equation}
whose overall coefficient is given by a divergent series in $l$, each term of which comes from projecting $l$ contracted derivatives onto the primary current component, while improvements can be dropped from the very beginning.

It was noted already in \cite{Sezgin:2003pt} that the Fronsdal currents are pseudo-local and we have obtained them explicitly. In \cite{Boulanger:2008tg} it was noted that only a pseudo-local redefinition will bring the unfolded equations into the standard form (i.e. with the least number of derivatives). The possibility that the corresponding higher derivative tails could be divergent was also mentioned therein. That the above couplings indeed require infinite resummations is one of the original results that we obtain in the present paper.\footnote{Note that the divergences we observed are completely different from those considered e.g. in \cite{Iazeolla:2008ix} that are caused by choosing elements outside the domain of the star-product. Indeed, while the exponential formula \eqref{expform} works fine for polynomials, the integral formula \eqref{Ystarproduct} has a larger domain of definition, but still one can easily find elements $C(Y)$ such that $C\star \pi(C)$ diverges. } Since the equations for $\omegatwo$ and $\Ctwo$ are related in a simple way, our result is valid for the $\mathcal{V}(\Omega,C,C)$-vertex too, which explains the puzzles observed in \cite{Giombi:2009wh, Giombi:2010vg}.

Generally, we can say that given a pseudo-local expression there are at least three equivalent ways to find out if it can be redefined to a local one. Here-below we assume that the spins of the three fields whose contribution to the pseudo-local expression we are looking at are fixed.
\begin{itemize}
  \item Thinking of the source on the right-hand side of the Fronsdal equations as coming from a cubic Lagrangian vertex, one can attempt to redefine the pseudo-local tail and reduce the successors to the primary canonical structures, which have the minimal number of derivatives. In doing so, the coefficients in the tail will be resummed into the factors in front of the few standard cubic vertices, whose number is known to be finite and the order of derivatives is known to be not greater than the sum of the three spins \cite{Metsaev:2005ar}. The same can be achieved by performing redefinition directly in the equations of motion;

  \item One can compute the holographic correlation functions. Again, on the example of three-point functions the number of different structures is finite (moreover, it coincides with that of cubic vertices). Therefore, all the terms of the pseudo-local vertex will have to contribute to the coefficients in front of the few three-point structures;

  \item One can perform the Fefferman-Graham analysis of the given equations to see if the corrections due to the interactions in the pseudo-local non-linearities can destroy the asymptotic behavior.
\end{itemize}
The first and the second ways are essentially equivalent: thinking of the correlation functions, we can either first reduce everything to the standard vertices in the bulk and then compute the correlators or compute the correlators for each term in the pseudo-local expression individually since pure improvements give no contribution. More details will be given elsewhere \cite{ST}.

A marginal example of a pseudo-local expression is given by the HS algebra invariant \cite{Colombo:2012jx, Didenko:2012tv},
\begin{align}
tr(C\star \pi(C)\star...)\label{traceinv}\,,
\end{align}
whose equation of motion counterpart was considered in Section \ref{subsec:SimplestExample}. This expression gives the correct answers for all the correlation functions of the free dual CFT's when evaluated on the boundary-to-bulk propagators, and makes sense as a bulk object. Moreover,
it can be decomposed into the components, the contractions of the derivatives of the Weyl tensors, each of those making sense as a local
bulk vertex ($C$ is gauge invariant to lowest order). However, as a whole \eqref{traceinv} is not only pseudo-local, but is completely non-local since it does not depend on the bulk point at all.\footnote{The proof is simple \cite{Didenko:2012tv}. Any solution to $\tadD C=0$ has the form $C(Y|x)=g\star C(Y)\star \pi(g^{-1})$, where $g=g(Y|x)$ is an invertible element, i.e. the $x$-dependence comes solely from $g(Y|x)$ while $C(Y)$ is a space-time constant. The trace is by definition invariant under the adjoint transformations. Therefore, $g$ drops out and the trace does not depend on the point in the bulk.}  Remarkably, this observation forbids the phase-function ambiguity $\theta=\theta_\star(B)$ present in the Vasiliev theory \cite{Vasiliev:1999ba} by requiring locality in the bulk, see also \cite{Vasiliev:2015mka} for the Fefferman-Graham type of argument.

The vertices we derived from the Vasiliev equations are much closer to the marginal example of the trace invariants \eqref{traceinv} rather than to the standard local vertices. This is because the former vertices are given by the star-products with some insertions of homotopy integrals that can only slightly change the $l\rightarrow\infty$ fall-off of the coefficients coming from the Moyal product. Indeed, the difference is only $l^{-3}$ as can be seen by comparing with Section \ref{subsec:SimplestExample}.

From the thorough study of the $3d$ case \cite{Kessel:2015kna} we know that solving the torsion constraint brings successors of primary $s$-derivative currents, which can be canceled by introducing higher successors and so on. There exists a pseudo-local backreaction \cite{Kessel:2015kna, ST} that is fine-tuned in such a way that all the successors cancel each other to only produce the primary canonical current as Fronsdal current. That pseudo-local series is radically different from the one given by star-products and Vasiliev's equations. Star-products yield a spin-independent fall-off of the coefficients which is $(l!l!)^{-1}$ for bare star-products and $(l!l!l^{-3})$ for the backreaction we obtained from Vasiliev's equations. The pseudo-local representative for a primary spin-$s$ current has a faster fall-off, $((l+s)!(l+s)!)^{-1}$, which is spin-dependent (see \cite{Kessel:2015kna} for the $3d$ example and \cite{ST} for the $4d$ one). Therefore, the conclusion that the backreaction collapses onto the primary currents with infinite coefficient can be reached quite easily just by comparing the Fronsdal currents with the pseudo-local form of the canonical ones.

\section{Solving Vasiliev Equations}\label{sec:SolvingVasiliev}

In the present section we explain how to extract HS equations in the unfolded form. This requires a good knowledge of Vasiliev's equations. Some of the readers may wish to proceed to the Discussion Section immediately. A gentle introduction in the perturbation theory of Vasiliev's equations can be found in Section 11 of \cite{Didenko:2014dwa}.

The unfolded equations are folded into Vasiliev's equations \cite{Vasiliev:1990en, Vasiliev:1990vu, Vasiliev:1999ba}. Once the Vasiliev's equations are partially solved with respect to the additional variables $Z_A$, the initial data for the $Z$-evolution are found to obey the unfolded equations \eqref{xpaceseq}. The idea behind Vasiliev's equations is that the unfolded equations \eqref{xpaceseq} can be embedded into a flat connection of a bigger algebra plus constraints that specify the embedding, which we need to solve for.

\subsection{Vasiliev Equations}\label{subsec:VasilievEqs}

The algebra \cite{Vasiliev:1992av} that all HS interactions can be embedded into as a flat connection is a twisted product of two copies of the HS algebra \cite{Alkalaev:2014nsa}. The doubling is achieved via $Y^A\rightarrow Y^A,Z^A$ and the twist is encoded in the following star-product \cite{Vasiliev:1992av}
\begin{equation}\label{StarYZ}
(f\star g)(Y,Z)=\int dU dV f(Y+U,Z+U) g(Y+V,Z-V) e^{iU_A V^A}\,,
\end{equation}
which is not a usual product on the tensor product of the two copies and corresponds to normal ordering for $Y\pm Z$, but for $Z$-independent functions \eqref{StarYZ} reduces to  \eqref{Ystarproduct}. All of the star-product computations with monomials can be done by applying the simple consequences of \eqref{StarYZ}:
\begin{align}
Y_A\star\bullet&=Y_A + i\pl^Y_A
-i \pl^Z_A\,,
& Z_A\star\bullet&=Z_A+i\partial^Y_A-i\partial^Z_A\,, \\
\bullet\star Y_A&=Y_A-i\partial^Y_A-i\partial^Z_A\,, &
\bullet\star Z_A&=Z_A+i\partial^Y_A+i\partial^Z_A\,.
\end{align}

The field content of the theory is given by one-form connection $W=W_\mm(Y,Z|x)\, dx^\mm$, an auxiliary field $S_A=S_A(Y,Z|x)$ that is an $sp(4)$-vector and a zero-form $\Phi=\Phi(Y,Z|x)$. The $4d$ Vasiliev equations in the $sp(4)$-covariant form read:
\besubeqs\label{4dVasiliev}
\begin{align}
&dW=W\star W\,,\\
&d(\Phi\star \klein)=[W, \Phi\star \klein]_\star\,,\\
&dS_{A}=[W,S_{A}]_\star \,, \\
&[S_A,S_B]_\star =-2i(C_{AB}+\Phi\star \Upsilon_{AB})\,,\\
&S_A\star \Phi+\Phi\star (\Upsilon\star S\star \Upsilon^{-1})_A=0\,.
\end{align}
\esubeqs
Here the $sp(4)$ symmetry is broken by the $so(3,2)$-compensator $V_{AB}$ via two projectors:\footnote{For example, one can choose $C_{AB}=\begin{pmatrix}
                                            \epsilon & 0 \\
                                            0 & \epsilon \\
                                          \end{pmatrix}$ and
                                          $V_{AB}=\begin{pmatrix}
                                            \epsilon & 0 \\
                                            0 & -\epsilon \\
                                          \end{pmatrix}$. }
\begin{align}
\Upsilon_{AB}&=\begin{pmatrix}
                 \epsilon_{\ga\gb}e^{i\theta} \klein & 0 \\
                 0 & \epsilon_{\gad\gbd}e^{-i\theta}\brklein \\
               \end{pmatrix}=
               \Pi^+_{AB}e^{i\theta} \klein+\Pi^-_{AB}e^{-i\theta}\brklein\,, && \Pi^{\pm}_{AB}=\frac12(C_{AB}\pm V_{AB})\,,
\end{align}
whose crucial constituents are a free parameter $\theta$ and two Klein operators
\begin{align}
\klein&=e^{iz_\ga y^\ga}\,,& \brklein&=e^{i\bar z_\gad \bar y^\gad}\,.
\end{align}
The fields need to obey kinematical constraints ensuring the theory to be bosonic:
\begin{align}
\label{4dbosonic}
[K,\Phi]_\star  & =0\;, & [K,W]_\star &=0\;, & \{K, S_A\}_\star &=0\;,
\end{align}
where $K=\klein\star \brklein$ is the total Klein operator. Operator $K$ behaves as $(-)^{N_Y+N_Z}$ where $N_Y$ and $N_Z$ are the number operators for $Y$ and $Z$, i.e. $K\star Y \star K=-Y$, etc. The holomorphic Klein operators $\klein$ and $\brklein$ do the same for $y_\ga,z_\ga$ and $\bry_\gad,\brz_\gad$, respectively. It is also useful to remember that
\begin{align}
f(y,z)\star \klein =f(-z,-y) \klein\,, && \klein\star\klein=1\,.
\end{align}
In practice, it is convenient to decompose the $sp(4)$-covariant equations to the $sl(2,\mathbb{C})$-components:
\besubeqs\label{VasAll}
\begin{align}
&dW=W\star W\,,\\
&d(\Phi\star \klein)=[W, \Phi\star \klein]_\star\,,\\
&dS_{\ga}=[W,S_{\ga}]_\star \,,
&&d\bar{S}_{\gad}=[W, \bar{S}_{\gad}]_\star \,,\\
&[S_{\ga},
S_{\gb}]_\star =-2i\epsilon_{\ga\gb}(1+e^{i\theta}\Phi\star \klein)\,,
&&[\bar{S}_{\gad},
\bar{S}_{\gbd}]_\star =-2i\epsilon_{\gad\gbd}(1+e^{-i\theta}\Phi\star \brklein)\,,\label{BBbar}\\
&\{S_{\ga}, \Phi\star \klein\}_{\star }=0\,,
&&\{\bar{S}_{\gad}, \Phi\star \brklein\}_{\star }=0\,,\\
&[S_{\ga}, \bar{S}_{\gad}]_\star =0\,.
\end{align}
\esubeqs
The prescription on how to extract the unfolded equations for HS fields is to solve perturbatively for the $Z$-dependence all but the first two equations, which we will do in the next section.

\subsection{Lorentz Covariant Perturbation Theory}\label{subsec:Perturbation}
Lorentz covariance imposes strong restrictions on the form of the vertices in the unfolded approach. The background Lorentz covariant derivative is already inside $\adD$ and $\tadD$, \eqref{AdjointDer}-\eqref{TwistedDer}, so we expect not to find it anywhere else, for example,  $\mathcal{V}(\Omega,\Omega,C,C)$ should depend on the background vierbein $h$ only, i.e. we have $\mathcal{V}(h,h,C,C)$. Notice that as in any nonlinear theory we are allowed to perform field-redefinitions, which can result in equations where Lorentz covariance is no longer manifest. Therefore, it is important to start in an appropriate, Lorentz covariant, frame. Within the Vasiliev framework this requires some dedicated efforts, which we shall briefly discuss below. A detailed exposition can be found in \cite{Prokushkin:1998vn, Vasiliev:1999ba,Sezgin:2002ru, Sezgin:2011hq, Didenko:2014dwa}.

First of all, let us list various generators that the Vasiliev theory is equipped with:
\besubeqs
\begin{align}
sp(4)&: &T_{AB}&=-\frac{i}{4}\{Y_A,Y_B\}\,, \\
sp(2)\oplus sp(2)&: &L^y_{\ga\ga}&=-\frac{i}{4}\{y_\ga,y_\ga\} \,, &\bL^y_{\gad\gad}&=-\frac{i}{4}\{\bry_\gad,\bry_\gad\}\,,\\
sp(2)\oplus sp(2)&: &L^z_{\ga\ga}&=+\frac{i}{4}\{z_\ga,z_\ga\}  \,, &\bL^z_{\gad\gad}&=+\frac{i}{4}\{\brz_\gad,\brz_\gad\}\,,\\
sp(2)\oplus sp(2)&: &K_{\ga\ga}&=+\frac{i}{4}\{S_\ga,S_\ga\}  \,, &\bar{K}_{\gad\gad}&=+\frac{i}{4}\{S_\gad,S_\gad\}\,.
\end{align}
\esubeqs
The $sp(4)$ generators $T_{AB}$ of the $AdS_4$-algebra can be split into two copies\footnote{The reality conditions $y^\dag_\ga=\bry_\gad$ makes two copies of $sp(2)$ into $sl(2,\mathbb{C})$, which is the $4d$ Lorentz algebra.} $sp(2)\oplus sp(2)$ of the Lorentz algebra generators $L^y_{\ga\ga}$ and $\bL^y_{\gad\gad}$. The complementary translation generators are
\begin{align}
P_{\ga\gad}&=-\frac{i}4 \{y_\ga, \bry_\gad\}\,.
\end{align}
There is another copy of $sp(4)$ and hence $sp(2)\oplus sp(2)$ that is due to $Z^A$. Moreover, a subset of Vasiliev's equations guarantees that $S_\ga$ and $S_\gad$ form two copies of $osp(1|2)$ algebra where $S_\ga$ and $S_\gad$ play the role of the odd generators. Therefore, $K_{\ga\ga}$ and $\bar{K}_{\gad\gad}$ deliver another $sp(2)\oplus sp(2)$.

Given the kinematical, $L^y$, $L^z$, and dynamical, $K$, generators of the Lorentz algebra one can identify the correct Lorentz generators with the help of the coset construction:
\begin{align}\label{RightLorentz}
\widehat{L}_{\ga\ga}&=L^y_{\ga\ga}+L^z_{\ga\ga}-K_{\ga\ga}\,, &
\widehat{\bL}_{\gad\gad}&=\bL^y_{\gad\gad}+\bL^z_{\gad\gad}
-\bar{K}_{\gad\gad}\,.
\end{align}
The Vasiliev equations are background independent and a vacuum solution to expand over is given by empty AdS space that is realized as $\Phi=0$, $W=\Omega$ and, most importantly, $S_A=Z_A$. Here $\Omega$ is an $sp(4)$ flat connection
\begin{align}\label{VacuumVas}
\Omega &=\tfrac12\,\varpi^{\ga\ga} L^y_{\ga\ga} +
h^{\ga\gad} P_{\ga\gad} +\tfrac12\,\bar{\varpi}^{\gad\gad} \bar{L}^y_{\gad\gad}\,,
\end{align}
which looks exactly the same as \eqref{Omegaconn}, but it takes values in the extended algebra of $Y$ and $Z$, which results in $[\Omega,\bullet]_\star$ having a form different from \eqref{AdjointDer} by $\pl^z_A$-terms. These additional terms do not spoil the consistency of the vacuum solution since it depends on $Y$ only. The background fields $h^{\ga\gad}$ and $\varpi^{\ga\ga}$, $\varpi^{\gad\gad}$ obey the same equations that followed from $d\Omega=\Omega\star\Omega$:\footnote{Let us note that we consider the perturbation theory that is Lorentz-covariant with respect to the background only, which is slightly different from \cite{Prokushkin:1998vn, Vasiliev:1999ba,Sezgin:2002ru, Didenko:2014dwa} where the spin-connection is taken to be the full spin-connection rather than only the background part of it.}
\besubeqs\label{flatadsconnection}
\begin{align}
R_{\ga\ga}&=d\varpi^{\ga\ga}-\varpi\fud{\ga}{\gamma}\wedge\varpi^{\gamma\ga}-h\fud{\ga}{\dot\gamma} \wedge h^{\ga\dot\gamma}=0\,,\\
\bar R_{\gad\gad}&=d\bvarpi^{\gad\gad}-\bvarpi\fud{\gad}{\dot\gamma}\wedge\bvarpi^{\dot\gamma\gad}-h\fdu{\gamma}{\gad} \wedge h^{\gamma\gad}=0\,,\\
T_{\ga\gad}&=dh^{\ga\gad}-\varpi\fud{\ga}{\gamma}\wedge h^{\gamma\gad}-\bvarpi\fud{\gad}{\dot\gamma} \wedge h^{\ga\dot\gamma}=0\,,
\end{align}
\esubeqs
i.e. they are connections of the empty $AdS_4$ space.

In accordance with the identification of the Lorentz generators in \eqref{RightLorentz} the correct background to yield a Lorentz-covariant perturbation theory is
\begin{align}\label{VacuumVasRight}
\hat{\Omega} &=\tfrac12\,\varpi^{\ga\ga} \widehat{L}_{\ga\ga} +
h^{\ga\gad} P_{\ga\gad} +\tfrac12\,\bar{\varpi}^{\gad\gad} \widehat{\bL}_{\gad\gad}\,,
\end{align}
which reduces to \eqref{VacuumVas} for the vacuum solution since $S_A=Z_A$ and hence $K=L^z$, i.e. $\widehat{L}=L^y$ at the zeroth order. At higher orders, as we will see, $\hat{\Omega}$ receives corrections via $\widehat{L}=\widehat{L}(S_A)$ in such a way as to make the perturbation theory manifestly Lorentz-covariant.

Next, we shift the fields by the background values thereof and insert $2i$ sometimes as to avoid cumbersome factors as much as possible:
\begin{align}\label{shift}
S_A&\longrightarrow Z_A+2i \aA_A\,, & W&\longrightarrow \Omega+W\,, &
\Phi&\longrightarrow2iB\,.
\end{align}
Let us first present the equations that form a closed subsector of differential equations with respect to auxiliary $Z$ variables and do not involve any space-time derivatives:
\besubeqs\label{VasAllB}
\begin{align}
&\partial_A B=\mathcal{A}{}_A\star B+B\star \begin{vmatrix}
                                              \klein\star\mathcal{A}_\ga\star\klein \\
                                              \brklein\star\mathcal{A}_\gad\star\brklein \\
                                            \end{vmatrix}\,,\\
&\pl_A \mathcal{A}_B-\pl_B \mathcal{A}_A= [\mathcal{A}_A,\mathcal{A}_B]_\star+\begin{vmatrix}
                                                                                \epsilon_{\ga\gb}e^{+i\theta} B\star\klein & 0 \\
                                                                                0 & \epsilon_{\gad\gbd}e^{-i\theta} B\star\brklein \\
                                                                              \end{vmatrix}\,,
\end{align}
\esubeqs
where $\pl_A=\pl^Z_A$, which came out of $[Z_A,\bullet]_\star=-2i \pl_A$. The solutions
can be obtained iteratively from\footnote{\label{defhomint}Indeed, the equations are of the form $\pl_A B=F_A$ and $\pl_A \mathcal{A}_B-\pl_B \mathcal{A}_A=F_{AB}$, where the right-hand side obey the Frobenius integrability constraints: $\pl_A F_B-\pl_B F_A\equiv0$ and analogously for $F_{AB}$. Therefore, we have a standard problem of solving $\mathrm{d} a=b$, $\mathrm{d}b=0$ for the exterior derivative $\mathrm{d}=dZ^A \pl_A $ in the $Z_A$-space. The solution is given by contracting homotopy, which is defined to be $\homo{n}{f}=\int_0^1 t^n\,dt\, f(zt)$. }
\begin{subequations}\label{solutionA}
\begin{align}
B&= C(Y)+z^\alpha\homo{0}{\aA_\alpha\star B+B\star\pi(\aA_\alpha)}+\brz^\gad\homo{0}{\aA_\gad\star B+B\star\pib(\aA_\gad)}\,,\label{solutionAA}\\
\aA_\alpha&=z_\alpha \homo{1}{\aA_\gamma\star \aA^\gamma}+\brz^\gbd\homo{1}{[\aA_\gbd, \aA_\alpha]}+z_\alpha \homo{1}{B\star \klein}e^{+i\theta}+\pl_\ga\epsilon\,,\label{solutionAB}\\
\aA_\gad &=z_\gad \homo{1}{\aA_\gdd\star \aA^\gdd}+z^\alpha\homo{1}{[\aA_\alpha, \aA_\gad]}+\brz_\gad \homo{1}{B\star \brklein}e^{-i\theta}+\pl_\gad\epsilon\,,\label{solutionAC}\end{align}
\end{subequations}
with the initial condition $B^{(1)}=C(Y)$. Above we defined $\pi(f)=\klein\star f \star \klein$ and $\pib(f)=\brklein\star f \star \brklein$, which are the extensions of the $\pi$-map of \eqref{trivialcocycles} to the $Y$-$Z$ space.

Note that the pure gauge modes of $\aA_A$ are given by $\epsilon=\epsilon(Y,Z)$. Any gauge choice gives a consistent system of unfolded equations. The simplest one, which is implied in \cite{Vasiliev:1999ba}, is to kill $\epsilon$ by the Schwinger-Fock gauge:
\be
Z^A \aA_A=0\,.\label{SFgauge}
\ee
We impose the Schwinger-Fock gauge in all our analysis and it is crucial to reproduce the formulas here-below. For the rest of the equations, those that do involve space-time $d=dx^\mm\pl_\mm$, we get
\begin{subequations}\label{VasAllA}
\begin{align}
\DO W&=W\star\wedge W-\frac12\left[h\fud{\ga}{\dot\gamma} \wedge h^{\ga\dot\gamma}(L^z_{\ga\ga}-K_{\ga\ga})+h.c.\right]\,,\label{VasAllAA}\\
\tadDO B&=W\star B-B\star \pi(W)\,,\label{VasAllAB}\\
\pl_A W&=-ad_h\aA_A- [W,\aA_A]+ \chi_A\,, \qquad\qquad Z^A\chi_A=0\,.\label{VasAllAC}
\end{align}
\end{subequations}
Here $\DO$ and $\tadDO$ are the AdS covariant derivatives defined as
\begin{align}
\DO f&=df-\Omega\star f\pm f\star \Omega=\nabla^{\rm{yz}}f-ad_h f\,,\\
\tadDO f&=df -\Omega\star f\pm f\star \pi(\Omega)=\nabla^{\rm{yz}}f-\widetilde{ad}_h f\,,
\end{align}
with respect to the effective connection
\begin{align}\label{VacuumVasRightYZ}
\Omega^{\rm{yz}} &=\tfrac12\,\varpi^{\ga\ga} {L}^{\rm{yz}}_{\ga\ga} +
h^{\ga\gad} P_{\ga\gad} +\tfrac12\,\bar{\varpi}^{\gad\gad} {\bL}^{\rm{yz}}_{\gad\gad}
\end{align}
in terms of the diagonal generators ${L}^{\rm{yz}}$:
\begin{align}
{L}^{\rm{yz}}_{\ga\ga}&=L^y_{\ga\ga}+L^z_{\ga\ga}\,, &
{\bL}^{\rm{yz}}_{\gad\gad}&=\bL^y_{\gad\gad}+\bL^z_{\gad\gad}\,.
\end{align}
The $h$-part of $\Omega^{\rm{yz}}$ is the same as for $\hat{\Omega}$ and acts as follows:
\besubeqs
\begin{align}
[h,\bullet]_\star&=\ad_h=h^{\ga\gad}\left((y_\ga-i\pl^z_\ga)\pl^y_\gad +(y_\gad-i\pl^z_\gad)\pl^y_\ga)\right)\,,\\
\{h,\bullet\}_\star&=\tad_h=-ih^{\ga\gad}\left((y_\ga-i\pl^z_\ga)(y_\gad-i\pl^z_\gad)-
\pl^y_\ga\pl^y_\gad\right)\,.
\end{align}
\esubeqs
Both $\DO$ and $\tadDO$ contain $Y$-$Z$ Lorentz-covariant derivative $\nabla^{\rm{yz}}$:
\begin{align}
\nabla^{\rm{yz}}\bullet=d+\varpi^{\ga\ga}(y_\ga\pl^y_\ga+z_\ga\pl^z_\ga)+\bvarpi^{\gad\gad}(\bry_\gad\pl^y_\gad+\brz_\gad\pl^z_\gad)\,.
\end{align}
Note that on the space of $Z$-independent function $\DO$ and $\tadDO$ coincide with $\adD$, \eqref{AdjointDer}, and $\tadD$, \eqref{TwistedDer}, respectively.

In order to derive \eqref{VasAllA} one has to use the equations of motion several times as well as the Schwinger-Fock gauge. There is one more $Z$-equation, \eqref{VasAllAC}, which can be solved as
\begin{align}\label{solutionB}
W&=\omega(Y)-z^\ga\homo{0}{\ad_h\aA_\ga}-\brz^\gad\homo{0}{\ad_h\aA_\ga}-
z^\ga\homo{0}{[W,\aA_\ga]}-\brz^\gad\homo{0}{[W,\aA_\gad]}\,,
\end{align}
where it was important that  $\chi_A$ obeys $Z^A\chi_A=0$ and hence disappears from the homotopy integrals. Also note that the redefinition of the Lorentz generators yields the last term in \eqref{VasAllAA}.

In order to get the unfolded equations \eqref{xpaceseq} for HS fields out of the Vasiliev equations one performs the following algorithm: one starts from \eqref{solutionAA} with the initial condition $B^{(1)}=C(Y)$, then one uses \eqref{solutionAB}-\eqref{solutionAC}. Next one gets $W$ from \eqref{solutionB} and repeats the cycle to get to the desirable order of perturbations. At every order, there are two integration constants, $C^{(n)}(Y)$ and $\omega^{(n)}(Y)$ that enter as dynamical variables in the perturbative expansion of \eqref{xpaceseq}. The last step to get \eqref{xpaceseq} is to plug the solutions for $W$ and $B$ to \eqref{VasAllAA}-\eqref{VasAllAB} and set $Z=0$, which results in the consistent unfolded equations in terms of $C^{(n)}(Y)$ and $\omega^{(n)}(Y)$ only. We make two iterations of this cycle in the next two sections.

\subsection{First Order}\label{subsec:FirstOrder}

The iterations begin with $\Bone=C(Y)$ and then using the algorithm we get:
\begin{subequations}
\begin{align}
\Bone&= C(Y)\,,\\
\aA^{(1)}_\mu&=\Aone_\mu=z_\mu \homo{1}{C\star \klein}e^{i\theta}=z_\mu \int_0^1 t\, dt\, C(-zt,\bry) e^{ityz+i\theta}\,,\\
\aA^{(1)}_\gmd&=\Aone_\gmd=\brz_\gmd \homo{1}{C\star \brklein}e^{-i\theta}=z_\gmd \int_0^1 t\, dt\, C(y,-t\brz) e^{it\bry\brz-i\theta}\,,\\
\Wone&=\omega(Y)+\Mone\,,
\end{align}
\end{subequations}
where we defined
\begin{equation}
\Mone=-z^\ga\homo{0}{\ad_h\Aone_\ga}-\brz^\gad\homo{0}{\ad_h\Aone_\gad}\,.
\end{equation}
It can be computed and simplified to\footnote{For $n\neq m$ the nested homotopy integrals can be resolved as $\Gamma_n\circ \Gamma_m=-(\Gamma_n-\Gamma_m)/(n-m)$.}
\begin{equation}
\Mone=-i h^{\ga\gad}z_\ga \pl^y_\gad \int_0^1 (1-t)\, dt\, C(-zt,\bry) e^{ityz+i\theta}+h.c.\,,
\end{equation}
The space-time equations at the first order are simply:
\begin{align}
\DO \Wone\big|_{Z=0}&=0\,,  &\tadDO \Bone\big|_{Z=0}&=0\,,
\end{align}
which results in equations we quoted in \eqref{xpaceseqQBA}:
\begin{align}
\adD \omega&=\mathcal{V}(\Omega,\Omega,C)\,, &\tadD C&=0\,,
\end{align}
where $\mathcal{V}(\Omega,\Omega,C)$, given at $\theta=0$ in \eqref{OMSTcocycle}, is
\be
\mathcal{V}(\Omega,\Omega,C)=\ad_h \Mone\big|_{Z=0}=-\frac12 H^{\gad\gad}\pl^y_\gad\pl^y_\gad C(0,\bry)e^{i\theta}+h.c.\,.
\ee
Therefore, we correctly reproduced the free equations for HS fields in the unfolded form.

\subsection{Second Order}\label{subsec:SecondOrder}
Continuing iterations we get at the second order:
\begin{subequations}
\begin{align}
\Btwo&=C^{(2)}(Y)+ B'^{(2)}\,,\\
\aA^{(2)}_\mu&=z_\mu \homo{1}{\Aone_\gamma\star \aA^{(1)\gamma}}+\brz^\gmd\homo{1}{[\Aone_\gmd, \Aone_\mu]}+z_\mu \homo{1}{\Btwo\star \klein}e^{i\theta}\,,\\
\aA^{(2)}_\gmd&=\brz_\gmd \homo{1}{\Aone_\gdd\star \aA^{(1)\gdd}}+z^\mu\homo{1}{[\Aone_\mu, \Aone_\gmd]}+\brz_\gmd \homo{1}{\Btwo\star \brklein}e^{-i\theta}\,,\\
\Wtwo&=\omega^{(2)}(Y)+M^{(2)}-z^\ga\homo{0}{[\omega,\Aone_\ga]}-\brz^\gad\homo{0}{[\omega,\Aone_\gad]}\\
&\qquad\qquad\qquad-z^\ga\homo{0}{[\Mone,\Aone_\ga]}-\brz^\gad\homo{0}{[\Mone,\Aone_\gad]}\,,
\end{align}
\end{subequations}
where
\begin{subequations}
\begin{align}
\Mtwo&=-z^\ga\homo{0}{\ad_h\aA^{(2)}_\ga}-\brz^\gad\homo{0}{\ad_h\aA^{(2)}_\gad}\,,\\
B'^{(2)}&=z^\nu\homo{0}{\Aone_\nu\star C+C\star\pi(\Aone_\nu)}+\brz^\gnd\homo{0}{\Aone_\gnd\star C+C\star\pib(\Aone_\gnd)}\,.
\end{align}
\end{subequations}
At the second order the space-time equations become
\begin{subequations}
\begin{align}
\DO\Wtwo&=(\omega+\Mone)\star (\omega+\Mone)-iH^{\ga\ga}\mathcal{A}^{(1)}_\ga\star \mathcal{A}^{(1)}_\ga-iH^{\gad\gad}\mathcal{A}^{(1)}_\gad\star \mathcal{A}^{(1)}_\gad\,,\\
\tadDO\Btwo&=\omega\star C-C\star \pi(\omega)+\Mone\star C-C \star\pi(\Mone)\,,
\end{align}
\end{subequations}
where we assumed that $Z=0$ is imposed after computing all star-products. The computations can be considerably simplified if one notices that the $Z$-dependent terms resulting
from homotopy integrals are always proportional to $Z$. Therefore, only the part of $\DO$ or $\tadDO$ that contains $\pl^Z_A$ needs to be taken into account:
\begin{equation}
\DO(Z f(Y,Z))\Big|_{Z=0}=-ad_h Z f(Y,Z)\Big|_{Z=0}\,,
\end{equation}
which can be expanded to (with analogous formulas with $\brz$)
\begin{subequations}\label{adzformula}
\begin{align}
\ad_h z^\nu f_\nu(y,z)\Big|_{z=0}&=-ih^{\nu\gad}\pl^y_\gad f_\nu(y,0)\,,\\
\tad_h z_\nu f(y,z)\Big|_{z=0}&=-h\fdu{\nu}{\gad}\left(\bry_\gad-i\pl^z_\gad\right)f(y,z)\Big|_{z=0}\,.
\end{align}
\end{subequations}
In this way one can extract various vertices:
\begin{subequations}\label{cocyclesgeneral}
\begin{align}
\mathcal{V}(\Omega,C,C)&=\Mone\star C-C\star \pi(\Mone)+\tad_h(B'^{(2)})\,,\\
\mathcal{V}(\Omega,\omega,C)&=\{\omega, \Mone\}_\star-\ad_h z^\nu\homo{0}{[\omega,\Aone_\nu]_\star}-\ad_h \brz^\gnd\homo{0}{[\omega,\Aone_\gnd]}\,,\\
\mathcal{V}(\Omega,\Omega,C,C)&=\Mone\star \Mone+iH^{\ga\ga} \Aone_\ga\star \Aone_\ga+iH^{\gad\gad} \Aone_\gad\star \Aone_\gad\\
&-\ad_h z^\nu \homo{0}{[\Mone,\Aone_\nu]_\star}-\ad_h \brz^\gnd \homo{0}{[\Mone,\Aone_\gnd]_\star}+\ad_h \Mtwo\,,\nonumber
\end{align}
\end{subequations}
where again $Z=0$ at the end. The expressions here-above provide all the information about the unfolded equations for HS fields at the second order. Details can be found in Appendix \ref{app:VerticesComputations}, where we computed, simplified and verified them, while the final expressions are collected in Section \ref{sec:SecondOrder}.

\section{Conclusions and Discussion}
\label{sec:Conclusions}

In this paper we worked out the second order equations for HS fields as they come out of the Vasiliev theory, i.e. in the unfolded form \eqref{xpaceseq}, which can be considered as a follow-up on \cite{Kessel:2015kna} where the $3d$ HS theory was studied. A dictionary between the unfolded equations and the Fronsdal equations was established and the corrections to the free propagation were explicitly found. The general feature of such corrections is that they are arranged in infinite sums of derivatives. The most important result of our study was to quantify the degree of non-locality that the HS equations exhibit: it is too high to enable one to interpret the equations as providing a local field theory in $AdS_4$.

A considerable part of the problem is about giving a precise definition of what HS theories are. At present, there seem to be two different definitions.

The first definition is that the HS equations are the unfolded equations of type \eqref{xpaceseq}. The boundary condition is to reproduce free HS fields dynamics when linearized around $AdS$ background. An additional constraint is on the classes of functions that the vertices $\mathcal{V}(\omega,\omega,C,...,C)$ belong to and on the redefinitions that preserve such a class. The old definition of what constitutes a good functional class can be found in \cite{Vasiliev:1988sa}. It states that $\mathcal{V}(\omega,\omega,C,...,C)$ should yield a polynomial once the arguments are polynomials in the HS algebra generating element $Y$.

The second definition, inspired by the AdS/CFT correspondence and specific conjectures relating HS theories to certain CFT's \cite{Sezgin:2002rt,Klebanov:2002ja,Sezgin:2003pt,Gaberdiel:2010pz,Chang:2012kt}, is that HS theories should reproduce the correlation functions of these CFT's. There seem to be no obstructions in reconstructing HS theories this way, see \cite{Bekaert:2014cea} for some results in this direction. While the problem of redefinitions that do not change the holographic S-matrix still remains, the holographic reconstruction can provide at least an example of the dual HS theory.

We would like to separate the two definitions from Vasiliev Equations that provide a tool to generate unfolded HS equations upon solving for the $Z$-dependence. The map from Vasiliev's equations to the unfolded HS equations is not unique: the ambiguity is due to the gauge choice that needs to be made when solving for the $Z$-dependence. The only choice available at present in the literature that reproduces the free dynamics around $AdS$ is the Schwinger-Fock gauge \eqref{SFgauge}.

The important point is to distinguish between the two definitions given above. As our results show, well-defined unfolded equations may lead to problems when standard field theory methods are applied. Since the zero-form $C$ encodes all the on-shell nontrivial derivatives of the fields, it contains an unbounded number of derivatives. As a result, an expression that is seemingly well-behaved in the formal expansion scheme, for example, $C\star C$, turns out to be pseudo-local even after fixing the spins. Moreover, it is not guaranteed that it can be reduced to the standard interaction with a finite coefficient in front of it.

We would like to add another definition of what is a HS theory in $AdS$ that relates the above two: There should exist a well-defined action principle or equations of motion formulated in terms of the minimal base of structures so as to avoid any infinities. The infinity we found is due to the usage of non-orthogonal base of structures: primaries and successors, where successors have a nonvanishing overlap with the primaries. The simplest example is that while the equations ($\Phi$ and $\Phi'$ are two different scalar fields)
\begin{align}
\square \Fron'=\sum_l a_l \nabla_{\mm(l)}\Fron\nabla^{\mm(l)}\Fron
\end{align}
are seemingly well-defined for any $a_l$, an attempt to derive them from the cubic vertex
\begin{align}
\sum_l a_l\int \Fron' \nabla_{\mm(l)}\Fron\nabla^{\mm(l)} \Fron\sim\sum_l a_l C_l \int \Fron'\Fron^2
\end{align}
tells us that all the terms are proportional to each other, and the question is whether $a=\sum_l a_l C_l$ is convergent or not, where $C_l$ are the numbers produced by integrations by parts and commuting covariant derivatives. The same result can be obtained by redefinitions $\Fron'\rightarrow \Fron'+\sum_l b_l \nabla_{\mm(l)}\Fron\nabla^{\mm(l)}\Fron$ directly in the equations of motion, resulting in the same resummed coupling $a=\sum_l a_l C_l$. While it is easy to imagine a pseudo-local coupling such that $a$ is finite, our result \eqref{spintwoinfinity} shows that it is not the case for the HS equations obtained from Vasiliev's equations in the Schwinger-Fock gauge.

Therefore, we see that the first definition is broader than the other two. There are unfolded equations for HS fields that yield naked infinities whenever one tries to bring them into a standard form by redefining higher derivatives or to compute correlation functions.\footnote{In the flat space limit one would reconstruct $\tfrac{1}{\square}\sim\tfrac10$ singularities discussed in \cite{Taronna:2011kt}.}

As we mentioned, there is a freedom in Vasiliev Equations that should allow one to resolve the problem: a gauge choice that needs to be made when solving for the $Z$-dependence and that would restore the locality in the unfolded equations. So far we have blindly followed the prescription to impose the Schwinger-Fock gauge. This choice leads to the simplest form of the free equations, but it allows for an ambiguity already at the second order. Unfortunately, the ambiguity is a functional one --- one can set $\epsilon$ in \eqref{solutionAB}-\eqref{solutionAC} to be any function that is bilinear in $C$. Different gauge choices in the Vasiliev equations would lead to different, possibly inequivalent, unfolded equations.

Despite the functional ambiguity hidden in the map from Vasiliev's equations to unfolded ones, the problem of fixing the gauge as to match the required correlation functions seems to be more tractable than that of writing down the most general ansatz for the Lagrangian and imposing agreement with AdS/CFT. The reason is that the HS symmetry is more easily taken into account by the unfolded equations.

Another possible resolution would be to find a regularization that preserves HS symmetries
and relates the standard frame with the minimal number of derivatives to a pseudo-local one via some regularization scheme. An indication that such a scheme does exist comes from \cite{Giombi:2010vg} where the correct correlation functions were obtained via a simple contour prescription after certain ill-defined changes of variables in the vertices, see also \cite{Colombo:2012jx} for the details and a  generalization of the prescription. Let us note that adding super-symmetry cannot improve the situation since all cubic vertices are independent, i.e. the vertices we have found are left unchanged if a super-symmetric extension is considered. Accordingly, the infinity we observed comes from the sum over the derivatives for fixed three spins rather than from the sum over the spins.

It would, of course, be much better to have a frame in which the HS theory experiences no problems in trying to interpret it as a field-theory in $AdS$ space. To this end the gauge function $\epsilon(C)$ that defines such a frame has to be found.

One of the general lessons we have learned by studying locality in the unfolded approach is that whenever the interaction vertices in the unfolded form are very close to the HS algebra structure constants and are at least bilinear in $C$ (an  example is $\mathcal{V}=\omega\star\omega\star C\star \pi(C)$ or the vertices we derived from the Vasiliev equations in the Schwinger-Fock gauge) they do not have any straightforward field-theoretical interpretation. We expect this to be a generic phenomenon of the HS algebra. The divergences we observed will appear as nested subdivergences at higher orders. It may well be that resolving the paradox at the second order will eliminate the divergences at higher orders as well.

A conservative viewpoint is to expect HS theory to be a meaningful field-theory in $AdS$ since there seems to be no conceptual problem in writing down the ansatz for the Lagrangian and fixing the coefficients by requiring it to reproduce the correct correlation functions\footnote{It has been suggested in \cite{Colombo:2012jx} that Vasiliev's theory is actually \emph{not} dual to the free $O(N)$ model, but instead to a deformed version thereof, where additional nonlinear terms in the sources are present.} \cite{Bekaert:2014cea} or by directly using the HS symmetry \cite{Kessel:2015kna}. The frame without higher derivative tails has to be unique (up to terms that vanish when going from the unfolded frame to the Fronsdal one).

A perhaps less conservative approach that we mentioned before is to try to regularize Vasiliev's theory. This appears as a rather natural procedure, if one entertains the idea that Vasiliev's theory emerges upon quantization of a microscopical topological open string, as was proposed in \cite{Engquist:2005yt,Arias:2015wha}.

\section*{Acknowledgments}
\label{sec:Aknowledgements}

N.B. would like to thank Xavier Bekaert, Axel Kleinschmidt, Malcolm Perry, Mitya Ponomarev, Tomas Prochazka, Martin Schnabl, Ergin Sezgin, Per Sundell for useful discussions.
P.K. would like to thank Stefan Fredenhagen  for useful discussions.
E.S. would like to thank Kostya Alkalaev, Slava Didenko, Maxim Grigoriev, and Mikhail Vasiliev for useful discussions.
M.T. would like to thank Rakibur Rahman, Charlotte Sleight, Per Sundell and Mikhail Vasiliev for useful discussions.
We are also grateful to the organizers of the Higher Spin Theory and Holography conference for hospitality during the realization of part of this work. P.K. would like to thank Service de M\'ecanique et Gravitation, Universit\'e de Mons for the hospitality during the final stage of this work. N.B. wants to thank the kind hospitality of AEI, Golm, where part of this work was done. N.B. is associate researcher of the F.R.S.-FNRS, Belgium. His work is supported by the ARC contract N$^{\rm o}$ AUWB-2010-10/15-UMONS-1. The research of E. Skvortsov and M. Taronna was supported by the Russian Science Foundation grant 14-42-00047 in association with Lebedev Physical Institute.

\begin{appendix}
\renewcommand{\thesection}{\Alph{section}}
\renewcommand{\theequation}{\Alph{section}.\arabic{equation}}
\setcounter{equation}{0}\setcounter{section}{0}

\section{Notation and Conventions}
\label{app:notation}
\setcounter{equation}{0}
The indices $\mm,\nn,...=0,...,d$ are those of $AdS_d$ in some local coordinates. The indices $a,b,c,...=0,...,d$ are those of the tangent space at a point. For most of the paper we have $d=4$, but formulas written with $a,b,...$ or $\mm,\nn,...$ work in any dimension. The indices $\ga,\gb,...=1,2$ and $\gad,\gbd,...=1,2$ are those of the fundamental and anti-fundamental representation of the $4d$ Lorentz algebra $sl(2,\mathbb{C})$. Lastly, the indices $A,B,...,=1,...,4$ are those of the vector representation of $sp(4,\mathbb{R})\sim so(3,2)$. We adopt a symmetrization convention such that all the indices in which a tensor is symmetric or to be symmetrized are denoted by the same letter with the number of indices indicated in brackets,  e.g.
\be\pl_\mm \xi_{\mm(s-1)}\equiv \pl_{\mm_1}\xi_{\mm_2...\mm_s}+\text{$(s-1)$ permutations; } \qquad \xi_{\mm_2...\mm_s}\text{ is already symmetric}\,. \ee
All the symplectic indices, i.e. $A,B,...$, $\ga,\gb,...$ and $\gad,\gbd,...$ are raised and lowered according to
\begin{align}
y^\ga&=\epsilon^{\ga\gb}y_\gb\,, && y_\gb=y^\ga\epsilon_{\ga\gb}\,,
\end{align}
where $\epsilon^{\ga\gb}=-\epsilon^{\gb\ga}$, $\epsilon^{12}=1$ and $\epsilon=i\sigma_2$ with $\sigma_i^{\ga\gad}$, $i=1,2,3$ being the Pauli matrices. Often we omit the indices in the scalar products, e.g. $y\xi\equiv y^\ga \xi_\ga$. The background vierbein $h^{\ga\gad}=h^{\ga\gad}_\mm\, dx^\mm$ and its inverse $h^\mm_{\gb\gbd}$ are defined as to obey
\begin{align}
h_\mm^{\ga\gad}h^\mm_{\gb\gbd}&= \epsilon\fdu{\gb}{\ga}\epsilon\fdu{\gbd}{\gad}\,, && h_\mm^{\ga\gad}h^\nn_{\ga\gad}=\delta^\nn_\mm\,.
\end{align}
For example, in Poincare coordinates one can take
\begin{align}
h_\mm^{\ga\gad}\,dx^\mm&=\frac1{2z} \sigma_\mm^{\ga\gad} dx^\mm\,, && h^\mm_{\ga\gad}=z \sigma^\mm_{\ga\gad}\,, && g_{\mm\nn}=\frac1{2z^2} \eta_{\mm\nn}dx^\mm dx^\nn\,.
\end{align}
The extra factor of $2$ in $g_{\mm\nn}$ explains the appearance of $2$ in front of all the mass-like terms that we derived in the spinorial language. We prefer to work with such a non-canonical normalization of the cosmological constant as to avoid cumbersome factors in \eqref{flatadsconnection}. The background vierbein can be used to define the base of two-, three- and four-forms:
\begin{align}
H^{\gad\gad}&=h\fdu{\nu}{\gad}\wedge h^{\nu\gad}\,, &
H^{\ga\ga}&=h\fud{\ga}{\gnd}\wedge h^{\ga\gnd}\,,
& h\fud{\ga}{\gnd}\wedge H^{\gbd\gnd}&=\hat{h}^{\ga\gbd}\,,
\end{align}
which obey certain useful identities:
\begin{align}\label{hidentities}
\begin{aligned}
h^{\ga\gad}\wedge h^{\gb\gbd}&=\frac12  H^{\ga\gb}\epsilon^{\gad\gbd}+\frac12H^{\gad\gbd}\epsilon^{\ga\gb}\,, \qquad\qquad &   H^{\ga\ga}\wedge H^{\gad\gad}&=0\,,\\
h^{\ga\gad}\wedge H^{\gb\gb}&=-\frac13 \epsilon^{\ga\gb}\hat{h}^{\gb\gad}\,,  &
h^{\ga\gad}\wedge \bar H^{\gbd\gbd}&=+\frac13 \epsilon^{\gad\gbd} \hat{h}^{\ga\gbd}\,,\\
h^{\ga\gad}\wedge \hat{h}^{\gb\gbd}&=-\frac14 \epsilon^{\ga\gb}\epsilon^{\gad\gbd}H_{\ga\ga}\wedge H^{\ga\ga}\,,\\
H_{\ga\ga}\wedge H^{\ga\ga}&=-H_{\gad\gad}\wedge H^{\gad\gad}\,, & H_{\ga\ga}\wedge H^{\ga\ga}&=-h_{\ga\gad}\wedge \hat{h}^{\ga\gad}\,.
\end{aligned}
\end{align}

We make extensive use of the so-called homotopy integrals in the paper which are integrals from $0$ to $1$, see footnote \ref{defhomint} and the formulas nearby. In most cases the integrals are implicit and the names we reserve for the integration variables are $t,q$ and $\tau,p$.
All homotopy integrals correspond to inverted number operators:
\begin{align}
(N+k)^{-1} f(y)=\int_0^1 t^{k-1}\,dt\, f(yt)
\end{align}
and they slightly improve the asymptotic of the Taylor coefficients of $f(y)$. Our convention for the Fourier transformed fields in their twistor variables is:
\begin{equation}
C(y,\bar y|x)=\int d^2\xi d^2\bar\xi\, e^{iy^\ga\xi_\ga+i\bar y^\gad\bar\xi_\gad}\,C(\xi,\bar\xi|x)=\int d^4\xi\,e^{iY\xi} C(\xi|x)\,.
\end{equation}
where in the last expression we used full $sp(4)$-vectors $Y^A$ and $\xi_A$. The $x$-dependence is usually implicit.

\section{HS Vertices}\label{app:VerticesComputations}\setcounter{equation}{0}

\subsection{Exponents} Let us give some details of the second-order computations of Section \ref{sec:SolvingVasiliev}. At the second order one finds terms of the form $C\star C$, $A\star C$, $C\star A$ and $A\star A$ with simple prefactors which are polynomials in $y$'s and $z$'s possibly followed by a differentiation that comes from $ad_h$. The star-products of these type are easy to evaluate. For example, given two polynomials $p_{1,2}(y,z)$,
\begin{align*}
&p_1(y,z) e^{i(ya+zb)}\star p_2(y,z) e^{i(yc+zd)}=\\
&\quad=\int p_1(y+u,z+u) p_2(y+v,z-v)e^{i((y+u)a+(z+u)b+(y+v)c+(z-v)d+vu)}=\\
&\quad=\int f(u,v) e^{i(uA+vB+vu)}\,,
\end{align*}
where in the last line we only highlighted the general structure of the integral over $u,v$. By shifting the variables appropriately we get
\begin{align}
\int f(u,v) e^{i(uA+vB+vu)}=e^{-iAB}\int f(u+B,v-A) e^{ivu}\,.
\end{align}
Bearing in mind that in our computations $f(u,v)$ is a simple polynomial, we can quickly integrate over $u$ and $v$ using
\begin{align}
\int e^{ivu}\, \underbrace{u_\ga..u_\ga}_n \underbrace{v_\gb...v_\gb}_m=\delta_{m,n}\, i^n n!\,\epsilon_{\ga\gb}...\epsilon_{\ga\gb}\,.\label{expintegral}
\end{align}

Let's collect various exponents that appear at the first and second order. At the first order we have just two (apart from pure $e^{iY\xi}=e^{iy\xi+i\bry\brxi}$ that is the image of $C(Y)$)
\begin{align}
\chi_\xi&=e^{iY\xi}\star \klein\, e^{+i\theta} =\exp{i\left(t(y+\xi)z+\bry \brxi+\theta\right)}\,,\\
\bar{\chi}_\xi&=e^{iY\xi}\star \brklein\, e^{-i\theta}=\exp{i\left(y\xi+t(\bry+\brxi)\brz-\theta\right)}\,,
\end{align}
where there is one homotopy parameter $t$ and we always drop out for ease of notation the integral signs over the homotopy parameters. At the second order we find several exponents:
\begin{align*}
R_1&=\exp{i\left( (y(1-t)-t\eta)\xi+(\bry-\breta)(\bry+\brxi)+\theta\right)}=e^{iY\xi}\star\chi_\eta\,,\\
R_2&=\exp{i\left( (y(1-t)-t\xi)\eta+(\bry-\breta)(\bry+\brxi)+\theta\right)}=\chi_\xi\star e^{iY\eta}\,,\\
S_1&=\exp{i\left( (y(1-t)+t\eta)\xi+(\bry-\breta)(\bry+\brxi)+\theta\right)}=e^{iY\xi}\star\pi(\chi_\eta)\,,\\
Q_1&=\exp{i\left((qt(y+\eta)(y+\xi)+(\bry-\breta)(\bry+\brxi)+2\theta\right)}=\chi_\xi\star\chi_\eta \,,\\
Q_2&=\exp{i\left((y(1-q)-q\eta)\xi+(\bry(1-t)-t\brxi)\breta\right)}=\bar{\chi}_\xi\star\chi_\eta\,, \\
P_1&=\exp{i\left((t(y+\eta)(y+\xi)+(\bry-\breta)(\bry+\brxi)+2\theta\right)}=Q_1|_{q=1}\,, \\
K&=\exp{i\left(t(qy+\eta)(qy+\xi)+(\bry-\breta)(\bry+\brxi)+2\theta\right)}=e^{i\theta}\homo{0}{R_2}\star \klein|_{Z=0}=e^{i\theta}\homo{0}{S_1}\star \klein|_{Z=0}\,,
\end{align*}
where we have set $Z=0$ after evaluating the star-products above, as it is this $Z$-slice that we want to extract the unfolded equations from.
There are at most two homotopy parameters $q$ and $t$. All the exponents listed here-above will be accompanied by the conjugate partners thereof, e.g.,
\begin{align}
\bar{Q}_1&=\exp{i\left((y-\eta)(y+\xi)+qt(\bry+\breta)(\bry+\brxi)-2\theta\right)}=\bar{\chi}_\xi\star\bar{\chi}_\eta\,, \end{align}
and the rule is to exchange barred and unbarred variables as well as to flip the sign of $\theta$. The first order fields are easily found to be
\besubeqs
\begin{align}
\Aone_\ga(\xi)&=z_\ga t \;\chi_\xi\,, & C(\xi)&=e^{iY\xi}\,,\\
\Aone_\gad(\xi)&=z_\gad t\; \bar{\chi}_\xi\,, &
\Mone(\xi)&=h^{\ga\gad}(1-t) z_\ga\brxi_\gad \chi_\xi+h^{\ga\gad}(1-t) \brz_\gad\xi_\ga \bar{\chi}_\xi\,,
\end{align}
\esubeqs
which are the Fourier images of the formulas in Section \ref{subsec:FirstOrder}.

\subsection{Raw Vertices}
Applying \eqref{expintegral} and \eqref{adzformula} repeatedly to \eqref{cocyclesgeneral} we easily derive all the components of various vertices. Here-below, the integration over the homotopy parameters $t$ and $q$ as well as the Fourier space integrals are implicit.
\paragraph{$\mathcal{V}(\Omega,C,C)$:}
\begin{align}
M\star C-C\star \pi(M)&=h^{\ga\gad}(1-t)\left[-\eta_\ga\brxi_\gad R_2-\xi_\ga \breta_\gad S_1 \right]C(\xi)C(\eta)+h.c.\,,\\
\widetilde{ad}_h z^\gc \homo{0}{A_\gc\star C+C\star \pi(A_\gc)}+h.c.&=-h^{\ga\gad}\bry_\gad t(-\eta_\ga R_2+\xi_\ga S_1)C(\xi)C(\eta)+h.c.\,.
\end{align}

\paragraph{$\mathcal{V}(\Omega,\omega,C)$:}
\begingroup\allowdisplaybreaks\begin{align}
\omega\star \Mone&= h^{\ga\gad}(1-t)(\xi_\ga \breta_\gad R_1+h.c.)\omega(\xi)C(\eta)\,,\\
\Mone\star \omega &=h^{\ga\gad}(1-t)(-\eta_\ga \brxi_\gad R_2+h.c.)C(\xi)\omega(\eta)\,,\\
-\ad_h z^C \homo{0}{[\omega,\mathcal{A}^{(1)}_C]}&=-h^{\ga\gad}t(\brxi+\breta)_\gad (-\xi_\ga R_1\omega(\xi)C(\eta)+\eta_\ga R_2C(\xi)\omega(\eta))+h.c.\,.
\end{align}\endgroup

\paragraph{$\mathcal{V}(\Omega,\Omega,C,C)$:} The Fourier space integrals for $C(\xi)C(\eta)$ is implied here-below:
\begingroup\allowdisplaybreaks\begin{align}\label{CCMM}
&\begin{aligned}
\Mone\star \Mone&=\frac12 (1-t)(1-q)qtQ_1(\brxi\breta)H^{\ga\ga} (y+\eta)_\ga(y+\xi)_\ga \\
&+(1-t)(1-q)Q_1H^{\gad\gad}\brxi_\gad\breta_\gad\Big[-i+\frac12qt(y+\eta)(y+\xi)\Big]+h.c.\,,
\end{aligned}
\\
\label{CCAA}
&iH^{\ga\ga} \Aone_\ga\star \Aone_\ga+iH^{\gad\gad} \Aone_\gad\star \Aone_\gad=iH^{\ga\ga}\,q^2t^2(y+\eta)_\ga(y+\xi)_\ga Q_1C+h.c.\,,
\\
\label{CCAM}&\begin{aligned}
-&\ad_h z^C \homo{0}{[\Mone,\Aone_C]}=\frac12 H^{\ga\ga}Q_1 (y+\xi)_\ga(y+\eta)_\ga (\brxi\breta)(q(1-t)+t(1-q))qt\\&+ H^{\gad\gad}Q_1(\brxi+\breta)_\gad[q(1-t)\brxi_\gad+t(1-q)\breta_\gad]
[-i+\frac12qt(y+\eta)(y+\xi)]
+h.c.\,,\end{aligned}
\\
\label{CCAtwo}
&\begin{aligned}
-\ad_h \Mtwo&= H^{\gad\gad}(\brxi+\breta)_\gad(\brxi+\breta)_\gad Q_1tq[-i+\frac12qt(y+\eta)(y+\xi)]\\
&-\frac12 H^{\gad\gad}(\brxi+\breta)_\gad(\brxi+\breta)_\gad (y(\xi-\eta))t K
+h.c.\,.\end{aligned}
\end{align}\endgroup
Let us note that all $Q_2$-terms, which should seemingly be present (for example, they are produced by $A_\ga\star A_\gad$-terms and $M\star A$-terms), disappeared from the raw expressions. This is due to a sort of holomorphic factorization that takes place at the second order in the theory, i.e. only purely (anti)holomorphic structures, e.g. $A_\ga\star A_\gb$, contribute to the final expressions.

\subsection{Simplified Vertices}
The vertices $\mathcal{V}(\Omega,\Omega,C)$, $\mathcal{V}(\Omega,\omega,C)$ and $\mathcal{V}(\Omega,C,C)$ are already simple enough. The raw expressions for $\mathcal{V}(\Omega,\Omega,C,C)$ can be considerably simplified by integrating by parts over $t,q$ and then collecting similar terms. Indeed, all $(y+\eta)(y+\xi)Q_1$-terms make a total derivative. However, since $Q_1$ is symmetric with respect to $t\leftrightarrow q$, there is an ambiguity on how to integrate by parts: either represent it as $-it^{-1}\pl_qQ_1$
or as $-iq^{-1}\pl_tQ_1$. We choose either of the two forms as to collapse everything onto the boundary terms. This strategy is successful for \eqref{CCAtwo} and \eqref{CCAM}, while a small boundary term remains for \eqref{CCMM}. The $K$-term in \eqref{CCAtwo} is a total derivative since $(y(\xi-\eta))t K=-i\pl_q K$, which results in two
boundary terms proportional to $\funat{K}{q=1}=\funat{Q_1}{q=1}=P_1$ and $\funat{K}{q=0}$. The simplest form of the cocyles can be found in Section \ref{sec:SecondOrder}.

\subsection{Checking Consistency}\label{app:Consistency} It still useful to check the consistency of the vertices found
and it is instructive to see the unfolded approach at work. In $4d$ due to propagating nature of HS fields,
various vertices are related to each other via the Frobenius integrability condition, \eqref{GrandJacobi}, which does not happen in $3d$, \cite{Kessel:2015kna}, where all vertices are just $\adD$-conserved $\adD\mathcal{V}(...)=0$.

First of all, the Vasiliev equations are consistent when all the fields are taken to be matrix-valued. Therefore,
terms of the form $\omega C$ should cancel independently of $C\omega$-terms. More precisely, there are different functions, e.g. $R_1$, $S_1$, etc., that enter the expressions for the vertices. These functions are independent
except possibly for boundary terms, i.e. some of the functions coincide at $t=0$ or $t=1$. Therefore, the bulk contributions should cancel independently for each of the functions, while there can appear some total derivatives in the homotopy parameters $t$ and $q$ that can yield boundary terms that should also cancel independently. In addition, each expression appears in the form $X+h.c.$ and it is sufficient to check the consistency for one half of any given vertex.

In what follows we omit the integral over the Fourier space of $\xi$ and/or $\eta$ as well as integrals over the homotopy parameters $t,q$. We will also need the adjoint and twisted-adjoint covariant derivatives:
\begin{align}
\adD&=\nabla-h^{\ga\gad}O_{\ga\gad}(y)\,, & O_{\ga\gad}(y)&=\bry_\gad\pl^y_\ga+y_\ga \pl^y_\gad\,,\\
\tadD&=\nabla+ih^{\ga\gad}\widetilde{O}_{\ga\gad}(y)\,, & \widetilde{O}_{\ga\gad}(y)&=y_\ga \bry_\gad-\pl^y_\ga\pl^y_\gad\,.
\end{align}
The Fourier images of these operators coincide with themselves, i.e. when acting, for example, on $C(\xi)$ one
should use $\nabla+ih^{\ga\gad}\widetilde{O}_{\ga\gad}(\xi)$. The vertices are multi-linear maps and the action of $\adD$ or $\tadD$ is the usual action on the tensor product.

\paragraph{Vertex $\mathcal{V}(\Omega,\Omega,C)$.} It is the simplest vertex to check the consistency of:
\begin{align}
D\mathcal{V}(\Omega,\Omega,C)&
=-h^{\ga\gad}(O_{\ga\gad}(y)+i\widetilde{O}_{\ga\gad}(\xi))\tfrac12 H^{\gbd\gbd} \brxi_\gbd\brxi_\gbd e^{i(\bry\brxi+\theta)} C(\xi)+h.c.\\
&=-h^{\ga\gad}(y_\ga\pl^y_\gad+\bry_\gad\pl^y_\ga +i(\xi_\ga\brxi_\gad-\pl^\xi_\ga\pl^\xi_\gad))\tfrac12 H^{\gbd\gbd} \brxi_\gbd\brxi_\gbd e^{i(\bry\brxi+\theta)} C(\xi)+h.c.\,.
\end{align}
In the last line, the second term does not contribute because the expression it acts on is $y$-independent, the last term is a total derivative, the rest gives
\begin{align}
D\mathcal{V}(\Omega,\Omega,C)&
=-h^{\ga\gad}(iy_\ga \brxi_\gad \brxi_\gbd\brxi_\gbd+i\xi_\ga\brxi_\gad \brxi_\gbd\brxi_\gbd)\tfrac12 H^{\gbd\gbd} \brxi_\gbd\brxi_\gbd e^{i(\bry\brxi+\theta)} C(\xi)+h.c.\,.
\end{align}
Each of the terms above vanishes thanks to identities \eqref{hidentities} for the vierbeins that bring $\epsilon^{\gad\gbd}$.

\paragraph{Vertex $\mathcal{V}(\Omega,C,C)$.} The consistency condition follows by applying $\tadD$ to
\begin{align}
&\tadD \Ctwo=\omega\star C-C\star \pi(\omega)+\mathcal{V}(\Omega,C,C)\,,
\end{align}
which after some algebra gives
\begin{align}
0=(D\omega)\star C-C\star \pi(D\omega)+\tadD\mathcal{V}(\Omega,C,C)\,.
\end{align}
Substituting $\mathcal{V}(\Omega,\Omega,C)$ in place of $\adD\omega$ and noticing its invariance under $\pi$, we get
\begin{align}\label{omegaCCidentity}
0=\mathcal{V}(\Omega,\Omega,C) \star C-C\star \mathcal{V}(\Omega,\Omega,C)+\tadD\mathcal{V}(\Omega,C,C)\,.
\end{align}
In the first two terms it is easy to compute the star-products:
\begin{align}
\mathcal{V}(\Omega,\Omega,C) \star C&=\frac12 H^{\gad\gad}\brxi_\gad\brxi_\gad e^{i(\bry\brxi+\theta)}C(\xi)\star e^{iY\eta}C(\eta)+h.c.\\
&=\frac12 H^{\gad\gad}\brxi_\gad\brxi_\gad \exp i\left(y\eta+(\bry-\breta)(\bry+\brxi)+\theta\right)C(\xi)C(\eta)+h.c.\\
&=\frac12 H^{\gad\gad}\brxi_\gad\brxi_\gad\funat{R_2}{t=0}C(\xi)C(\eta)+h.c.\,,
\end{align}
where in the last line we anticipated the cancelation with the boundary terms that should be produced by $\tadD\mathcal{V}(\Omega,C,C)$ and identified the exponent as $R_2$ at $t=0$. Analogously,
\begin{align}
-C\star \mathcal{V}(\Omega,\Omega,C)&=-\frac12 H^{\gad\gad}\breta_\gad\breta_\gad\funat{S_1}{t=0}C(\xi)C(\eta)+h.c.\,.
\end{align}
With the help of the Fourier space images of the equations of motion the third term of \eqref{omegaCCidentity} reads:
\begin{align}\label{omegaCCoperator}
ih^{\ga\gad}h^{\gb\gbd}\left(\widetilde{O}_{\ga\gad}(y)-\widetilde{O}_{\ga\gad}(\xi)-\widetilde{O}_{\ga\gad}(\eta)\right)
K_{\gb\gbd}(Y,\xi,\eta)C(\xi)C(\eta)\,.
\end{align}
The cocycle $K_{\gb\gbd}$ consists of two terms, the one with $R_2$ and another one with $S_1$, and the conjugates thereof. As it was discussed, these should cancel independently up to some boundary terms. Applying \eqref{omegaCCoperator} to the $R_2$ terms one gets a term proportional to $\pl_t R_2$ and a bare $R_2$ term after some straightforward algebra. Integrating by parts everything cancels except for the boundary terms:
\begin{align}
\frac12H^{\gad\gad}\left(\bry_\gad\bry_\gad \funat{R_2}{t=1}-\brxi_\gad\brxi_\gad \funat{R_2}{t=0}\right)C(\xi)C(\eta)\,.
\end{align}
The same story is with the $S_1$ terms, the boundary remnant being
\begin{align}
-\frac12H^{\gad\gad}\left(\bry_\gad\bry_\gad \funat{S_1}{t=1}-\breta_\gad\breta_\gad \funat{S_1}{t=0}\right)C(\xi)C(\eta)\,.
\end{align}
Now, taking into account that $\funat{R_2}{t=1}=\funat{S_1}{t=1}$, we see that everything cancels.

\paragraph{Vertex $\mathcal{V}(\Omega,\omega,C)$ and $\mathcal{V}(\Omega,\Omega,C,C)$.} The two vertices are related and cannot be considered separately. The Frobenius integrability condition is found by applying $\adD$ to
\begin{align}
\adD \omegatwo=\mathcal{V}(\Omega,\Omega,\Ctwo)+\mathcal{V}(\omega,\omega)+
\mathcal{V}(\Omega,\omega,C)+
\mathcal{V}(\Omega,\Omega,C,C)\,,
\end{align}
which results in
\begin{align}
0=\adD\mathcal{V}(\Omega,\Omega,\Ctwo)+\adD\omega\star \omega-\omega\star \adD\omega+
\adD\mathcal{V}(\Omega,\omega,C)+
\adD\mathcal{V}(\Omega,\Omega,C,C)\,.
\end{align}
The zero-form $\Ctwo$ disappears since it enters exactly the same way as $C$ at the linear level, but it leaves some aftertaste since $\tadD\Ctwo\neq0$. With the right-hand side of  $\tadD\Ctwo$ plugged in we get
\begin{align*}
0&=\mathcal{V}(\Omega,\Omega,\omega\star C-C\star \pi(\omega))+\mathcal{V}(\Omega,\Omega,\mathcal{V}(\Omega,C,C))
+\mathcal{V}(\Omega,\Omega,C)\star \omega-\omega\star \mathcal{V}(\Omega,\Omega,C)+\\&\qquad\qquad\qquad\qquad+
\adD\mathcal{V}(\Omega,\omega,C)+
\adD\mathcal{V}(\Omega,\Omega,C,C)\,.
\end{align*}
There are two groups of terms that should cancel independently: $\Omega\Omega \omega C $ and $\Omega\Omega\Omega CC$. Let us first have a look at $\Omega\Omega \omega C $, which themselves split into $\Omega\Omega \omega C $ and $\Omega\Omega C\omega $ since fields can be matrix-valued. This part of the integrability constraint reads:
\begin{align}
0&=\mathcal{V}(\Omega,\Omega,\omega\star C-C\star \pi(\omega))
+\mathcal{V}(\Omega,\Omega,C)\star \omega-\omega\star \mathcal{V}(\Omega,\Omega,C)+
\adD\mathcal{V}(\Omega,\omega,C)+O(C^2)\,,
\end{align}
where we assumed that any $C^2$-terms resulting from the last term need to be dropped. It is easy to see that
\begin{align}
\mathcal{V}(\Omega,\Omega,\omega\star C-C\star \pi(\omega))&=
-\frac12 H^{\gad\gad}e^{i\theta}\pl^y_\gad\pl^y_\gad(\omega\star C-C\star \pi(\omega))+h.c.\\
&=\frac12 H^{\gad\gad}(\brxi_\gad+\breta_\gad)(\brxi_\gad+\breta_\gad)\left(\funat{R_1}{t=1}\omega(\xi)C(\eta)-
\funat{R_2}{t=1}C(\xi)\omega(\eta)\right)\notag\,.
\end{align}
The second and the third terms are similar to the one computed in checking $\mathcal{V}(\Omega,C,C)$:
\begin{align}
\mathcal{V}(\Omega,\Omega,C)\star \omega-\omega\star \mathcal{V}(\Omega,\Omega,C)&=
-\frac12 H^{\gad\gad}\breta_\gad\breta_\gad \funat{R_1}{t=0}\omega(\xi)C(\eta)+ \frac12 H^{\gad\gad}\brxi_\gad\brxi_\gad \funat{R_2}{t=0}C(\xi)\omega(\eta)\notag\,.
\end{align}
The most tedious is the last term, for which the Fourier space image of $\adD$ is
\begin{align}
-h^{\ga\gad}\left(O_{\ga\gad}(y)+O_{\ga\gad}(\xi)+i\widetilde{O}_{\ga\gad}(\eta)\right)
h^{\gb\gbd}L_{\gb\gbd}(Y,\xi,\eta)\omega(\xi)C(\eta)\,,\\
-h^{\ga\gad}\left(O_{\ga\gad}(y)+i\widetilde{O}_{\ga\gad}(\xi)+O_{\ga\gad}(\eta)\right)
h^{\gb\gbd}\bar{L}_{\gb\gbd}(Y,\xi,\eta)C(\xi)\omega(\eta)\,.
\end{align}
The rest works the same way as for $\mathcal{V}(\Omega,C,C)$, i.e. there are terms that combine together to form a $\pl_t$-derivative of the functions involved and some other term. After integrating by parts the bulk contribution
cancels and only the boundary terms remain, which are, respectively,
\begin{align}
\Omega^2 \omega C&: && -\frac12 H^{\gad\gad}(\brxi_\gad+\breta_\gad)(\brxi_\gad+\breta_\gad)\funat{R_1}{t=1}+
\frac12H^{\gad\gad}\breta_\gad\breta_\gad \funat{R_1}{t=0}+h.c.\,,\\
\Omega^2 C\omega&: && +\frac12 H^{\gad\gad}(\brxi_\gad+\breta_\gad)(\brxi_\gad+\breta_\gad)\funat{R_2}{t=1}-
\frac12H^{\gad\gad}\brxi_\gad\brxi_\gad \funat{R_2}{t=0}+h.c.\,.
\end{align}
Now, everything cancels. Coming back to the $\Omega^3C^2$-terms, the integrability implies
\begin{align}\label{OOOCCident}
0&=\mathcal{V}(\Omega,\Omega,\mathcal{V}(\Omega,C,C))
-\mathcal{V}(\Omega,\mathcal{V}(\Omega,\Omega,C),C)+
\adD\mathcal{V}(\Omega,\Omega,C,C)\,,
\end{align}
where we used the $\omega$ equations of motion to simplify the second term. The third term should vanish
up to some boundary terms that are then canceled by the first and the second terms. The Fourier-space image of $\adD$ for the third term is
\begin{align}
\adD=-h^{\ga\gad}\left(O_{\ga\gad}(y)+i\widetilde{O}_{\ga\gad}(\xi)+i\widetilde{O}_{\ga\gad}(\eta)\right)\,.
\end{align}
The following identities are useful ($F_{\ga\gad}=(y+\eta)_\ga\breta_\gad-(y+\xi)_\ga\brxi_\gad$)
\begin{align*}
\adD H^{\gb\gb}(y+\xi)_\gb(y+\eta)_\gb Q_1&=\hh^{\ga\gad}F_{\ga\gad}Q_1\Big(1-\frac{i(1-qt)}{3}(y+\eta)(y+\xi)\Big)\,,\\
\adD H^{\gb\gb}(y+\xi)_\gb(y+\eta)_\gb Q_1(\brxi\breta)&=\hh^{\ga\gad}F_{\ga\gad}Q_1\Big(\Big(1-\frac{i(1-qt)}{3}(y+\eta)(y+\xi)\Big)(\brxi\breta)
-i+\frac{qt}3(y+\eta)(y+\xi)\Big)\,,\\
\adD H^{\gbd\gbd}\brxi_\gbd\breta_\gbd Q_1&=\hh^{\ga\gad}F_{\ga\gad}Q_1\Big(qt-\frac{i(1-qt)}3(\brxi\breta)\Big)\,.
\end{align*}
The same identities can be applied to the boundary term $P_1$ upon setting $q=1$ since it equals $Q_1$ at $q=1$. Denoting the structures on the left-hand side of the above identities as $A$, $B$ and $C$, respectively, and taking the coefficients from \eqref{SimplifiedVertices}, the cancelation amounts to
\begin{align}
\adD\left(iq^2 t^2 A+\frac{qt(1-qt)}2 B-\frac{i}2 C +\frac{i}2(1-t)\funat{C}{q=1}\right)=0\,,
\end{align}
which is indeed satisfied, but in order to see this one has to integrate some of the terms by parts using $(y+\eta)(y+\xi)=-it^{-1}\pl_qQ_1$. Then, the bulk terms cancel while the boundary terms so produced cancel with the boundary term that is already present.

In $\mathcal{V}(\Omega,\Omega,C,C)$ there is also another type of boundary term, $\funat{K}{q=0}$. This one cancels with the first term of \eqref{OOOCCident}. Indeed, one observes that
\be\funat{K}{q=0}=e^{i\theta}\funat{R_2}{y=0}=e^{i\theta}\funat{S_1}{y=0}\,.\ee
The straightforward algebra yields for the nested vertices (each of the two vertices can be written in the form $X+h.c.$, which after nesting gives four terms and those that are not (anti)-holomorphic vanish identically thanks to $h^{\ga\gad}\wedge H^{\gb\gb}\eta_\ga\eta_\gb\eta_\gb\equiv0$-like identities):
\begin{align}\label{nestedcocyle}
\mathcal{V}(\Omega,\Omega,\mathcal{V}(\Omega,C,C))=\hh^{\ga\gad}(\xi-\eta)_\ga(\brxi+\breta)_\gad \funat{K}{q=0}
\Big(-\frac{i}2t+\frac13 (-(\brxi\breta)+t(\bry-\breta)(\bry+\brxi))\Big)\,,
\end{align}
while the Fourier-space image of $\adD$ applied to the $K$-term of $\mathcal{V}(\Omega,\Omega,C,C)$ yields
\begin{align}
\adD H^{\gbd\gbd}(\brxi+\breta)_\gbd(\brxi+\breta)_\gbd\funat{K}{q=0}&=-2i\eqref{nestedcocyle}\,.
\end{align}
The last expression with the coefficient $-\frac{i}2$ cancels the last but one.

The second nested vertex of \eqref{OOOCCident} vanishes identically. To see this, we need to rewrite $\mathcal{V}(\Omega,\Omega,C)$ in Fourier space:
\begin{align}
\adD \omega(\xi|x)&=\frac12 H^{\gad\gad} \brxi_\gad\brxi_\gad e^{i\theta} \delta^2(\xi) \int d^2\chi\, C(\chi,\brxi|x)+h.c.\,.
\end{align}
Substituting this in place of $\omega$ into $\mathcal{V}(\Omega,\omega,C)$-cocyle we see that all terms vanish either because of $\delta^2(\xi)\xi_\ga\equiv0$ or due to $h^{\ga\gad}\wedge H^{\gb\gb}\eta_\ga\eta_\gb\eta_\gb\equiv0$.

\section{Fronsdal Currents}\setcounter{equation}{0}\label{app:FronsdalCurrents}
In this section we provide more technical details on the evaluation of the Fronsdal currents.
The recipe how to obtain Fronsdal currents from the unfolded equation can be found in Section \ref{subsec:FronsdalCurrents}.

\subsection{Fierz-Invariant Form}
There is an ambiguity in writing expressions in terms of spinorial variables  due to Fierz
identities\footnote{See Section C.1.4 of \cite{Kessel:2015kna} for the detailed discussion of Fierz transformations. The only difference with $3d$ is that we have two independent actions of the Fierz transformations on dotted and undotted indices.}. In order to solve the torsion constraint \eqref{realTorsion} we need to rewrite $\formJ^{s.t.}$ as
\begin{align}\label{appcanform}
\begin{aligned}
\formJ^{s.t.}(Y,\xi,\eta)&=H^{\ga\ga}\pl_\ga\pl_\ga J^{\pl\pl}+y^\ga H\fdu{\ga}{\gb}\pl_\gb J^{y\pl}+ y^\ga y^\ga
H_{\ga\ga}J^{yy}+\\&+H^{\gad\gad}\pl_\gad\pl_\gad \brJ^{\pl\pl}+y^\gad H\fdu{\gad}{\gbd}\pl_\gbd \brJ^{y\pl}+ \bry^\gad \bry^\gad H_{\gad\gad}\brJ^{yy}+h.c.\,.
\end{aligned}
\end{align}
This form has an advantage of having all the Fierz transformations fixed. Actually, it is expressed in terms of invariants of the Fierz transformations. To put expression in such a form we apply \eqref{niceidentityT}. Finally, the Fierz-invariant form of $\formJ^{s.t.}$ is
\begin{align}\label{fullcanonical}
\begin{aligned}
J^{\pl\pl}&=\Big[(y\xi)(y\eta) \tau G_{qt}\Big]T e^{2i\theta}\,,\\
J^{y\pl}&=\Big[-iqt G_{qt} \tau^2(\eta\xi)(y(\xi+\eta))+\frac{i}{qt} M_{qt}-2H_{qt}(y\eta)\Big]Te^{2i\theta}\,,\\
J^{yy}&=\Big[-q^2t^2G_{qt} (\eta\xi)^2\tau^2p+M_{qt}+2iqt H_{qt}(\eta\xi)\tau\Big]Te^{2i\theta}\,,\\
\brJ^{\pl\pl}&=\Big[(\bry\brxi)(\bry\breta)F\tau\Big]\bar{T}e^{2i\theta}\,,\\
\brJ^{y\pl}&=\Big[-iF \tau^2(\breta\brxi)(\bry(\brxi-\breta))\Big]\bar{T}e^{2i\theta}\,,\\
\brJ^{yy}&=\Big[F (\breta\brxi)^2\tau^2p\Big]\bar{T}e^{2i\theta}\,,\\
\end{aligned}
\end{align}
where
\begin{align}
\begin{aligned}
T&=L\Big(iq^2t^2+(\brxi\breta)\frac{qt(1-qt)}2\Big)\,, & L&=\exp{i\left((\bry-\breta)(\bry+\brxi)\right)}\,,\\
\bar{T}&=\frac{i}2\big(-M_{qt}+(1-t)M_t\big)\,, & M_x&=\exp{ix\left((y+\eta)(y+\xi)\right)}\,,\\
G_x&=\exp{ix\big(p\tau y(\xi-\eta)+\eta\xi\big)}\,, & F&=\exp{i\big(p\tau\bry(\breta+\brxi)-\breta\brxi\big)}\,,\\
H_x&=\exp{ix\big(\tau y(\xi-\eta)+\eta\xi\big)}\,.
\end{aligned}
\end{align}
and we have to introduce two more homotopy parameters $\tau$ and $p$, the integral left implicit as well as $C(\xi)C(\eta)$. To make contact with Section \ref{app:VerticesComputations} we note that $Q_1=M_{qt}L e^{2i\theta}$ and $P_1= M_{t}L e^{2i\theta}$. Note that $h.c.$ in \eqref{appcanform} means adding the usual h.c. tail with barred and unbarred variables exchanged (this may be confusing since the second line seems to be $h.c.$ of the first, but the actual expressions for the second line cannot be obtained via $h.c.$ from the first line, see here-above).

The canonical projection leads immediately to the required form:
\begin{align}\label{canformcan}
J^{\text{can},\pl\pl}&=\frac1{4} Q_1\Big(i+(\brxi\breta)\frac{(1-qt)}{2qt}\Big)\,,&
\brJ^{\text{can},\pl\pl}&=\frac{i}8 \Big(Q_1-(1-t) P_1\Big)\,,
\end{align}
where again everything needs to be supplemented with the $h.c.$ part.

One may also be interested in the pure improvement terms that account for the difference between the full backreaction \eqref{fullcanonical} and its canonical part \eqref{canformcan}. These read:
\begin{subequations}\label{improvements}
\begin{align}
J^{\text{impr.},yy}&=f_1 \left[Q_1+(\zeta^+\pl^y) Q_1^\tau+\tfrac4{\tau}(\zeta^+\pl^y)^2 Q_1^{\tau\sigma}\right]\,, & \bar{J}^{\text{impr.},yy}&=\frac1{4\tau}f_2(\brzeta^-\cdot\partial^\bry)^2 Q^{\tau\sigma}_1\,,\\
J^{\text{impr.},y\pl}&=f_1 \,(y\zeta^+)\left[Q_1^\tau+\tfrac{1+\tau^2}{4\tau}(\zeta^+\pl^y)Q_1^{\sigma\tau}\right]\,, & \bar{J}^{\text{impr.},y\pl}&=f_2 \tfrac{1+\tau^2}{4\tau}(\bry\brzeta^-)(\brzeta^-\pl^\bry)Q_1^{\bar{\sigma}\bar{\tau}}\,,\\
J^{\text{impr.}\pl\pl}&=\tfrac{\tau}4 f_1 (y\cdot\zeta^+)^2Q^{\sigma\tau}_1\,,
&\bar{J}^{\text{impr.},\pl\pl}&=\tfrac{\tau}4f_2 (\bry\cdot\brzeta^-)^2Q^{\bar{\sigma}\bar{\tau}}_1\,,
\end{align}
\end{subequations}
where: $\vartheta=\delta(q-1)$,
\begin{subequations}
\begin{align}
Q^x_1&=\exp i[qt ( xy\zeta^--\xi\eta)+(\bry\brzeta^++\brxi\breta)+2\theta]\,, &
f_1&=\left(i (qt)^2+(\brxi\breta)\frac{qt(1-qt)}{2}\right)\,,\\
Q^{\bar{x}}_1&=\exp i[qt (y\zeta^--\xi\eta)+( x\bry\brzeta^++\brxi\breta)+2\theta]\,,
 & f_2&=-\frac{i}{2}\left[1-(1-t)\vartheta(q-1)\right]\,,
\end{align}
\end{subequations}
and
\begin{align}
\zeta^{\pm}&=\xi \pm \eta\,,& \brzeta^{\pm}&=\brxi\pm\breta\,,\\
\zeta^-\partial^y Q_1^{x}&=-2iq t x\xi\eta\,,&\brzeta^-\partial^{\bry} Q_1^{\bar{x}}&=2i x\brxi\breta\,.
\end{align}
Note that the maximal power of $\zeta^+$, which measures how far the improvement is from the canonical current, is two, which is different from the $3d$ case \cite{Kessel:2015kna}, where all possible improvements contribute to the full Fronsdal current.

\subsection{Fronsdal Equations in Spinorial Language}\label{app:FronsdalOperator}
There is a nontrivial dictionary between the Fronsdal equations in vectorial and spinorial languages. We need to fix the normalization as to be able to translate the formulas for the Fronsdal currents in the language of tensors. Following the same algorithm as was used to derive the Fronsdal currents \eqref{FronsdalCurrentsDef} but for $\nabla e$ and adding the mass-like terms that are seen in \eqref{currentdef} we get for \eqref{FronsdalEmbedding}
\besubeqs\label{spinFronsdal}
\begin{align}
f&= Z_s\left(\square \Front +2(s^2-2s-2)\Front -\frac1{s}(y\nabla\bry)(\pl\nabla\bar{\pl})\Front+\frac1{s}(y\nabla\bry)(y\nabla\bry)\Front' \right)\,,\\
f'&=Z'_s\left(\square \Front' +2s^2\Front' +\frac{1}{2s-1}(y\nabla\bry)(\pl\nabla\bar{\pl})\Front'-\frac{1}{2s-1}(\pl\nabla\bar{\pl})(\pl\nabla\bar{\pl})\Front \right)\,,
\end{align}
\esubeqs
where $(y\nabla\bry)=y^\nu \nabla_{\nu\gnd}\bry^\gnd$,  $(\pl\nabla\bar{\pl})=\pl^\nu \nabla_{\nu\gnd}\bar{\pl}^\gnd$. The mass-like terms are twice the expected in $4d$ due to the normalization of the cosmological constant we use. The normalization factors $Z_s$ and $Z'_s$ are:
\begin{align}
Z_s&=\frac{1}{4(s-1)}\,, &Z'_s&=-\frac{(2s-1)}{4s(s+1)}\,.
\end{align}
The Bianchi identity \eqref{BianchiSpinorial}, which is a spinorial counterpart of \eqref{BianchiVectorial}, reads:
\begin{align}\label{spinBianchi}
(s-1)(\pl\nabla\bar{\pl})f+(s+1) (y\nabla\bry) f'\equiv0\,.
\end{align}
These formulas we need to compare with the Fronsdal operator $F$, \eqref{FronsdalEq}, that we decompose as $F_s=f_s+\frac1{\gamma}gf'_s$ and rewrite in terms of $\Fron_s=\phi_s+\frac1{\gamma}g\phi'_s$, where $\phi'$ and $\phi$ are trace and trace-free components of $\Fron_s$ and $\gamma=d+2s-4$. Next, we translate it in the spinorial language
\besubeqs\label{vecFronsdal}
\begin{align}
f&= \left(\square \Front +2(s^2-2s-2)\Front -\frac1{s}(y\nabla\bry)(\pl\nabla\bar{\pl})\Front+\frac1{2s^2}(y\nabla\bry)(y\nabla\bry)\Front' \right)\,,\\
f'&=\frac{(2s-1)}{s}\left(\square \Front' +2s^2\Front' +\frac{1}{2s-1}(y\nabla\bry)(\pl\nabla\bar{\pl})\Front'-\frac{2s}{2s-1}(\pl\nabla\bar{\pl})(\pl\nabla\bar{\pl})\Front \right)\,.
\end{align}
\esubeqs
Also, we keep in mind that $\nabla_\mm T_{\mm(s-1)}$ in vectorial base corresponds to $\tfrac{1}{s} (y\nabla\bry) T(y,\bry)$ in the spinorial one. Analogously, the Bianchi identity \eqref{FronsdalBianchi} is:
\begin{align}\label{vecBianchi}
(\pl\nabla\bar{\pl})f-\frac{1}{2s} (y\nabla\bry) f'\equiv0\,.
\end{align}
In order for the two sets of formulas to match we should rescale fields $\phi\rightarrow \alpha_s\phi_s$, $\phi'_s\rightarrow \alpha'_s\phi'_s$ as well as the components of the Fronsdal operator $f\rightarrow \beta_sf_s$, $f'_s\rightarrow \beta'_sf'_s$ because the spinorial Bianchi identity \eqref{spinBianchi} is different from the vectorial one \eqref{vecBianchi}. The matching of the Fronsdal operators \eqref{spinFronsdal} and \eqref{vecFronsdal} requires $\alpha'_s/\alpha_s=2s$. The overall rescaling can be fixed as $\alpha_s=Z^{-1}_s$ so that the Fronsdal operator starts canonically $f=\square \phi+...$. Then, the rescaling of $\compj$ and $\compj'$ can be found by requiring $f$ and $f'$ to obey the canonical Bianchi identity, which gives $\alpha_s/\beta_s=Z_s$ and $\tfrac{(2s-1)}{s}\tfrac{\alpha'_s}{\beta'_s}=Z'_s$. Finally,
\begin{align}
\tilde{a}_s&=\frac{a_s}{\beta_s}=\frac{1}{4(s-1)\alpha_s}a_s\,, & \tilde{c}_s&=\frac{a'_s}{\beta'_s}=-\frac{1}{8 s(s+1)\alpha_s}c_s\,.
\end{align}

\subsection{Coefficients of the Fronsdal Current}\label{app:Coefficients}
In this section we will write some explicit coefficients for the Fronsdal currents. Contrary to the $3d$ case considered in \cite{Kessel:2015kna}, in the $4d$ HS theory it is relatively simple to write the contribution of the scalar fields to the spin-$s$ field's equations of motion. The rules are almost identical to those of \cite{Kessel:2015kna}. We need to extract coefficients from Fourier space expressions whose general form is
\begin{align}
(y\xi)^{A'}(y\eta)^{B'}(\xi\eta)^{C'} \exp{i[(y\xi)a+(y\eta)b+(\xi\eta)c]} P(t,q,...) \, C(\xi) C(\eta) \,,
\end{align}
where $a,b,c$ are possibly functions of homotopy parameters, constants or zero and $A',B',C'$ are integers. Analogous decomposition
is true for the anti-holomorphic sector. Obviously, both sectors are completely independent for the problem at hand. Then, with the help of
\begin{align}
\left[\left.e^{iy\xi}\right|_{\xi=\pl^u}C(-iu)\right]\Big|_{u=0}=C(y)
\end{align}
the coefficient of
\begin{align}
\frac1{(n+m)!}C_{\ga(n)\nu(l)} C\fud{\nu(l)}{\ga(m)}\, y^{\ga(n+m)}
\end{align}
can be found to be
\begin{align}
f^{n,m,l}_{A',B',C'}=\frac{i^{l-A'-B'-C'}(m+n)!}{(n-A')!(m-B')!(l-C')!} \label{eq:coeffFunc}
\times\int d t\,d q\,...\, a^{n-A'}b^{m-B'}c^{l-C'} P(t,q,...)\,.
\end{align}

There is an alternative formulation to encode the relative coefficients in the backreaction in terms of a contour integral representation. The idea is based on the following identity:
\begin{align}
&\int dt\,dq\, f(q,t)e^{x(q,t)\xi\eta}=\oint d\tau F(\alpha)\big|_{\alpha=\tau^{-1}} e^{\tau \xi\eta}\,, && F(\alpha)=\int dt\,dq\, \frac{\alpha f(q,t)}{1-x(q,t)\alpha}\,,
\end{align}
where the $F(\alpha)$ must be considered as a formal series in $\alpha=0$ and when performing the contour integration in $\tau$ one should use $\alpha=\tau^{-1}$. It turns out to be convenient to work in terms of the generating functions $F(\alpha)$ upon performing all homotopy integrals since many operations and the covariant derivative translate into ordinary differential operators in the variable $\alpha$.

The canonical current sector is particularly easy to deal with since it is specified by two functions of four variables:
\begin{align}\begin{aligned}
\formJ^{\text{2form}}_{can}=-\oint d\tau_1\,d\tau_2\,ds\,dr\, \big[&\beta^2\, g^{\pl\pl}_1(\alpha_1,\alpha_2,\beta,\gamma)H^{\ga\ga}\pl_\ga\pl_\ga\\&+\gamma^2\, g^{\pl\pl}_2(\alpha_1,\alpha_2,\beta,\gamma)H^{\gad\gad}\pl_\gad\pl_\gad\big]e^{i(s y\zeta^-+\tau_1\xi\eta+r \bry\brzeta^++\tau_2\brxi\breta)}\,,
\end{aligned}\end{align}
that we will represent as a 2d vector:
\begin{equation}
\vec{g}^{\,\pl\pl}=\begin{pmatrix}
g_1^{\pl\pl}\\
g_2^{\pl\pl}
\end{pmatrix}\,.
\end{equation}
Here:
\begin{align}
\alpha_1&=\tau_1^{-1}\,,& \alpha_2&=\tau_2^{-1}\,,&\beta&=s^{-1}\,,&\gamma&=r^{-1}\,.
\end{align}
Finally, the contour integration can be performed monomial by monomial using the following dictionary:
\begin{equation}
\alpha_2^{m+1} \alpha_1^{n+1} \beta ^{s_1+1} \gamma ^{s_2+1}\rightarrow \frac{(i\brxi\breta){}^{m} (i\xi\eta){}^{n} (i\bry\brzeta^+){}^{s_2} (iy\zeta^-){}^{s_1}}{m! n! s_1! s_2!}\,.
\end{equation}
As an example, the whole Fronsdal current of spin-$s$ that is bilinear in two scalar fields is
\begin{align}
j_{\mm(s)}&=2\cos2\theta \sum_k \tilde{a}_{s,k,l}\nabla_{\mm(s-k)\nn(l)}\Fron\, \nabla\fud{\nn(l)}{\mm(k)}\Fron\,,\\
j'_{\mm(s-2)}&=2\cos2\theta \sum_k \tilde{c}_{s,k,l}\nabla_{\mm(s-k-2)\nn(l)}\Fron\, \nabla\fud{\nn(l)}{\mm(k)}\Fron\,,
\end{align}
where the coefficients are found to be
\begin{align}
&\begin{aligned}
a_{s,l,k}&=\frac{ s!s!}{l!l!k!k!(s-k)!(s-k)!}\frac{i (-1)^k\big( 2 r k (s-1) (k-s)+s^2 \left(2 s^2 (l+s) (l+s+1)-r\right)\big)}{4 (s-1) s^2 (l+s)^2 (l+s+1)^2 (l+s+2)}\,,\\
r&=2 s^3+(4 l-1) s^2+l (2 l-1) s+2 l\,,
\end{aligned}\\
&\begin{aligned}
c_{s,l,k}&=\frac{ (s-2)!(s-2)!}{l!l!k!k!(s-k-2)!(s-k-2)!}\frac{i (-1)^k r'}{2 s^2 (s+1) (l+s-1)^2 (l+s) (l+s+1)}\,,\\
r'&=3 s^5+(10 l-6) s^4+\left(11 l^2-15 l+3\right) s^3+l \left(4 l^2-11 l+8\right) s^2+l \left(-2 l^2+7 l-5\right) s-2 (l-1) l\,.
\end{aligned}\notag
\end{align}
The data completely determines the contribution to the $\langle 00s \rangle$ correlator. The canonical projection
of the formulas here-above can be found in Section \ref{subsec:SpinTwo}. Moreover, as was explained in Section \ref{subsec:Canonical} for every spins $s$ there exists a conserved current of rank $2s+k$ that is built with $k$ derivatives of the Weyl tensor $C_{s}$ for $\Phi_{s}$, for the example the Bel-Robinson tensor. Since Weyl tensors in the vectorial language have a symmetry of the two-row rectangular Young diagram, the expressions for the currents are simpler in the spinorial language, where we find
\begin{align*}
\compj&= \sum_{s,k,j}a_{s,l} \frac{i (-1)^k}{j!(k-j)!l!l!(2s+k-j)!(2s+j)!} C_{\ga(2s+k-j)\nu(l),\gad(k-j)\gnd(l)}C\fdudu{\ga(j)}{\nu(l)}{,\gad(2s+j)}{\gnd(l)} y^{\ga(2s+k)}\bry^{\gad(2s+k)}
\end{align*}
and the canonical projection gives \eqref{canonicalFrCurrents}, which results in
\begin{align}
a_{s,l}&=2 \cos 2\theta\frac{ S (2 l (S-1)+S (2 S-1))}{8 (S-1) (l+S)^2 (l+S+1)^2}\,,
\end{align}
where $S=2s+k$ is the total spin of the current. In principle, it is easy to give the full coefficients of the canonical projection by Taylor expanding \eqref{canonicalFrCurrents}. Other components correspond to various "super-currents" built of fields with different spins.

\subsection{Details on Canonical Sector}\label{app:genfuncdetails}
For one-forms there is only one function that upon applying $\adD$ produces just the canonical structure above and it is given by:
\begin{align}
\formJ^{\text{1form}}_{can}&=-\oint_{\tau_i,s,r} (\beta\gamma)\, g^{\pl\bar\pl}(\alpha_1,\alpha_1,\beta,\gamma)h^{\ga\gad}\pl_\ga\pl_\gad e^{i(s y\zeta^-+\tau_1\xi\eta+r \bry\brzeta^++\tau_2\brxi\breta)}\,.%
\end{align}
Other tensor structures are not relevant to study improvements in the canonical current sector.
We can then obtain the action of $\adD$ on the above ansatz in closed form:
\begin{align}\label{Dgdd}
\adD g^{\pl\bar\pl}=\begin{pmatrix}
\frac1{4 \alpha_1 \beta }\,\left[\gamma  (\alpha_1 \gamma +\beta ) \partial_\gamma +\alpha_2 \beta  (\alpha_1 \alpha_2+1) \partial_{\alpha_2}\right]\\
\frac{1}{4 \alpha_2 \gamma }\left[\beta  (\alpha_2 \beta -\gamma ) \pl_\beta-\alpha_1 \gamma  (\alpha_1 \alpha_2+1) \pl_{\alpha_1}\right]
\end{pmatrix}g^{\pl\bar\pl}\,,
\end{align}
and
\begin{multline}\label{DgddB}
\adD\vec{g}^{\,\pl\pl}=-\frac{1}{2 \alpha_1 \alpha_2}\Big[\alpha_1 (\alpha_1 \alpha_2+1) (\alpha_1\partial_{\alpha_1})g^{\pl\pl}_1+\alpha_1 \gamma^{-1}  (\gamma -\alpha_2 \beta ) (\beta\partial_{\beta}) g^{\pl\pl}_1+\alpha_1 g^{\pl\pl}_1\\+\alpha_2 (\alpha_1 \alpha_2+1) (\alpha_2\partial_{\alpha_2})g^{\pl\pl}_2+\alpha_2\beta^{-1}  (\alpha_1 \gamma +\beta )(\gamma \partial_{\gamma}) g^{\pl\pl}_2+\alpha_2 g^{\pl\pl}_2\Big]\,.
\end{multline}
One can explicitly check that the above representation squares to zero. Furthermore, it is easy to compute the source to the Fronsdal tensor sitting in $R_{+1}$. One can check that the above produces conserved currents.
For later convenience we give below the operator that computes the Fronsdal current on the canonical sector:
\begin{align}
J_{+1}^{\text{Fr.}}&=+\frac{(\beta\gamma)^{s+1}}{s-1}\,\alpha_2^{-1}\left[\alpha_1 (\alpha_1 \alpha_2+1) \partial_{\alpha_1}k_1(\alpha_1,\alpha_2)+(s+1) k_1(\alpha_1,\alpha_2)+\alpha_2 (s-1) k_2(\alpha_1,\alpha_2)\right]\,,
\end{align}
where $k_1$ and $k_2$ are certain components of the backreaction $J_0$ and $J_{+1}$ defined as:
\begin{align}
\vec{J}_{0}&=(\beta\gamma)^{s}\begin{pmatrix}
k_1(\alpha_1,\alpha_2)\\
\cdots
\end{pmatrix}\,,
&\vec{J}_{+1}&=(\beta\gamma)^{s}\left(\frac{\beta}{\gamma}\right)\begin{pmatrix}
\cdots\\
k_2(\alpha_1,\alpha_2)
\end{pmatrix}\,.
\end{align}
and $\cdots$ correspond to the components that have no effect no the result. Note that prefactor $(\beta\gamma)^{s+1}$, when expanded, gives canonical currents involving fields of different spins.

\subsection{Generating Function of the Fronsdal Currents}\label{app:Components}
The expressions for the full Fronsdal current are quite involved. Instead, let us present the generating function
for the canonical projection, which follows from \eqref{FronsdalCurrentsDef} and \eqref{HSstCanonical},
\begin{align}\label{canonicalFrCurrents}
j=\compJ^{\pl\pl}+\frac{1}{\bar N+1}f\,,
\end{align}
where $\compJ^{\pl\pl}$ is the corresponding component of the canonical backreaction \eqref{canformcan} and $f$ is:
\begin{align}
&f=-\frac{i}{4}
\begin{pmatrix}
\Big[\frac{1}{4} (1-\vartheta  (3 t-1))-\frac{1}{2} i q t (\xi\eta)+\frac{1}{4} i (\brxi\breta) (-4 q t+\vartheta  (t-1)+3)+\frac{1}{4} (\brxi\breta) (\xi\eta) (q t-1)\Big]{Q}_1\\
+\Big[\frac{1}{2} (\vartheta  (t-1)+1)\left(1-\frac{i (\xi\eta)}{2 q t}+\frac{i}{2} (y-\eta)(y+\xi)\right)\Big]\bar Q_1
\end{pmatrix}\,.\notag
\end{align}
Recall that $\vartheta=\delta(q-1)$.

\section{Resummation of Fronsdal Currents}\setcounter{equation}{0}\label{app:Resummation}

Let us consider the spin-two case and decompose all terms with contracted derivatives, i.e. successors, into their primary component and improvements, reducing thereby a pseudo-local canonical current to a local one without any contracted derivative plus an improvement. In the spin-two case the one-form canonical ansatz from which we can build the most general exact two-form via \eqref{Dgdd} reads:
\begin{align}\label{exact}
\alpha_1\alpha_2\left[(\beta\gamma)^2 k(\alpha_1\alpha_2)+\alpha_2\beta^3\gamma k_2(\alpha_1\alpha_2)+\alpha_1\beta\gamma^3 k_3(\alpha_1\alpha_2)\right]\,.
\end{align}
Note that $k_2$ and $k_3$ do not affect the Fronsdal current. The most general form of the spin-two improvement in the canonical sector that can be obtained by solving torsion from \eqref{exact} via \eqref {DgddB} reads:
\begin{align}
\delta \compj\sim\frac{(\beta\gamma)^2}{4} \left[\tau  (\tau +1) \left(\tau  (\tau +1) k''(\tau )+(3 \tau +7) k'(\tau )\right)+(\tau +3)^2 k(\tau )\right]\,,
\end{align}
where $\tau=(\alpha_1\alpha_2)$ and $\tau^{l}$ corresponds to $\square^{l-1}$. The redefinition that removes higher-powers of $\Box$ from the source to the Fronsdal tensor can then be found by requiring that each single $\Box$ term is removed independently by a \emph{local} redefinition. This boils down to fixing the coefficient $C_l$ by requiring the particular solutions of the following ODE:
\begin{align}
\tau  (\tau +1) \left[\tau  (\tau +1) k''(\tau )+(3 \tau +7) k'(\tau )\right]+(\tau +3)^2 k(\tau )=C_l \tau +\alpha-\tau ^l\,,
\end{align}
to be polynomial.
Using standard techniques in ODE we can set to zero the leading singularity of the solution given by a dilog. Furthermore we can show that it is always possible to find $C_l$ and\footnote{Notice that the choice of $\alpha$ does not affect the solution since this term vanishes upon performing the contour integration.} $\alpha$ such that the corresponding solution is polynomial. The corresponding value of $C_l$, that corresponds to the projection of a term of the type $\Box^l$ in the current to its primary part, is given by:
\begin{align}
C_{l}=\frac{1}{12} (-1)^{l} (l+1) (l+2)^2 (l+3)\,.
\end{align}
Combining the above result with the explicit coefficient detailed above, we conclude that the 4d theory produces results that are divergent upon resummation and require regularization. The corresponding divergent coefficient of the primary current reads:
\begin{align}
\compj_{s=2}=-\frac{i}{12} \cos(2 \theta)\left(\sum_{l=1}^\infty l\right)\frac{\bry\brzeta^+{}^2 y\zeta^-{}^2}{4}+\text{improvements}\,,
\end{align}
where the improvements can be removed by a field redefinition making the current local.

\end{appendix}

\ifpdf%
{\begingroup%
\linespread{1}\selectfont%
\setlength{\emergencystretch}{8em}%
\printbibliography%
\endgroup%
}%
\else%
\bibliographystyle{JHEP-2}%
\bibliography{megabib}%
\fi%
\end{document}